\documentclass[sigconf]{acmart}
\AtBeginDocument{%
  }

\usepackage{subcaption}

\usepackage{rotating}

\usepackage{multirow}
\usepackage{siunitx}
\usepackage{calc}

\newcolumntype{Y}{>{\raggedright\arraybackslash}X}
\newcolumntype{L}[1]{>{\raggedright\arraybackslash\hsize=#1\hsize}X}

\usepackage{textcase} %

\usepackage{xcolor}         %
\usepackage{booktabs}
\usepackage{tabularx}

\copyrightyear{2026}
\acmYear{2026}
\setcopyright{cc}
\setcctype{by}
\acmConference[IUI '26]{31st International Conference on Intelligent User Interfaces}{March 23--26, 2026}{Paphos, Cyprus}
\acmBooktitle{31st International Conference on Intelligent User Interfaces (IUI '26), March 23--26, 2026, Paphos, Cyprus}
\acmDOI{10.1145/3742413.3789149}
\acmISBN{979-8-4007-1984-4/2026/03}

\begin{document}

\title{Adaptive Prompt Elicitation for Text-to-Image Generation}

\author{Xinyi Wen}
\orcid{0009-0006-3814-5039}

\affiliation{%
  \institution{Aalto University \& \\
  University of Helsinki \& \\
  ELLIS Institute Finland}
  \city{Helsinki}
  \country{Finland}
}
\email{xinyi.wen@aalto.fi}

\author{Lena Hegemann}
\orcid{0000-0001-9000-7916}
\affiliation{%
  \institution{Aalto University}
  \city{Helsinki}
  \country{Finland}
}
\email{lena.hegemann@aalto.fi}

\author{Xiaofu Jin}
\orcid{0000-0002-7239-3769}
\affiliation{%
  \institution{Aalto University}
  \city{Helsinki}
  \country{Finland}
}
\email{xjinao@connect.ust.hk}

\author{Shuai Ma}
\orcid{0000-0002-7658-292X}
\affiliation{%
  \institution{Aalto University}
  \city{Helsinki}
  \country{Finland}
}
\email{mashuai@iscas.ac.cn}

\author{Antti Oulasvirta}
\orcid{0000-0002-2498-7837}
\affiliation{%
  \institution{Aalto University \& \\
  ELLIS Institute Finland}
  \city{Helsinki}
  \country{Finland}
}
\email{antti.oulasvirta@aalto.fi}

\renewcommand{\shortauthors}{Wen et al.}

\begin{abstract}
Aligning text-to-image generation with user intent remains challenging, as users frequently provide ambiguous inputs and struggle with model idiosyncrasies. We propose \textsc{Adaptive Prompt Elicitation} (APE), a technique that adaptively poses visual queries to help users refine prompts without extensive writing. Our technical contribution is a formulation of interactive intent inference under an information-theoretic framework. APE represents latent user intent as interpretable feature requirements using language model priors, adaptively generates visual queries, and compiles elicited requirements into effective prompts. Evaluation on IDEA-Bench and DesignBench shows that APE achieves stronger alignment with improved efficiency. A user study with 128 participants on user-defined tasks demonstrates 19.8\% higher perceived alignment without increased workload. Our work contributes a principled approach to prompting that offers an effective and efficient complement to the prevailing prompt-based interaction paradigm with text-to-image models.
\end{abstract}



\begin{teaserfigure}
  \includegraphics[width=\textwidth]{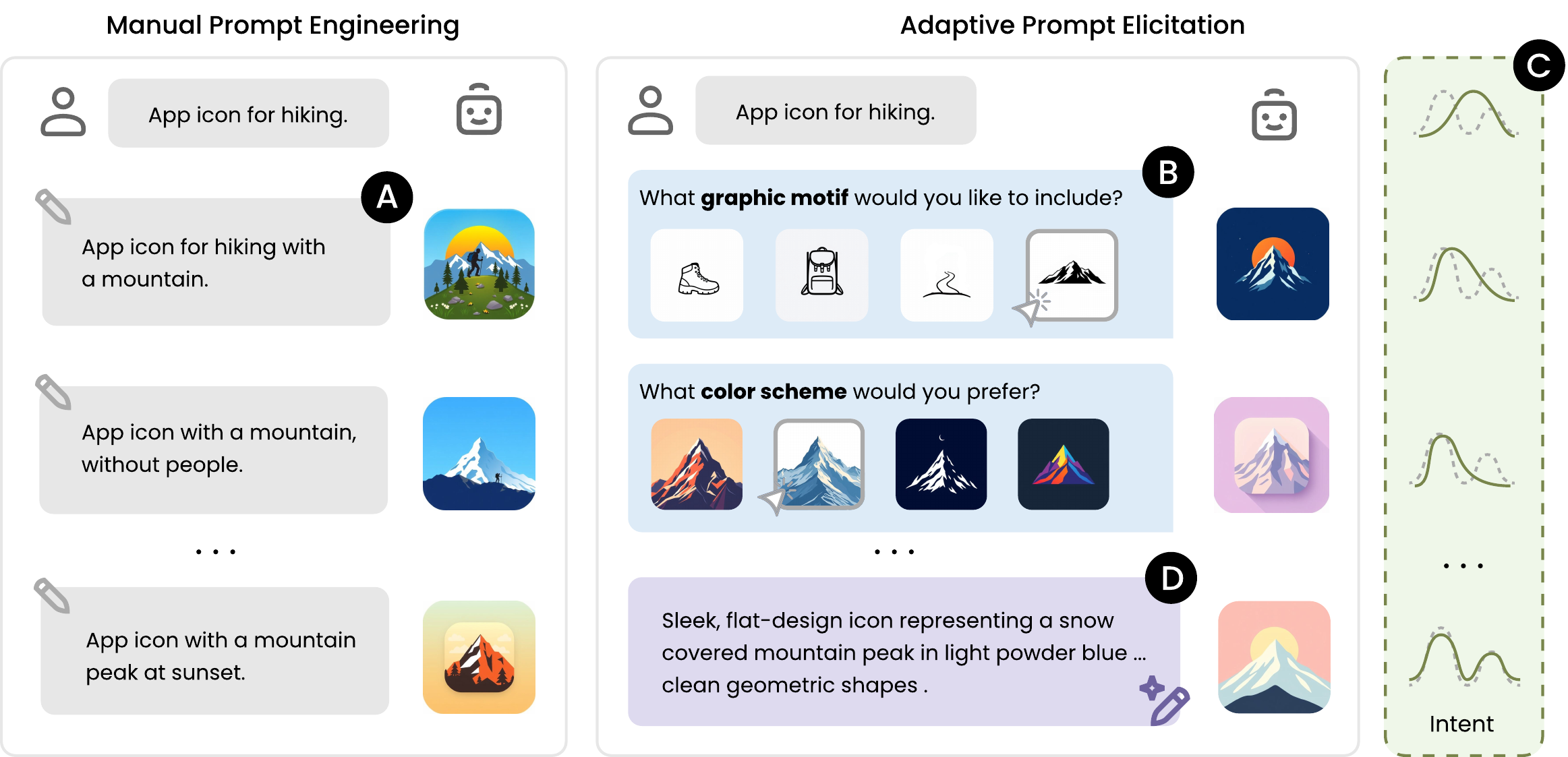}
  \caption{Manual prompt engineering (A) requires users to iterate extensively. \textsc{Adaptive Prompt Elicitation} (APE) addresses this by (B) presenting visual queries, (C) updating its belief about user intent (solid curve) to align with the user's latent intent (dotted curve) based on responses, and (D) generating optimized prompts from its beliefs.}
  \label{fig:teaser}
  \Description{Side-by-side comparison of two prompting approaches. Left (Manual Prompt Engineering): A user iterates through three text prompts with increasing detail ("App icon for hiking", "App icon for hiking with a mountain", "App icon with a mountain, without people"), generating different results each time. Right (Adaptive Prompt Elicitation): Shows (B) visual queries presenting image options for graphic motif and color scheme choices; (C) a graph depicting system belief (solid curve) updating toward user's latent intent (dotted curve) based on responses; (D) the final optimized detailed prompt generated from accumulated beliefs, producing a refined app icon with mountain and hiking theme.}
\end{teaserfigure}

\maketitle

\section{Introduction}

Text-to-image generation models have achieved remarkable advances in synthesizing imagery from natural language descriptions~\cite{rombach2022high, saharia2022photorealistic, tian2024Visual}, however, \textit{aligning model outputs with user intent} remains an outstanding challenge. This stems from the difficulty of articulating rich visual concepts precisely through text~\cite{mahdavigoloujeh2024It, palani2024Evolving, sanchez2023Examining} besides inherent limitations in model capacity. This problem is particularly pronounced for users who lack familiarity with prompt engineering conventions and the characteristics of the specific model ~\cite{liu2022Design, zamfirescu2023johnny,xie2023Prompt}. Even experienced users struggle to anticipate how specific linguistic choices influence generation outcomes, leading to iterative trial-and-error cycles that are time-consuming and cognitively taxing~\cite{oppenlaender2024Taxonomy, wang2023DiffusionDB}.

Our work converges on two lines of prior research. One line aims to guide users to write better prompts, via keyword-rich descriptions, style modifiers, artist references, or quality tokens~\cite{liu2022Design, oppenlaender2024Taxonomy}. These approaches, however, require familiarity with technical vocabulary or interaction styles that may not be accessible to non-expert users ~\cite{zamfirescu2023johnny}. A second line focuses on developing automatic prompt optimization methods that translate user input into high-quality prompts without requiring manual refinement. These methods implicitly assume that user intentions are clearly articulated, may leave users to refine the prompt iteratively through trial and error~\cite{mahdavigoloujeh2024It, rezwana2022designing}.

In this paper, we introduce \textsc{Adaptive Prompt Elicitation} (APE), a computational method that inverts the traditional prompting paradigm by actively eliciting user's latent intent and compiling it into effective prompts. From the user's perspective, APE acts like a human interviewer experienced in the domain of interest and asks guiding questions that direct the conversation: ``Would you like this or that more?''. As shown in Figure~\ref{fig:teaser}, after receiving an initial prompt that sets the theme for interaction, APE presents visual queries that help users progressively clarify their intent. These queries require only visual judgments rather than linguistic articulation, reducing reliance on domain knowledge and linguistic ability. Users respond through simple selections, and APE synthesizes these responses into refined prompts that capture their clarified intent. 
APE is designed for scenarios where users have latent visual preferences but struggle to articulate them precisely through text, a documented challenge in text-to-image generative interaction~\cite{mahdavigoloujeh2024It, han2025Understanding}.

Technically, to bridge the gap between user intent and the model, we establish an explicit set of interpretable visual features as a partial observation of users' latent intent, operationalized through language model priors that map natural language to structured feature requirements. APE maintains a belief distribution over possible intents and selects visual queries grounded in an information-theoretic framework ~\cite{chaloner1995bayesian, rainforth2023modern}, providing a powerful model-based framework for choosing designs automatically. Specifically, it ensures that each query maximally reduces misalignment about intent while respecting constraints on user interaction time. By grounding the elicitation process in a coherent probabilistic framework, APE adaptively tailors queries to individual users and tasks, achieving well-aligned images within a few iterations.

Our complementary evaluations provide convergent evidence for APE's effectiveness and efficiency. On two text-to-image generation benchmarks, IDEA-Bench~\cite{liang2025IDEABench} and DesignBench~\cite{hahn2025Proactive}, APE achieves stronger alignment between generated images and ground-truth intents compared to baseline prompting approaches, while requiring significantly fewer interaction steps. A user study with 128 participants with varying experience levels further validates these findings in ecologically valid settings with self-defined creative tasks. Participants reported a 19.8\% increase in perceived alignment between generated outputs and their intention compared to manual prompting. Qualitative analysis reveals that APE's interactive elicitation process helps users discover and articulate preferences they could not initially verbalize, suggesting that the approach not only elicits existing intent but facilitates intent formation through guided exploration~\cite{louie2020novice, rezwana2022designing}.

Our primary contributions are as follows:
\begin{itemize}
    \item We present a novel adaptive prompt elicitation method that formalizes prompt optimization as interactive intent inference over interpretable feature spaces under an information-theoretic framework, integrating principled query generation with visual interaction design for text-to-image generation.
    \item We demonstrate through systematic evaluations that APE achieves superior alignment with improved efficiency across both controlled benchmarks and user-defined tasks, showing that visual query interactions can serve as an effective and accessible complement to text-based prompting.
\end{itemize}

Our work demonstrates that reframing prompting as bi-directional communication, where systems actively elicit preferences rather than passively interpret text, can address alignment challenges that model improvements alone fall short of. By combining information-theoretic query selection with transparent, editable representations, APE makes text-to-image generation more accessible to general users. 
Source code is available at \url{https://github.com/e-wxy/Adaptive-Prompt-Elicitation}.

\section{Related Work}

Our work draws upon research across three primary areas: prompt optimization for text-to-image generation, interactive techniques in text-to-image systems, and preference learning from human feedback.

\subsection{Prompt Engineering for Text-to-Image Generation}

Early research established foundational principles for prompt design. Liu et al.~\cite{liu2022Design} provided comprehensive guidelines on prompt keywords and model hyperparameters, while Oppenlaender~\cite{oppenlaender2024Taxonomy} systematically categorized prompt modifiers through an ethnographic study, identifying six key types: subject terms, style modifiers, image prompts, quality boosters, repeating terms, and magic terms. These guidelines consistently suggest that more detailed and specific prompts tend to produce higher-quality images.

Complementing this line of work, automatic prompt optimization techniques have been proposed to help users write effective prompts without extensive learning. A prominent approach involves training specialized language models as prompt optimizers. Promptist~\cite{hao2023Optimizing} and BeautifulPrompt~\cite{cao2023BeautifulPrompt} combine supervised learning on curated preference data with reinforcement learning from AI feedback to improve semantic alignment and aesthetic quality. Prompt Expansion~\cite{datta2024Prompt} and UF-FGTG~\cite{hei2024UserFriendly} train deep generative models to transform input prompts into high-quality, detailed descriptions using large-scale prompt expansion datasets, thereby enhancing the aesthetic quality and diversity of generated images. More recently, large language models have been explored as black-box prompt optimizers~\cite{liu2024Language, he2025automated, oscar2024improving, chen2024Tailored}, leveraging their natural language understanding capabilities to refine user inputs.

While manual prompt design offers more precise control and user agency, it suffers from steep learning curves and strong model-specific knowledge requirements, limiting scalability across users and use cases \cite{han2025Understanding, mahdavigoloujeh2024It}. Automatic prompt optimization reduces the need for user expertise but often overlooks the fundamental challenge of enabling users to express their intentions precisely, as well as the personalized and nuanced nature of user preferences \cite{mahdavigoloujeh2024It, hahn2025Proactive, wang2025Twin}.

 \subsection{Interaction Techniques in Text-to-Image Generation}

Interaction techniques have emerged as a critical middle ground between manual prompt engineering and fully automatic optimization. By integrating human feedback into the generation loop, these techniques seek to bridge the gap between abstract user intent and concrete model outputs. We categorize prior work into three primary paradigms: interface-driven exploration, mixed-initiative control, and query-based ambiguity resolution.

To mitigate the trial-and-error nature of prompting, several systems provide structured interfaces for exploring the prompt space. Promptify~\cite{brade2023Promptify}, PromptMagician~\cite{feng2024PromptMagician}, and PrompTHis~\cite{guo2024PrompTHis} support iterative refinement through prompt suggestions and history visualization. PromptMap~\cite{adamkiewicz2025PromptMap} visualizes large collections of prompts as semantic maps, helping users navigate high-dimensional design spaces by clustering similar aesthetic outcomes. ThematicPlane~\cite{lee2025ThematicPlane} enables users to navigate and manipulate high-level semantic concepts within an interactive thematic design plane. While these tools reduce the memory burden of tracking iterations, they still rely on users to manually evaluate and synthesize multiple prompt candidates, which can become cognitively overwhelming as the search space grows.

Moving beyond simple text entry, recent work has introduced mixed-initiative systems that combine automated optimization with fine-grained user control. PromptCharm~\cite{wang2024PromptCharm} exemplifies this approach by integrating an automated prompt optimizer (Promptist~\cite{hao2023Optimizing}) with explainable AI features, such as visualizing token-level attention to help users understand how specific words influence the generated image and enable targeted refinement. Other systems expand the interaction modalities: PromptPaint~\cite{chung2023PromptPaint} and SketchFlex~\cite{lin2025SketchFlex} allow users to constrain generation through spatial sketches, while AdaptiveSliders~\cite{jain2025AdaptiveSliders} enables continuous adjustment of semantic concepts. StyleFactory~\cite{zhou24StyleFactory} facilitates style alignment by offering an interactive interface for granular style-strength adjustment and systematic evaluation. Brickify~\cite{shi2025Brickify} supports expressive design intent specification through direct manipulation of design tokens. Although these methods offer expressive control, they often raise the barrier to entry, as interpreting attention maps or specifying spatial constraints may require technical expertise or artistic skills that novice users may lack.

Most closely related to our work, Proactive Agent~\cite{hahn2025Proactive} parses user prompts into belief graphs to identify underspecified attributes or entities, and generates clarifying questions to fill these gaps. 
Twin-Co~\cite{wang2025Twin} uses image captioning to detect discrepancies between user descriptions and generated outputs, then asks targeted questions to resolve mismatches.
These pioneering works establish query-based interaction as a powerful paradigm for eliciting intent. However, query efficiency and the articulation burden of textual interaction remain underexplored.
APE addresses these limitations by grounding query generation in information-theoretic principles and shifting from textual to visual interaction, enabling more efficient and accessible preference elicitation with potential to generalize to more complex professional design tasks.

\subsection{Preference Learning from Human Feedback}
 
Preference learning from human feedback has emerged as a powerful paradigm for aligning generative models with user intent. 
Reinforcement Learning from Human Feedback (RLHF) enables the training of models using comparative human feedback without requiring task-specific reward engineering~\cite{christiano2017Deep}. 
Optimization methods such as Proximal Policy Optimization (PPO)~\cite{schulman2017Proximal, ouyang2022training} and Direct Preference Optimization (DPO)~\cite{rafailov2023direct} have been proposed to align language models with human preferences. 
Recent extensions of RLHF have been applied to image generation, where large-scale preference signals are leveraged to train text-to-image models, leading to improvements in aesthetic quality, prompt alignment, and visual fidelity~\cite{black2024training, xu2023ImageReward, lee2023aligning}.

Beyond offline population-level training, Interactive Machine Learning (IML)~\cite{fails2003interactive} facilitates real-time adaptation to user feedback. 
Given the high cost of human attention, sample efficiency is paramount.
Information-theoretic frameworks, such as Bayesian experimental design, provide principled foundations for making optimal decisions in this context~\cite{chaloner1995bayesian, rainforth2023modern, huang2024Amortized, son2022BIGexplore}. 
Such principles underpin active learning strategies for acquiring labeled data~\cite{houlsby2011bayesian, kirsch2019batchbald, gal2017deep, ji2025reinforcement, ma2019smarteye}, and manifest specifically as active preference learning, which focuses on efficiently eliciting user preferences through strategic query selection. Related ideas have been widely explored within Preferential Bayesian Optimization (PBO)~\cite{gonzalez2017Preferential}, which aims to find the optimum of black-box functions based on preference feedback, and applied to assist users in various design tasks, including visualization, photography, and typography~\cite{koyama2022BO, yamamoto2022Photographic, tatsukawa2025FontCraft, koyama2020sequential, brochu2010interactive}. BayesGenie~\cite{cai2025Bayesian} further integrates Bayesian Optimization with large language models (LLMs) to optimize image editing parameters based on user feedback. However, these methods typically operate over predefined continuous parameter spaces, limiting their applicability to open-ended tasks where user requirements are heterogeneous, not known a priori, and difficult to parametrize numerically.

More recently, LLM-based preference elicitation methods have emerged that use language models to learn user preferences through natural language dialogue~\cite{li2025eliciting, kobalczyk2025active, mahmud2025MAPLEa}, enabling more flexible design spaces and a more natural alignment with human intent.

Building upon these insights, this work extends active preference learning to text-to-image generation by grounding information-theoretic query selection in interpretable visual feature spaces. Unlike prior methods that operate over fixed low-dimensional parameter spaces~\cite{cai2025Bayesian}, APE uses dynamically generated visual queries to reduce articulation burden while maintaining principled optimization of intent alignment.

\section{Adaptive Prompt Elicitation}

This section presents \textsc{Adaptive Prompt Elicitation} (APE), our interactive method for aligning text-to-image generation with user intent through strategic visual queries. We first illustrate the system through a walkthrough scenario, then formalize the alignment problem and describe our technical approach.

\subsection{System Walkthrough}

\begin{figure*}[htb]
\centering
\includegraphics[width=\linewidth]{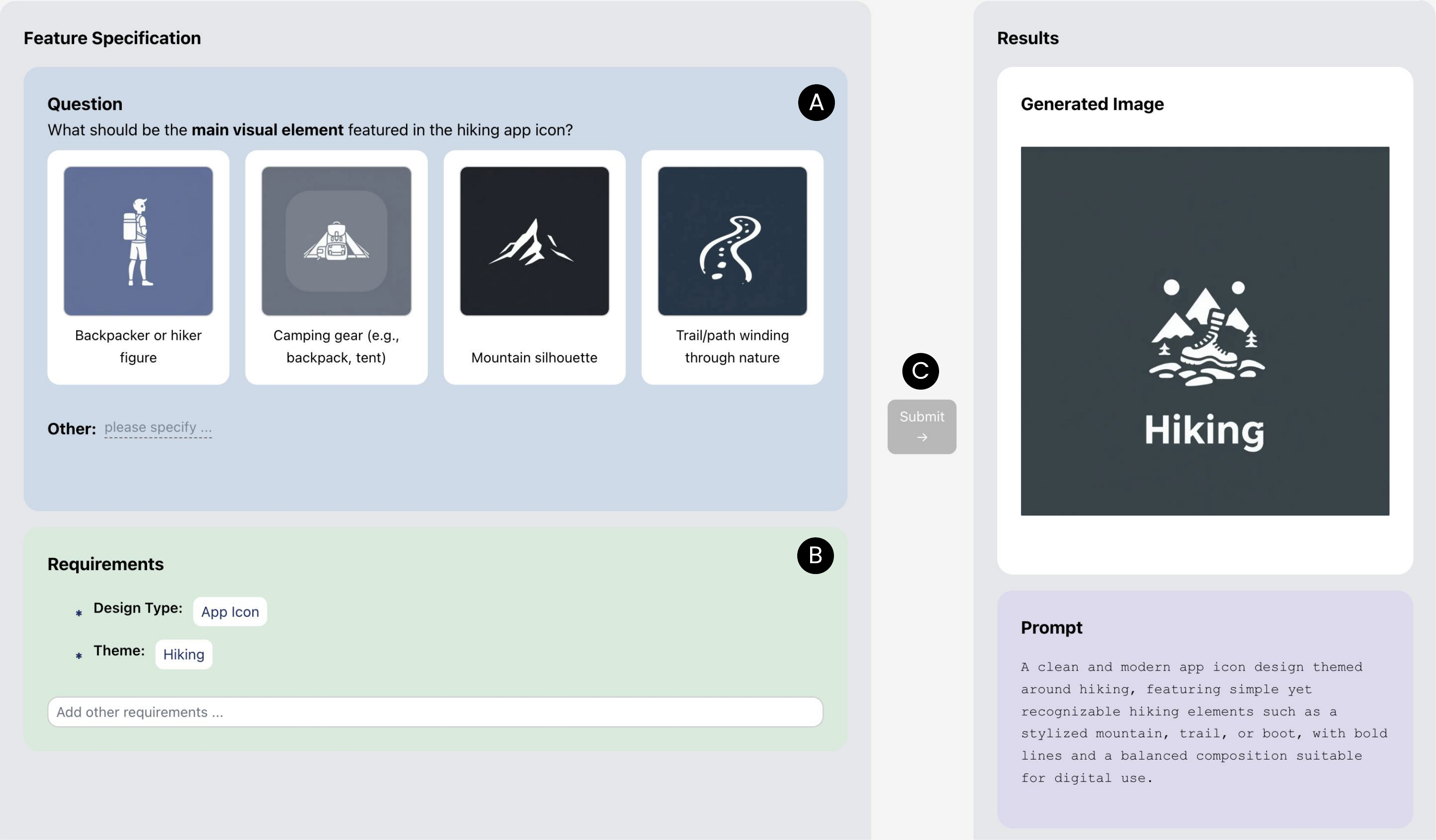}
\caption{User interface design. The left area functions as feature specification, consisting of (A) question panel that presents visual queries, and (B) requirements panel that presents visual feature requirements specified during the interactions. (C) The user can generate an image by clicking the ``Submit'' button, which translates the requirements into effective prompts to generate an image.}
\label{fig:walkthrough}
\Description{User interface screenshot with three labeled sections. (A) Question Panel: Displays "What should be the main visual element?" with four image options (backpacker figure, camping gear, mountain silhouette, winding trail) plus "Other" text input. (B) Requirements Panel: Shows accumulated specifications as tags ("Design Type: App Icon", "Theme: Hiking") with input for additional requirements. (C) Submit button to generate an image. The right side shows the generated dark blue hiking app icon with a mountain graphic and the corresponding detailed prompt text.}
\end{figure*}

We introduce APE through a realistic scenario where Sarah, a hiking enthusiast with no prompt engineering experience, wants to create an app icon for her club's upcoming events.
Sarah begins with a simple, underspecified prompt: \textit{``an app icon for hiking.''} 
Based on this initial input, the system presents a visual query — ``What should be the main visual element?'' — accompanied by several image options, including mountains, backpacks, boots, and trails (Figure~\ref{fig:walkthrough}A). Sarah selects the mountain option, which best aligns with her design intent. APE then updates its interpretation of her intent and generates the next visual query accordingly. If none of the options are satisfactory, Sarah can specify her preferences using the ``Other'' option.

Each selection incrementally refines Sarah's design intent, which is continuously reflected and updated in the requirements panel (Figure~\ref{fig:walkthrough}B). In addition to responding to system-generated visual queries, Sarah can directly add or modify the textual requirements in this panel to further articulate her desired outcome. Once the requirements are finalized, she clicks ``Submit'' (Figure~\ref{fig:walkthrough}C) to generate an image. This process can be repeated iteratively until she achieves a satisfactory result.

\subsection{Problem Formulation}

Let $\pi$ represent a text-to-image model that maps prompts $x$ to distributions over images $y$. Each user has a latent visual intent $\theta^*$ that characterizes their preferences across visual features like style, composition, mood, and subject matter. The goal is to find a prompt $x^*$ that generates images maximally aligned with the user's true intent:
\begin{equation}
\label{eq:alignment}
x^* = \arg\max_{x \in \mathcal{X}} \mathbb{E}_{y \sim \pi(\cdot | x)} [U_{\theta^*}(y)]
\end{equation}
where $U_{\theta^*}$ is the alignment utility function parameterized by $\theta^*$. This formulation assumes access to the user's true intent $\theta^*$, which is unknown in practice.

In manual prompt engineering, users attempt to learn a mapping $f_\pi: \Theta \rightarrow \mathcal{X}$ that translates latent intent $\theta^*$ into effective prompts, despite limited knowledge of the model $\pi$ and the difficulty of expressing abstract visual concepts in natural language. This results in inefficient trial-and-error search in a high-dimensional prompt space.

Automatic prompt optimization instead learns a transformation $g_\pi: \mathcal{X} \rightarrow \mathcal{X}$ that refines an initial prompt. While effective at capturing population-level preferences through curated data, these methods may fail to adapt to individual users and assume that the initial prompt sufficiently reflects the user's latent intent.

Both paradigms suffer from complementary forms of information asymmetry: users do not know how the model interprets prompts, while the system does not know the user's true intent. We address these limitations via a bi-directional interactive optimization, which jointly optimizes prompts while adaptively eliciting $\theta^*$ through user interaction, bridging the gap between what users envision and what models produce.

\subsection{System Overview}

\begin{figure*}[htb]
\centering
\includegraphics[width=0.8\linewidth]{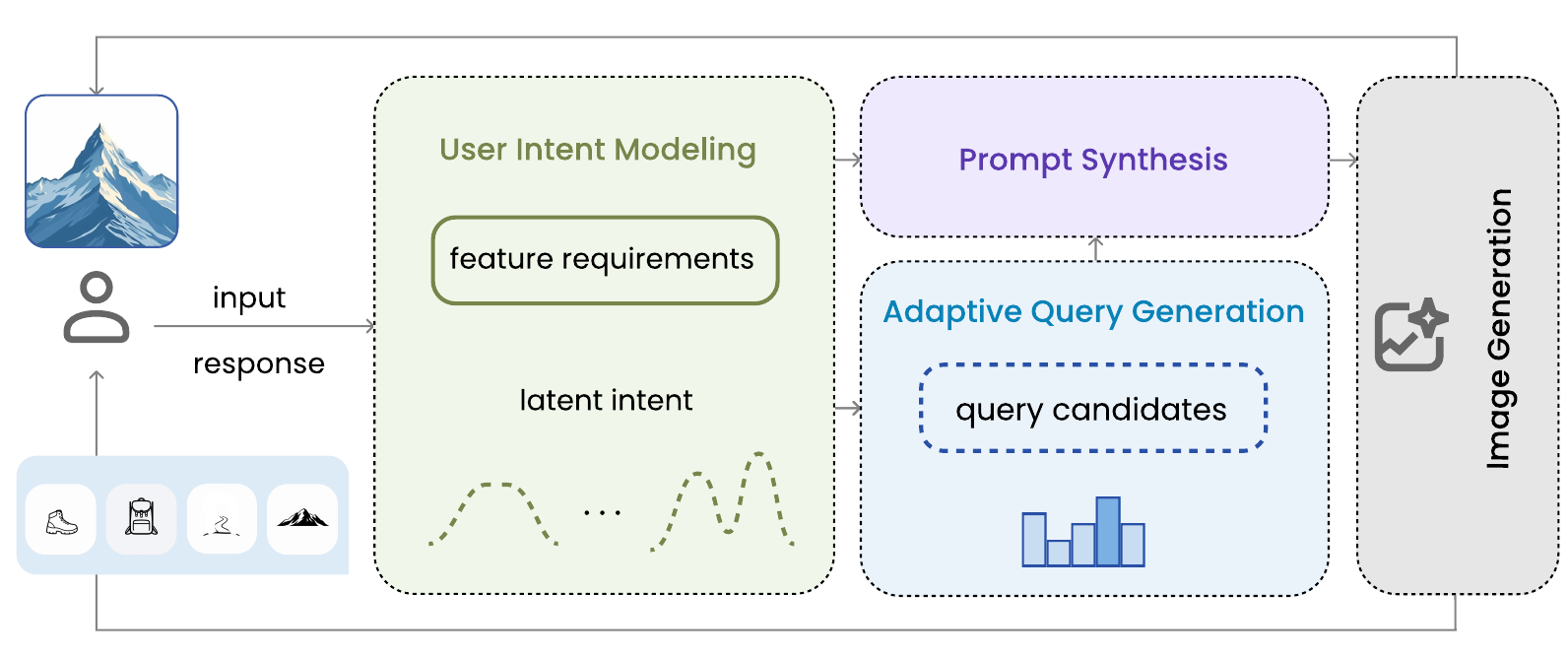}
\caption{APE's computational pipeline. The system maintains a user intent model over visual features, adaptively generates visual queries to infer the user's intent, and then synthesizes optimized prompts for text-to-image generation models.}
\label{fig:method_overview}
\Description{System architecture diagram showing APE's three-component pipeline. On the left, a user icon provides input and responses. The central area contains three interconnected modules: (1) User Intent Modeling (green box) maintains feature requirements and models latent intent as probability distributions shown by curved lines; (2) Adaptive Query Generation (blue box) maintains query candidates and uses information-theoretic selection shown by a bar chart; (3) Prompt Synthesis (purple box at top) translates requirements into prompts. Arrows indicate flow: user input enters User Intent Modeling, which connects bidirectionally with Adaptive Query Generation; queries return to the user for responses; final prompts flow to Image Generation (right side). The architecture implements an iterative loop where user responses progressively refine the intent model, generate more targeted queries, and ultimately produce aligned prompts.}
  
\end{figure*}

APE consists of three main components that work in concert to elicit user intent and generate optimized prompts (Figure~\ref{fig:method_overview}): (1) \textbf{User Intent Modeling}, which represents and updates beliefs about user preferences; (2) \textbf{Adaptive Query Generation}, which selects informative visual queries based on current uncertainty; and (3) \textbf{Prompt Synthesis}, which translates elicited intent into effective prompts. We describe each component in detail below.

\subsection{User Intent Modeling}

The intent modeling component maintains and refines APE's understanding of the user's latent visual intent $\theta^*$ as responses are collected. We parameterize the user's intent through visual features. Let $\mathcal{V}$ denote the space of relevant visual features (e.g., "artistic style," "lighting," "color palette"), where each feature $v \in \mathcal{V}$ has an associated value domain (e.g., "impressionist" or "photorealistic" for style). The user's true intent $\theta^*$ specifies desired values for each feature. Rather than committing to a single hypothesis in the full feature space, APE represents uncertainty through a partial specification $R_t$ and models plausible completions using language model priors.

\paragraph{Partial Intent Specification.} At iteration $t$, APE maintains a structured specification $R_t = \{(v_i, s_i)\}$ consisting of feature-value pairs. Each pair $(v_i, s_i)$ indicates that visual feature $v_i$ (e.g., ``artistic style'', ``lighting'') should take value $s_i$ (e.g., ``watercolor'', ``golden hour''). Initially, $R_0$ summarizes features explicitly stated in the user input. As users respond to queries, $R_t$ grows incrementally: $R_{t+1} = R_t \cup \{(v_{q_t}, a_t)\}$ where $v_{q_t}$ is the queried feature and $a_t$ is the user's selected value. This representation serves as a transparent interface between user and model, allowing the system to update its belief adaptively. Additionally, it provides an interpretable trace of elicited preferences that users can review and modify if needed.

\paragraph{Dynamic Feature Space.} To better adapt to the user's needs, instead of fixing the feature taxonomy a priori, APE's feature space $\mathcal{V}_t$ evolves dynamically based on context. Specifically, we utilize the LLM's in-context learning ability to propose relevant features conditioned on the current specification $R_t$. For example, if $R_t$ contains (subject, ``portrait''), the LLM might propose features like ``facial expression'', ``gaze direction'', or ``background style'' that are semantically relevant to portraits but not to landscape scenes. This adaptive approach ensures queries remain contextually appropriate and generalizable to various tasks. We prompt the LLM with $R_t$ and request: ``Given these specified visual features, what other important features would help clarify the user's vision?'' The LLM returns a list of candidate features $\mathcal{V}_t = \{v_1, v_2, \ldots, v_n\}$, which forms the basis for query generation.

\paragraph{LLM Persona for Uncertainty Modeling.} When $R_t$ is underspecified, many complete intents $\theta$ remain consistent with the specified features. APE reasons about this uncertainty by modeling the distribution over plausible complete intents through the \textit{LLM persona}, denoted $p_{\mathcal{L}}(\theta \mid R_t)$. The LLM persona captures prior knowledge of visual intent based on the LLM's training on diverse visual descriptions.

\paragraph{Handling User Response.} When the user selects value $a_t$ for queried feature $v_{q_t}$, we update the specification to $R_{t+1} = R_t \cup \{(v_{q_t}, a_t)\}$ and update the posterior of user intent via $p_{\mathcal{L}}(\cdot \mid R_{t+1})$. This resampling incorporates the user's feedback, causing the LLM to generate samples consistent with both previous and newly specified features. Through iterative refinement, the sampled intents progressively converge toward the user's true latent intent $\theta^*$.

\subsection{Adaptive Query Generation}

The query generation component determines which visual feature to query next. Although a straightforward approach that enumerates all features would yield complete knowledge of user intent, it risks expending user effort on irrelevant or low-impact dimensions. We formulate query generation as a decision-making problem within an information-theoretic framework~\cite{chaloner1995bayesian, rainforth2023modern, huang2024Amortized}, with the objective of proposing queries that are expected to maximally increase alignment utility.

\paragraph{Candidate Query Generation.} At iteration $t$, we first generate a set of candidate queries $\mathcal{Q}_t$ from the dynamic feature space $\mathcal{V}_t$. For each candidate feature $v \in \mathcal{V}_t$ that has not yet been specified in $R_t$, we generate diverse value options that partition the feature's domain.
To ensure complete coverage when finite options cannot exhaustively represent all possibilities, we include a residual option (``Other'') that allows users to indicate their preference lies outside the presented choices. If selected, this triggers a follow-up free-form query where users can specify their desired value in natural language.

\paragraph{Adaptive Query Selection.} For each candidate query $q$ about feature $v_q$ with options $\{s_1, \ldots, s_m\}$, we define the Expected Alignment Utility Gain (EAUG) as:
\begin{equation}
\begin{aligned}
\text{EAUG}(q; R_t)
&= \alpha_{v_q} \cdot H[p(\cdot \mid v_q, R_t)] \\
&= -\alpha_{v_q} \sum_{i=1}^{m} p(s_i \mid v_q, R_t)
   \log p(s_i \mid v_q, R_t)
\end{aligned}
\label{eq:eaug}
\end{equation}
where $H[p(\cdot \mid v_q, R_t)]$ denotes the Shannon entropy quantifying our uncertainty about which value the user prefers for feature $v_q$, which collapses to zero upon observing a definitive response, assuming user feedback is sufficiently reliable and can be calibrated. The weighting factor $\alpha_{v_q}$ captures the extent to which feature $v_q$ influences overall alignment. For computational tractability, we approximate these quantities using the LLM persona samples $\{\theta^{(k)}\}_{k=1}^K$ drawn from the user intent model.

APE selects the query that maximizes the improvement in alignment quality:
\begin{equation}
q_t^* = \arg\max_{q \in \mathcal{Q}_t} \; \text{EAUG}(q; R_t)
\label{eq:query_selection}
\end{equation}
This strategy prioritizes queries about features that both have uncertain outcomes and substantially impact final alignment quality.

\paragraph{Visual Query Presentation.} Once query $q_t^* = (v_{q_t}, \{s_1, \ldots, s_m\})$ is selected, APE generates visual exemplars to present each option. For each value option $s_i$, we:
\begin{enumerate}
    \item Construct a prompt that emphasizes the target feature-value pair: $x_i = f( \{(v_{q_t}, s_i)\} \mid R_t)$, where $f$ is a prompt synthesis function as described in Section~\ref{sec:prompt}
    \item Generate an exemplar image: $y_i \sim \pi(\cdot \mid x_i)$ using the base text-to-image generation model
    \item Present images $\{y_1, \ldots, y_m\}$ side-by-side with descriptive labels
\end{enumerate}

To emphasize the queried feature, we use a fixed random seed and identical generation parameters across all options, varying only the feature-specific prompt component. This controlled setup isolates the effect of the target feature while keeping other visual aspects consistent.

The query interface presents a simple multiple-choice question: ``Which [feature name] best matches your vision?'' followed by the labeled exemplar images. Users select their preferred option through a single click, instead of articulating preferences textually. This leverages recognition rather than recall, substantially reducing the articulation burden.

\subsection{Prompt Synthesis}
\label{sec:prompt}

The prompt synthesis component translates the elicited intent specification $R_T$ into an optimized natural language prompt. We formalize synthesis as a function $f: \mathcal{R} \to \mathcal{X}$ that maps explicit intents to text prompts. 
\begin{equation}
x^* = f(R_T) = \mathcal{L}(m, c_\pi, R_T)
\label{eq:synthesis}
\end{equation}
where $\mathcal{L}$ is a large language model, $m$ is a meta-prompt that guides synthesis, and $c_\pi$ is the context regarding the targeted image generation model. The meta-prompt $m$ orchestrates synthesis in two stages: We first ask $\mathcal{L}$ to generate prompt engineering guidelines specific to the target model $\pi$. These elicited guidelines $g$ encode model-specific best practices. Using $g$ as context, we instruct $\mathcal{L}$ to synthesize a prompt incorporating all features from $R_T$.

This adaptive design ensures synthesis is both faithful to user intent and optimized for model-specific conventions. By explicitly eliciting guidelines, we make the LLM's prompt engineering knowledge transparent and consistent.

We now validate its effectiveness through systematic evaluation. The following two sections assess APE's performance through complementary lenses: controlled technical evaluation with ground-truth benchmarks (Section~\ref{sec:technical_eval}) establishes whether the approach achieves its core objectives of improving alignment and efficiency, while subsequent user studies (Section~\ref{sec:user_study}) examine how these technical capabilities translate to real-world user experience.

\section{Technical Evaluation}
\label{sec:technical_eval}

To establish APE's technical capabilities, we conducted a controlled evaluation on benchmarked tasks with ground-truth references. This evaluation addresses fundamental questions about APE's core mechanisms: whether adaptive query elicitation improves alignment quality, and how efficiently information-theoretic query selection converges compared to alternative strategies. We employed LLM-based user simulators that responded to visual queries according to predefined targets, enabling us to (1) evaluate at scale across diverse tasks and (2) systematically compare strategies under identical conditions.

Our evaluation addresses two research questions:
\begin{itemize}
\item \textbf{RQ1}: Does APE improve alignment between user intent and generated images compared to existing prompt optimization approaches?
\item \textbf{RQ2}: How efficiently does APE converge to high-quality prompts, and how does information-theoretic query selection compare to alternative strategies?
\end{itemize}

We evaluate these questions across two complementary benchmarks: DesignBench for users with visual preferences but limited articulation ability, and IDEA-Bench for users with clear conceptual goals requiring translation into effective prompts. Together, these benchmarks provide comprehensive coverage of text-to-image use cases and establish APE's technical foundations before examining real-world user experience in Section~\ref{sec:user_study}.

\subsection{Method}

\subsubsection{Datasets}

We evaluated APE on two complementary benchmarks that represent different user scenarios and task complexities:

\begin{itemize}
    \item \textbf{DesignBench}~\cite{hahn2025Proactive}: comprises 30 diverse cases for daily usage, including hand-drawn cartoons, surreal photorealistic scenes, and artistic photographs. Each case provides an initial prompt, a target image, and a detailed description. %
    \item \textbf{IDEA-Bench}~\cite{liang2025IDEABench}: encompasses comprehensive professional design and art tasks for image generation, distilled from real-world design cases across various platforms. From the complete dataset containing both text-to-image and image-to-image tasks, we selected 29 text-to-image generation cases spanning architectural style, business cards, sculptures, tickets, landmarks, logos, posters, and interior design. Each case includes a concise task description and a ground truth text prompt. We generated reference images using the ground truth prompts with FLUX.1~\cite{labs2025FLUX1a} (5 images per case to establish robust references). %
\end{itemize}

Together, these benchmarks cover the spectrum from intuitive visual preferences to structured design requirements, providing comprehensive coverage of text-to-image use cases.

\subsubsection{User Simulation}

To enable systematic evaluation at scale while maintaining experimental control, we developed LLM-based user simulators that represent different user profiles:

\begin{itemize}
    \item \textbf{Vision-Based Simulator (DesignBench).} We employed an LLM as a multi-modal user simulator that receives the target image as its ``unarticulated vision.'' The simulator responds to visual queries by analyzing the reference image, mimicking users who know what they want when they see it but cannot describe it initially. This approach simulates the common scenario where users specify their preferences through visual feedback.
    \item \textbf{Intent-Based Simulator (IDEA-Bench).} For professional design tasks, we used an LLM simulator that adopts the ground truth prompt as its internal ``creative brief.'' This configuration represents users with well-defined conceptual goals who need assistance translating intent into effective model input.
\end{itemize}

To ensure response quality and prevent hallucination, we designed simulation prompts using Chain-of-Thought reasoning \cite{wei2022Chainofthought}, requiring simulators to analyze options before selection. To prevent leaking the ground truth to APE, only the question answer updates the feature specification $R_t$, with explanatory reasoning discarded. 

While simulated users cannot fully replicate human preference formation or capture the complexity of human perception, they provide essential advantages for technical evaluation: perfect experimental control, reproducibility across conditions, elimination of learning effects, and the ability to test at scale across diverse scenarios. These controlled conditions allow us to isolate APE's technical contributions.

\subsubsection{Baselines}

We compare APE against three conditions:

\begin{itemize}
    \item \textbf{Unoptimized}: The initial prompts, establishing the starting point and demonstrating overall improvement.
    \item \textbf{Automatic Prompt Optimization (APO)}: A non-interactive automatic optimization approach that uses language models to refine initial prompts provided by the user~\cite{hao2023Optimizing}. This isolates the value of interactive elicitation.
    \item \textbf{In-Context Query}: An interactive baseline that generates queries based on LLM in-context reasoning without principled information-theoretic selection. This isolates the contribution of our adaptive query selection strategy.
\end{itemize}

\subsubsection{Evaluation Metrics}

We employed multi-modal alignment metrics to comprehensively assess different aspects of generation quality:

\textbf{Image-Image Similarity} measures visual fidelity between generated and reference images using DreamSim~\cite{fu2023DreamSim}, a perceptual similarity metric trained on human judgments, and DINOv2~\cite{oquab2024DINOv2}, a self-supervised vision transformer capturing semantic visual features. These metrics assess whether the generated images visually resemble the intended target.

\textbf{Text-Text Similarity} evaluates prompt quality by measuring semantic similarity between synthesized prompts and ground truth descriptions. We used E5~\cite{wang2024multilingual}, a state-of-the-art text embedding model with strong cross-task transfer, and OpenAI's text-embedding-3-large~\cite{2025Vector}. High text-text similarity indicates that APE successfully elicits and encodes the intended semantic content.

\textbf{Text-Image Similarity} assesses cross-modal consistency between generated images and textual descriptions using VQAScore~\cite{lin2024Evaluating}, which leverages visual question answering models to compute alignment by measuring the probability of affirmative responses to ``Does this figure show [text]?'' This metric captures whether the visual output matches the linguistic specification.

This multi-metric approach captures complementary aspects of alignment: visual fidelity (image-image), semantic correctness (text-text), and cross-modal consistency (text-image).

\subsubsection{Apparatus}

Our evaluation employed the following configuration: We used the FLUX.1-schnell text-to-image model~\cite{labs2025FLUX1a} accessed via the Fal API\footnote{https://fal.ai/, accessed on Oct 10, 2025} with eight inference steps, providing a balance between generation quality and computational efficiency. The user simulator is based on GPT-4o-mini~\cite{2025Vector}, accessed through the OpenAI API. Each simulation proceeded for a maximum of 15 iterations, ensuring sufficient exploration while maintaining realistic interaction limits. In each round, up to five candidate queries were generated, each with a maximum of five options, offering diverse yet cognitively manageable choices. Finally, we performed five independent experimental runs per case to establish reliable confidence intervals.

\subsection{Improved Alignment Quality (RQ1)}

\begin{table*}[tbh]
\centering
\caption{Alignment quality on DesignBench and IDEA-Bench (mean $\pm$ 95\% CI). Best results per row bolded.}
\label{tab:tech_results}
\begin{tabularx}{\linewidth}{p{5em} p{6em} p{7em} *{4}{>{\centering\arraybackslash}X}}
\toprule
\textbf{Dataset} & \textbf{Modality} & \textbf{Metric$\uparrow$} & \textbf{Unoptimized} & \textbf{APO} & \textbf{In-Context} & \textbf{APE (Ours)} \\
\midrule
\multirow{5}{*}{\rotatebox[origin=c]{90}{DesignBench}}
& \multirow{2}{*}{Text-Text} 
& \small{E5} & 0.887\,\textcolor{gray}{\scriptsize$\pm$\,0.05} & 0.894\,\textcolor{gray}{\scriptsize$\pm$\,0.05} & 0.898\,\textcolor{gray}{\scriptsize$\pm$\,0.02} & \textbf{0.899\,\textcolor{gray}{\scriptsize$\pm$\,0.02}} \\
&& \small{OpenAI} & 0.732\,\textcolor{gray}{\scriptsize$\pm$\,0.12} & 0.789\,\textcolor{gray}{\scriptsize$\pm$\,0.10} & \textbf{0.811\,\textcolor{gray}{\scriptsize$\pm$\,0.04}} & 0.797\,\textcolor{gray}{\scriptsize$\pm$\,0.04} \\
\cmidrule{2-7}
& \multirow{2}{*}{Image-Image} 
& \small{DINOv2} & 0.664\,\textcolor{gray}{\scriptsize$\pm$\,0.32} & 0.668\,\textcolor{gray}{\scriptsize$\pm$\,0.33} & 0.676\,\textcolor{gray}{\scriptsize$\pm$\,0.15} & \textbf{0.679\,\textcolor{gray}{\scriptsize$\pm$\,0.16}} \\
&& \small{DreamSim} & 0.639\,\textcolor{gray}{\scriptsize$\pm$\,0.16} & 0.633\,\textcolor{gray}{\scriptsize$\pm$\,0.15} & 0.636\,\textcolor{gray}{\scriptsize$\pm$\,0.08} & \textbf{0.654\,\textcolor{gray}{\scriptsize$\pm$\,0.08}} \\
\cmidrule{2-7}
& Text-Image & \small{VQAScore} & 0.891\,\textcolor{gray}{\scriptsize$\pm$\,0.14} & 0.890\,\textcolor{gray}{\scriptsize$\pm$\,0.15} & 0.901\,\textcolor{gray}{\scriptsize$\pm$\,0.04} & \textbf{0.904\,\textcolor{gray}{\scriptsize$\pm$\,0.03}} \\
\midrule
\multirow{5}{*}{\rotatebox[origin=c]{90}{IDEA-Bench}}
& \multirow{2}{*}{Text-Text} 
& \small{E5} & 0.888\,\textcolor{gray}{\scriptsize$\pm$\,0.06} & 0.886\,\textcolor{gray}{\scriptsize$\pm$\,0.04} & \textbf{0.903\,\textcolor{gray}{\scriptsize$\pm$\,0.01}} & 0.901\,\textcolor{gray}{\scriptsize$\pm$\,0.02} \\
&& \small{OpenAI} & 0.785\,\textcolor{gray}{\scriptsize$\pm$\,0.09} & 0.758\,\textcolor{gray}{\scriptsize$\pm$\,0.12} & 0.788\,\textcolor{gray}{\scriptsize$\pm$\,0.04} & \textbf{0.792\,\textcolor{gray}{\scriptsize$\pm$\,0.04}} \\
\cmidrule{2-7}
& \multirow{2}{*}{Image-Image} 
& \small{DINOv2} & 0.471\,\textcolor{gray}{\scriptsize$\pm$\,0.32} & 0.513\,\textcolor{gray}{\scriptsize$\pm$\,0.30} & 0.596\,\textcolor{gray}{\scriptsize$\pm$\,0.14} & \textbf{0.613\,\textcolor{gray}{\scriptsize$\pm$\,0.14}} \\
&& \small{DreamSim} & 0.554\,\textcolor{gray}{\scriptsize$\pm$\,0.21} & 0.557\,\textcolor{gray}{\scriptsize$\pm$\,0.18} & 0.611\,\textcolor{gray}{\scriptsize$\pm$\,0.09} & \textbf{0.621\,\textcolor{gray}{\scriptsize$\pm$\,0.09}} \\
\cmidrule{2-7}
& Text-Image & \small{VQAScore} & 0.675\,\textcolor{gray}{\scriptsize$\pm$\,0.23} & 0.675\,\textcolor{gray}{\scriptsize$\pm$\,0.23} & 0.674\,\textcolor{gray}{\scriptsize$\pm$\,0.11} & \textbf{0.678\,\textcolor{gray}{\scriptsize$\pm$\,0.10}} \\
\bottomrule
\end{tabularx}
\end{table*}

\subsubsection{Overall Performance}
Table~\ref{tab:tech_results} presents comprehensive alignment results across both datasets. 
APE achieves the best performance in 8 of 10 metric-dataset combinations, demonstrating consistent improvements across diverse scenarios and evaluation modalities. This multi-metric consistency suggests that APE effectively bridges the gap between user intent and model interpretation.

For IDEA-Bench's challenging professional design scenarios, APE demonstrates substantial gains over the baseline. Image similarity metrics show marked improvements (DINOv2: +30.2\%, DreamSim: +12.1\%), indicating that interactive elicitation provides considerable value for complex creative tasks requiring domain-specific vocabulary and nuanced specifications. These gains are particularly pronounced compared to the more modest improvements observed on DesignBench, suggesting that APE's benefits scale with task complexity.

Comparing APE and In-Context Query against APO reveals the critical value of bidirectional communication. APO shows minimal improvement over baseline across both datasets, highlighting a fundamental challenge of non-interactive approaches: discovering user preferences that remain unexpressed in the initial prompt. Even sophisticated language models operating on user input alone lack sufficient signal to infer latent visual intentions, underscoring the necessity of explicit preference elicitation. 

APE's advantages over In-Context Query demonstrate the practical benefits of principled query generation. While both methods engage users interactively, APE's information-theoretic approach systematically identifies more informative and impactful features, translating to measurably better alignment outcomes. This advantage appears across multiple metrics and task types, validating the theoretical foundation of our query selection strategy.

\subsubsection{Task-Specific Performance}

\begin{table*}[htb]
\centering
\caption{Task-specific alignment quality on IDEA-Bench across design categories
(mean $\pm$ 95\% CI). Best results per row bolded.}
\label{tab:task_specific}
\begin{tabularx}{\linewidth}{l l *{4}{>{\centering\arraybackslash}X}}
\toprule
\textbf{Design Type} & \textbf{Metric$\uparrow$} & \textbf{Unoptimized} & \textbf{APO} & \textbf{In-Context} & \textbf{APE (Ours)} \\
\midrule

\multirow{2}{*}{Architectural Style}
& \small{DINOv2}   & 0.432\,\textcolor{gray}{\scriptsize$\pm$\,0.07} & 0.481\,\textcolor{gray}{\scriptsize$\pm$\,0.08} & 0.503\,\textcolor{gray}{\scriptsize$\pm$\,0.10} & \textbf{0.560\,\textcolor{gray}{\scriptsize$\pm$\,0.06}} \\
& \small{DreamSim} & 0.586\,\textcolor{gray}{\scriptsize$\pm$\,0.04} & 0.576\,\textcolor{gray}{\scriptsize$\pm$\,0.01} & 0.520\,\textcolor{gray}{\scriptsize$\pm$\,0.03} & \textbf{0.598\,\textcolor{gray}{\scriptsize$\pm$\,0.02}} \\

\cmidrule{1-6}
\multirow{2}{*}{Business Card}
& \small{DINOv2}   & 0.474\,\textcolor{gray}{\scriptsize$\pm$\,0.09} & 0.453\,\textcolor{gray}{\scriptsize$\pm$\,0.07} & \textbf{0.577\,\textcolor{gray}{\scriptsize$\pm$\,0.09}} & 0.499\,\textcolor{gray}{\scriptsize$\pm$\,0.11} \\
& \small{DreamSim} & 0.477\,\textcolor{gray}{\scriptsize$\pm$\,0.05} & 0.471\,\textcolor{gray}{\scriptsize$\pm$\,0.04} & \textbf{0.527\,\textcolor{gray}{\scriptsize$\pm$\,0.03}} & 0.508\,\textcolor{gray}{\scriptsize$\pm$\,0.02} \\

\cmidrule{1-6}
\multirow{2}{*}{Interior Design}
& \small{DINOv2}   & 0.362\,\textcolor{gray}{\scriptsize$\pm$\,0.04} & 0.619\,\textcolor{gray}{\scriptsize$\pm$\,0.08} & 0.622\,\textcolor{gray}{\scriptsize$\pm$\,0.05} & \textbf{0.671\,\textcolor{gray}{\scriptsize$\pm$\,0.07}} \\
& \small{DreamSim} & 0.576\,\textcolor{gray}{\scriptsize$\pm$\,0.04} & 0.634\,\textcolor{gray}{\scriptsize$\pm$\,0.03} & 0.684\,\textcolor{gray}{\scriptsize$\pm$\,0.01} & \textbf{0.707\,\textcolor{gray}{\scriptsize$\pm$\,0.02}} \\

\cmidrule{1-6}
\multirow{2}{*}{Painting}
& \small{DINOv2}   & 0.386\,\textcolor{gray}{\scriptsize$\pm$\,0.06} & 0.372\,\textcolor{gray}{\scriptsize$\pm$\,0.06} & 0.451\,\textcolor{gray}{\scriptsize$\pm$\,0.03} & \textbf{0.456\,\textcolor{gray}{\scriptsize$\pm$\,0.05}} \\
& \small{DreamSim} & 0.555\,\textcolor{gray}{\scriptsize$\pm$\,0.06} & 0.554\,\textcolor{gray}{\scriptsize$\pm$\,0.07} & 0.577\,\textcolor{gray}{\scriptsize$\pm$\,0.05} & \textbf{0.601\,\textcolor{gray}{\scriptsize$\pm$\,0.05}} \\

\cmidrule{1-6}
\multirow{2}{*}{Sculpture}
& \small{DINOv2}   & 0.601\,\textcolor{gray}{\scriptsize$\pm$\,0.07} & 0.611\,\textcolor{gray}{\scriptsize$\pm$\,0.04} & 0.667\,\textcolor{gray}{\scriptsize$\pm$\,0.04} & \textbf{0.737\,\textcolor{gray}{\scriptsize$\pm$\,0.02}} \\
& \small{DreamSim} & 0.544\,\textcolor{gray}{\scriptsize$\pm$\,0.02} & 0.545\,\textcolor{gray}{\scriptsize$\pm$\,0.03} & \textbf{0.635\,\textcolor{gray}{\scriptsize$\pm$\,0.02}} & 0.630\,\textcolor{gray}{\scriptsize$\pm$\,0.04} \\

\cmidrule{1-6}
\multirow{2}{*}{Ticket Design}
& \small{DINOv2}   & 0.560\,\textcolor{gray}{\scriptsize$\pm$\,0.04} & \textbf{0.561\,\textcolor{gray}{\scriptsize$\pm$\,0.06}} & 0.551\,\textcolor{gray}{\scriptsize$\pm$\,0.05} & 0.545\,\textcolor{gray}{\scriptsize$\pm$\,0.06} \\
& \small{DreamSim} & 0.570\,\textcolor{gray}{\scriptsize$\pm$\,0.03} & 0.580\,\textcolor{gray}{\scriptsize$\pm$\,0.01} & 0.609\,\textcolor{gray}{\scriptsize$\pm$\,0.05} & \textbf{0.621\,\textcolor{gray}{\scriptsize$\pm$\,0.03}} \\

\cmidrule{1-6}
\multirow{2}{*}{Landmark Building}
& \small{DINOv2}   & 0.788\,\textcolor{gray}{\scriptsize$\pm$\,0.04} & 0.813\,\textcolor{gray}{\scriptsize$\pm$\,0.05} & \textbf{0.854\,\textcolor{gray}{\scriptsize$\pm$\,0.01}} & 0.845\,\textcolor{gray}{\scriptsize$\pm$\,0.02} \\
& \small{DreamSim} & 0.756\,\textcolor{gray}{\scriptsize$\pm$\,0.02} & \textbf{0.793\,\textcolor{gray}{\scriptsize$\pm$\,0.02}} & 0.768\,\textcolor{gray}{\scriptsize$\pm$\,0.01} & 0.782\,\textcolor{gray}{\scriptsize$\pm$\,0.03} \\

\cmidrule{1-6}
\multirow{2}{*}{Logo Design}
& \small{DINOv2}   & 0.396\,\textcolor{gray}{\scriptsize$\pm$\,0.03} & 0.567\,\textcolor{gray}{\scriptsize$\pm$\,0.05} & 0.625\,\textcolor{gray}{\scriptsize$\pm$\,0.03} & \textbf{0.631\,\textcolor{gray}{\scriptsize$\pm$\,0.06}} \\
& \small{DreamSim} & 0.480\,\textcolor{gray}{\scriptsize$\pm$\,0.03} & 0.489\,\textcolor{gray}{\scriptsize$\pm$\,0.03} & \textbf{0.574\,\textcolor{gray}{\scriptsize$\pm$\,0.05}} & 0.568\,\textcolor{gray}{\scriptsize$\pm$\,0.08} \\

\cmidrule{1-6}
\multirow{2}{*}{Poster Design}
& \small{DINOv2}   & 0.319\,\textcolor{gray}{\scriptsize$\pm$\,0.06} & 0.465\,\textcolor{gray}{\scriptsize$\pm$\,0.06} & 0.548\,\textcolor{gray}{\scriptsize$\pm$\,0.02} & \textbf{0.586\,\textcolor{gray}{\scriptsize$\pm$\,0.03}} \\
& \small{DreamSim} & 0.545\,\textcolor{gray}{\scriptsize$\pm$\,0.03} & 0.569\,\textcolor{gray}{\scriptsize$\pm$\,0.02} & \textbf{0.605\,\textcolor{gray}{\scriptsize$\pm$\,0.02}} & 0.592\,\textcolor{gray}{\scriptsize$\pm$\,0.03} \\

\bottomrule
\end{tabularx}
\end{table*}

Table~\ref{tab:task_specific} reveals heterogeneous performance across IDEA-Bench's nine design categories, providing insights into how task characteristics interact with our elicitation approach.

Three categories demonstrate substantial gains: Interior Design (DINOv2: +85.4\%, DreamSim: +22.7\% over baseline), Poster Design (DINOv2: +83.7\%, DreamSim: +8.6\%), and Logo Design (DINOv2: +59.3\%, DreamSim: +18.3\%). These tasks require coordinating multiple interdependent visual elements where initial prompts underspecify numerous critical attributes. Architectural Style (DINOv2: +29.6\%, DreamSim: +2.0\%) and Painting (DINOv2: +18.1\%, DreamSim: +8.3\%) show moderate gains with notable metric divergence—substantial DINOv2 improvements coupled with minimal DreamSim changes may reflect limitations in DreamSim's ability to detect fine-grained visual refinements. Landmark Building shows modest gains (DINOv2: +7.2\%, DreamSim: +3.4\%), with APE and In-Context Query performing comparably. We hypothesize this occurs because models possess strong priors for well-documented landmarks from training data, making additional elicitation less critical. Interestingly, In-Context Query (DINOv2: 0.577) slightly outperforms APE (DINOv2: 0.499) in Business Card design. This anomaly suggests that for highly constrained tasks with limited degrees of freedom, strategic query selection may provide minimal advantage over simpler interactive approaches.

It is important to note that image similarity metrics may not fully capture fine-grained professional design quality. For Sculpture, where the underlying text-to-image model is known to struggle with geometrically accurate three-dimensional forms, and Ticket Design, where the model notoriously fails at rendering coherent text~\cite{liang2025IDEABench}, similarity metrics may primarily capture whether the correct object category is present rather than professional design quality such as anatomical accuracy or typography legibility. The observed improvements in these categories, while statistically meaningful, should be interpreted with caution as metric gains may not correspond to perceptually meaningful quality differences for humans.

\subsection{Improved Query Efficiency (RQ2)}

\begin{figure*}[htb]
\centering
\includegraphics[width=\textwidth]{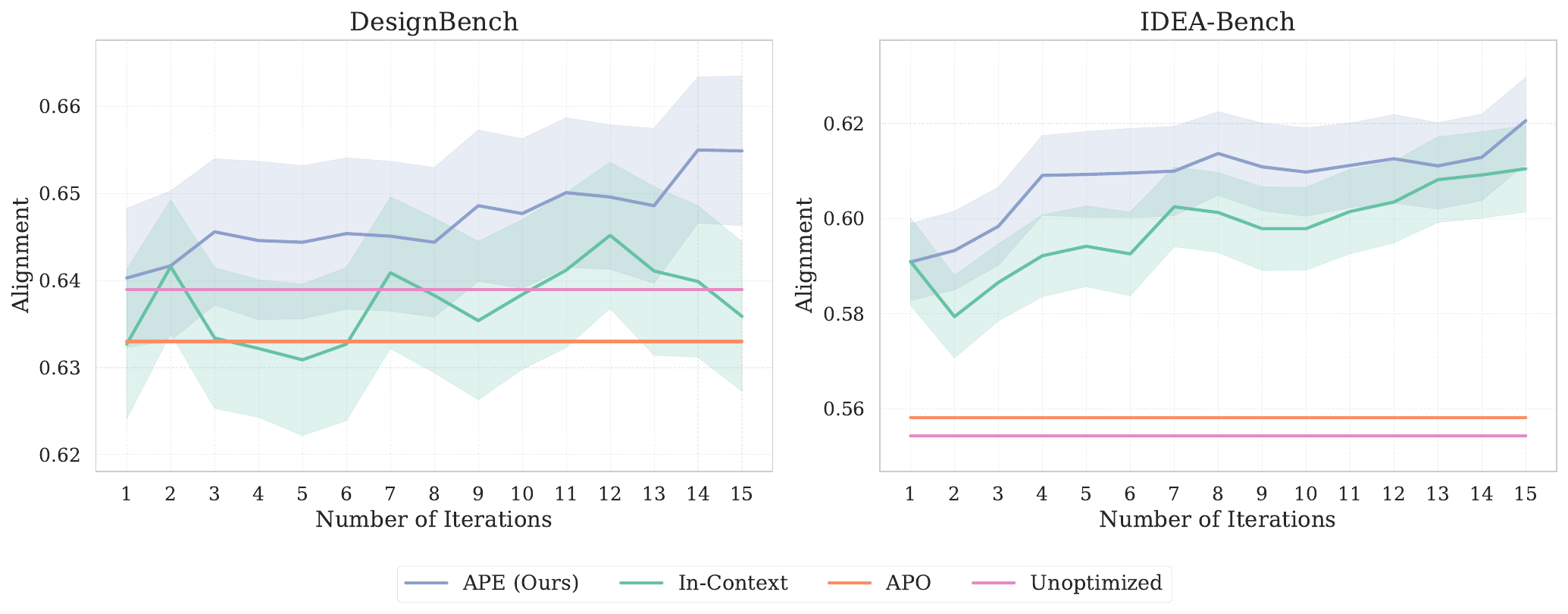}
\caption{The progress of alignment over iterations, evaluated by image-image similarity (DreamSim$\uparrow$) with 95\% CI. APE demonstrates faster convergence and higher final alignment compared to In-Context Query.}
\label{fig:interaction_rounds}
\Description{Two line plots comparing alignment scores (DreamSim similarity, y-axis) across 15 iterations (x-axis) for four methods: APE (blue), In-Context (green), APO (orange), and Unoptimized (red), with 95\% confidence intervals. Left panel (DesignBench): APE reaches 0.655, In-Context 0.65, while APO and Unoptimized plateau around 0.635-0.64. Right panel (IDEA-Bench): More pronounced gains with APE reaching 0.62, In-Context 0.61, while APO and Unoptimized remain flat at 0.56. Both panels show that APE achieves faster convergence and higher final alignment.}
\end{figure*}

Figure~\ref{fig:interaction_rounds} tracks alignment quality evolution across interaction rounds, revealing APE's superior convergence properties. On both datasets, APE reaches higher final alignment scores while requiring fewer interactions, with efficiency gains particularly pronounced in early iterations.
This pattern suggests that APE's principled query strategy identifies high-impact features during this critical early phase, enabling rapid convergence toward user intent. This efficiency advantage is more pronounced on IDEA-Bench, suggesting that professional design tasks with richer feature spaces amplify the value of strategic query selection. APE demonstrates steeper initial improvement followed by gradual refinement, indicating that it effectively prioritizes foundational design decisions—those with broad downstream impact—before addressing fine-grained details.

The observed efficiency gains translate to meaningful practical benefits. Reducing the number of required interactions decreases both user effort and cumulative interaction time. For text-to-image systems where each generation cycle incurs computational cost and user attention, this efficiency improvement represents substantial value in real-world deployment scenarios.

\section{User Evaluation}
\label{sec:user_study}

While the technical evaluation (Section~\ref{sec:technical_eval}) demonstrates APE's performance on controlled benchmarks with ground-truth intents, real-world deployment requires understanding how the system performs with actual users pursuing open-ended creative goals. To address this, we conducted a between-subjects user study with 128 participants recruited from Prolific\footnote{https://www.prolific.com/, accessed on Oct 10, 2025}. Unlike the technical evaluation, where ground truth was available for objective comparison, the user study focuses on subjective alignment quality, cognitive workload, and interaction efficiency as perceived by users with varying levels of prior experience engaging with self-defined creative tasks. 

The study evaluates APE using a prototyped user interface (Figure~\ref{fig:walkthrough}) that reflects how the system would be deployed in practice, and compares it against manual prompt engineering, a baseline condition representing the current standard in which users iteratively refine textual prompts.  This comparison is guided by three research questions:

\begin{itemize}
\item \textbf{RQ3:} How do users perceive the alignment of the generated image with their design intent when using APE compared to manual prompt engineering?
\item \textbf{RQ4:} How does APE affect users' perceived cognitive workload relative to traditional prompting approaches?
\item \textbf{RQ5:} How does APE affect interaction efficiency in terms of iterations required to achieve aligned results?
\end{itemize}

These questions complement the technical evaluation by examining whether APE's computational advantages translate to meaningful improvements in user experience. The following subsections detail our methodology, present quantitative and qualitative findings.

\subsection{Method}
\subsubsection{Participants}

We recruited 128 participants from Prolific meeting the following criteria: at least a 99\% historical approval rate, more than 100 historical submissions, proficiency in English, and having normal or corrected-to-normal vision.
The sample included 48 female, 74 male, and 6 non-binary participants ($M_{age} = 34.4$, $SD = 10.8$).
We did not filter by design expertise, aiming to capture a naturalistic distribution of users who might encounter these tools. This resulted in a diverse sample spanning varied levels of prior experience with LLMs, text-to-image systems, and design activities (see Table~\ref{tab:pre}). 
Each session lasted approximately 25 minutes, and participants received £4 as compensation.

\subsubsection{Task and Apparatus}

Each participant was given a task scenario from a set of eight design types: interior design, painting, photography, app icon design, poster design, logo design, fashion design, and architectural style design. Given the scenario, participants decided their design idea and interacted with either our method or the baseline method to generate an image. The task descriptions are shown in Table~\ref{tab:tasks}.  
We selected these scenarios because they combine commonly used benchmarks in text-to-image generation research with familiar, everyday creative tasks.  
By covering a wide range of contexts—from artistic creation (painting, fashion design) to practical design (logos, app icons, posters) and imaginative visualization (dream interiors, ideal architecture)—we ensured both task diversity and participant familiarity.  
This balance encourages natural engagement and enables us to assess how generative models perform across a broad spectrum of creative objectives.  

The user study was deployed on a web-based interface.  
The server side integrated our APE algorithm, supporting real-time interaction with the text-to-image model.

\subsubsection{Experimental Design}

We compared two conditions:

\begin{itemize}
    \item \textbf{Baseline (Manual Prompt Engineering).}  
    Participants could freely edit prompts and iteratively generate images. They could refine the prompt as many times as they wished within a 15-minute limit.  
    This setup mirrors commonly used text-to-image tools.
    
    \item \textbf{APE (\textsc{Adaptive Prompt Elicitation}).}  
    Participants completed the task using our proposed APE system. Instead of manually editing prompts, the system engaged participants through adaptive, informative questions and automatically translated their elicited requirements into prompts.  
    The interaction time was matched to the Baseline condition (15 minutes).
\end{itemize}

To avoid learning and ordering effects, we adopted a between-subjects design where each participant was assigned one condition and one task. Each task was used equally often per condition.  
We conducted a power analysis using G*Power~\cite{faul2009statistical} for a two-condition between-subjects design.
Assuming a moderate effect size ($f = 0.5$), $\alpha = 0.05$, and statistical power of 0.8, the required sample size was estimated at 128 participants.

\subsubsection{Procedure}

The experiment followed a structured sequence:

\begin{enumerate}
    \item \textbf{Introduction.}  
    After reading an introduction to the study, participants provided informed consent and completed a pre-task survey assessing their prior experience with LLM tools, text-to-image systems, and design-related activities.  
    They then received a short tutorial on the assigned system, followed by a brief comprehension quiz.

    \item \textbf{Generation Task.}  
    Participants were assigned one of the eight generation tasks and were encouraged to think about their desired outcome.  
    They then used the assigned system for up to 15 minutes to generate images, with the option to stop earlier once being satisfied with the result.

    \item \textbf{Post-task Questionnaires.}  
    After completing the task, participants filled out questionnaires assessing the alignment, task workload and effectiveness of questions.
\end{enumerate}

\subsubsection{Measurements}

We measured three aspects: \textbf{alignment}, \textbf{workload} and \textbf{efficiency}.

\paragraph{\textbf{Alignment}}

Since no standardized measure exists for assessing AI-alignment with user intentions in open-ended tasks, we deployed a custom questionnaire that draws conceptually from prior discussion of alignment gaps in these scenarios \cite{subramonyam2024bridging}. The questionnaire focuses on four dimensions: the \textit{capability gap}, the \textit{instruction gap}, and the \textit{intentionality gap}, along with an \textit{overall} assessment of envisioning-support of the system. For each dimension, participants responded to one pair of positively and negatively valenced statements reflecting potential system support or limitations (see Appendix~\ref{sec:alignment_ques}). All items were rated on a 7-point Likert scale (1 = strongly disagree, 7 = strongly agree), and negatively-worded items were reverse-scored before averaging across each dimension.

\paragraph{\textbf{Task workload}} was measured using the NASA Task Load Index (NASA-TLX)~\cite{hart1988development}, a widely validated multidimensional measure of subjective workload.

\paragraph{\textbf{Efficiency}} was measured through:
\begin{itemize}
    \item \textbf{Iteration Count:} The number of iterations required for participants to obtain a satisfactory image.  
    \item \textbf{Useful Question Ratio:} The proportion of visual questions generated by the system that participants rated as useful.
\end{itemize}

\paragraph{\textbf{Open-ended feedback}}
In addition, we collected open-ended, qualitative feedback about participants' experiences.

\subsubsection{Data Analysis}

We employed a mixed-methods approach. For quantitative measures, we used non-parametric Mann–Whitney~U tests \cite{mann1947Test} with Bonferroni correction for multiple comparisons.  
For qualitative feedback, we conducted a thematic analysis~\cite{hsieh2005three}.  
We report key themes supported by representative participant quotes.

\subsection{Perceived Alignment Quality (RQ3)}

\subsubsection{Quantitative Results}

\begin{figure}[htb]
\centering
\includegraphics[width=1\linewidth]{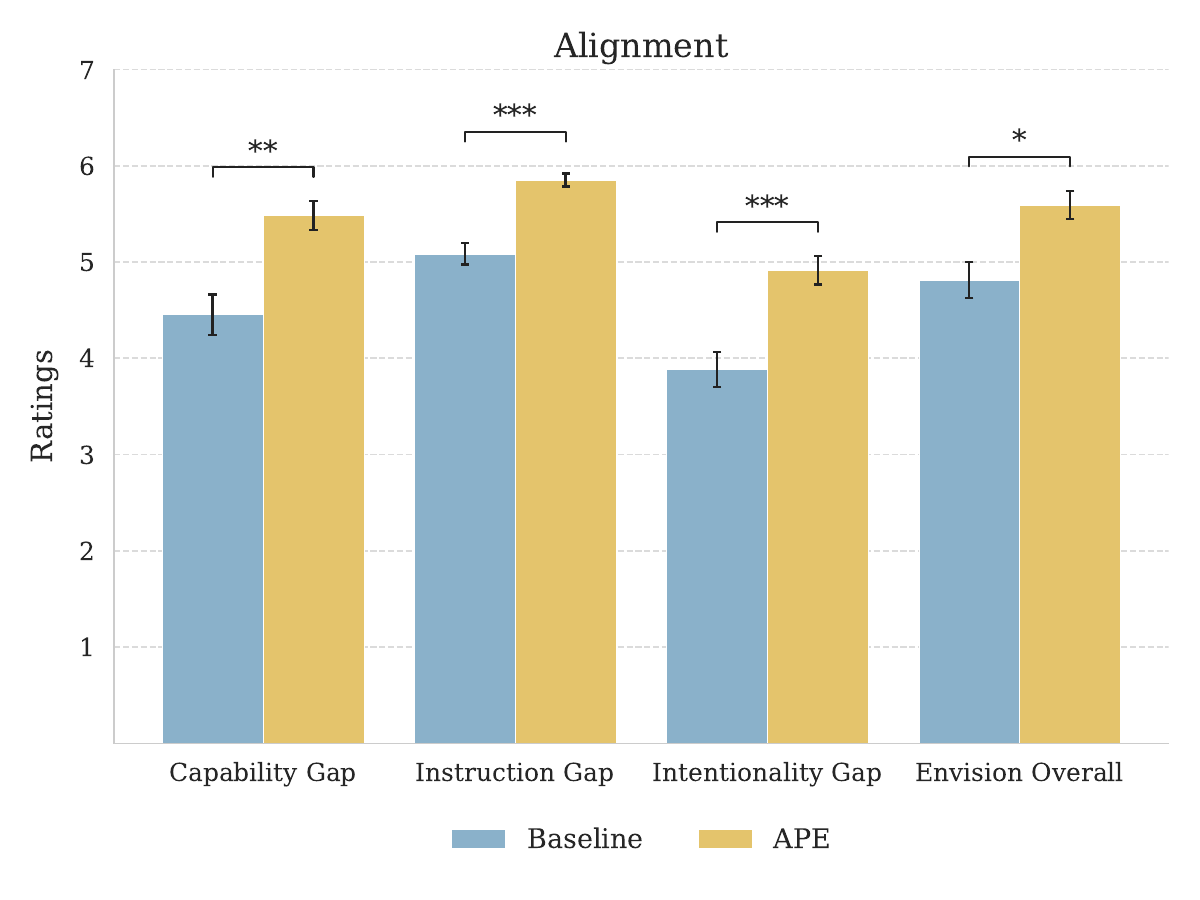}
\caption{Alignment ratings (± SE) for four dimensions across conditions. Participants using APE reported significantly higher scores for all dimensions compared to Baseline. Asterisks indicate significance levels (* $p < 0.05$, ** $p < 0.01$, *** $p < 0.001$).}
\label{fig:alignment}
\Description{Grouped bar chart showing mean alignment ratings (1-7 scale) with standard errors for four dimensions, comparing Baseline (blue) vs APE (yellow). APE shows significantly higher scores: Capability Gap (5.48 vs 4.45, p<0.01), Instruction Gap (5.85 vs 5.09, p<0.001), Intentionality Gap (4.91 vs 3.88, p<0.001), and Envision Overall (5.59 vs 4.81, p<0.05). Largest improvements appear in Instruction and Intentionality gaps, indicating APE better helps users communicate preferences and clarify intentions.}
\end{figure}

Figure~\ref{fig:alignment} presents participant ratings across four dimensions of alignment between their envisioned concepts and the generated images. APE demonstrated statistically significant improvements across all four dimensions (Mann-Whitney U tests, Bonferroni-corrected, all $p < 0.05$). 
For capability gap, participants rated APE significantly higher ($M = 5.48, SE = 0.15$) than Baseline ($M = 4.45, SE = 0.21$; $U = 1309.5, p = 0.0016, r = 0.36$), indicating that APE's interactive elicitation process helped participants feel more capable of achieving their desired outcomes. 
The instruction gap showed the strongest improvement, with APE ($M = 5.85, SE = 0.07$) substantially outperforming Baseline ($M = 5.09, SE = 0.11$; $U = 1033.0, p < 0.0001, r = 0.50$). This suggests that APE effectively helps users communicate preferences they struggle to articulate textually.
The intentionality gap showed marked improvement with APE ($M = 4.91, SE = 0.15$) over Baseline ($M = 3.88, SE = 0.18$; $U = 1205.0, p = 0.0002, r = 0.41$). This gap of over one full rating point suggests that APE's structured elicitation process helps users clarify their own intent, enabling them to evaluate the result better. 
Finally, envision overall ratings confirmed this pattern, with APE ($M = 5.59, SE = 0.15$) significantly exceeding Baseline ($M = 4.81, SE = 0.19$; $U = 1428.5, p = 0.0118, r = 0.30$).

Computing overall alignment as the sum of ratings across all four dimensions, APE achieved significantly higher total scores ($M = 21.84, SE = 0.41$) compared to Baseline ($M = 18.23, SE = 0.58$), corresponding to a 19.8\% improvement.
These results provide convergent evidence that APE addresses a genuine need: helping users bridge the gap between their visual intentions and textual articulation. The consistent improvements across all alignment dimensions, with a medium effect size, suggest the approach successfully reduces both communication barriers and expectation mismatches.

\subsubsection{Qualitative Feedback}
Participants' primary complaints in the Baseline condition centered on the lack of alignment between their intent and the generated outputs, as well as the difficulty of achieving high-quality, fine-grained designs. Many reported that the system failed to fully interpret their prompts even after repeated refinements: ``\textit{I keep adding [to the] prompt, but the generated image is not getting better at understanding the prompt.}'' Participants expressed frustration that the Baseline could not support detailed or precise adjustments: ``\textit{I was only about 65 percent satisfied because even though it made the image I was going for, it wouldn't let me tweak it further to take it to perfection.}'' Additionally, users frequently observed inconsistent adherence to instructions—such as incorrect numbers of people, unstable elements across iterations, or ignored constraints—which further undermined trust in the generation process: ``\textit{It felt like some of the prompts I wrote were ignored. Thus the image I had in mind seemed to be impossible to be created.}''

By contrast, in the APE condition, participants often described stronger alignment with their envisioned outcomes, crediting the elicited visual queries for surfacing overlooked or implicit design details: ``\textit{The questions presented to me were very helpful!}'' and ``\textit{My first prompt wasn't the best, but with the option to make edits, my vision was brought to life.}'' Several noted that the final images closely matched their mental image. However, a minority still reported occasional inconsistencies in low-level details (e.g., colors, small elements) or regressions during refinement, and some wished for greater output variety or finer-grained control after elicitation. This residual inconsistency, similar to what was observed in the Baseline condition, may stem from inherent limitations of the current text-to-image models rather than the elicitation approach itself.

\subsubsection{The Effect of Prior Experience}

\begin{figure*}[ht]
\centering
\includegraphics[width=0.85\linewidth]{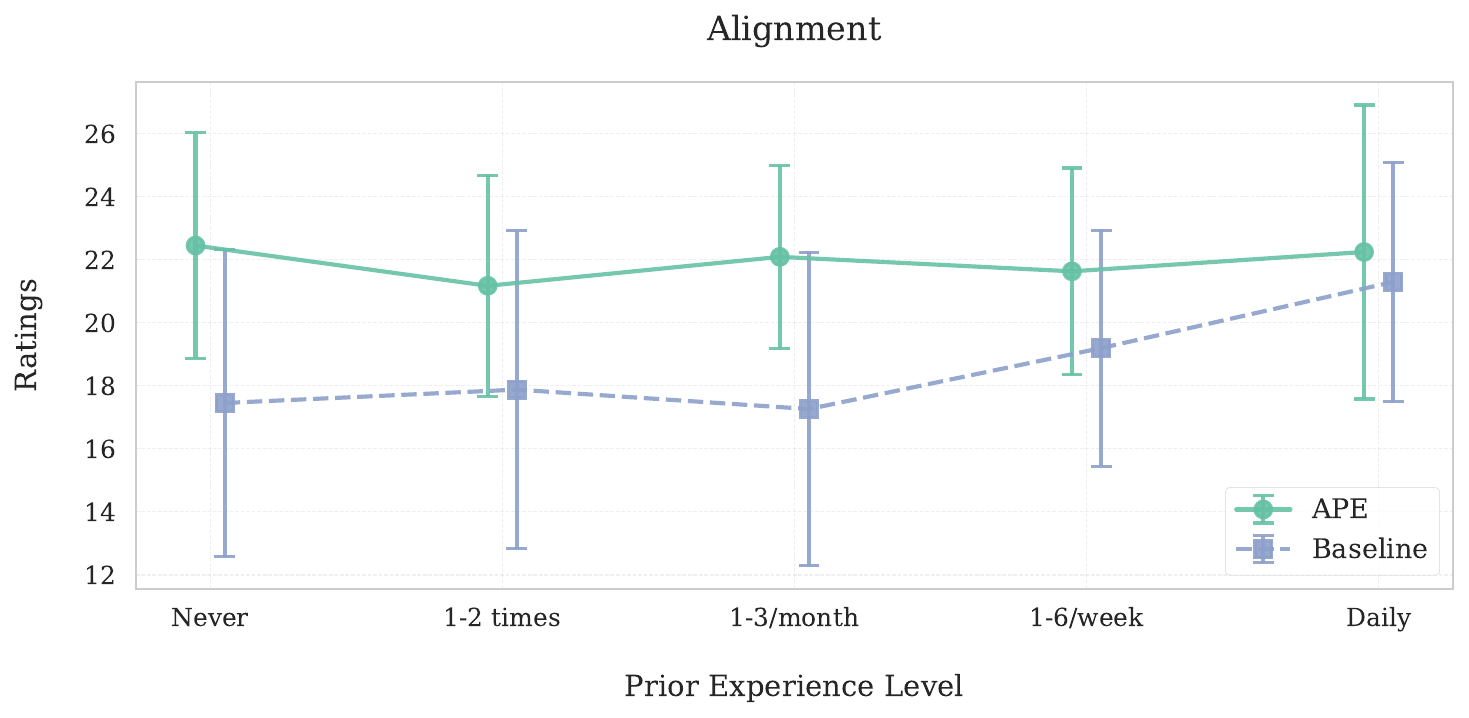}
\caption{Alignment ratings (sum $\pm$ SD) by condition and prior experience level.}
\label{fig:exp_interaction}
\Description{Line plot showing total alignment ratings (y-axis, 12-26 range) versus prior experience levels (x-axis: Never through Daily). APE (green solid line) maintains consistently high ratings (21-22) across all experience levels with minimal variation. Baseline (purple dashed line) shows lower ratings for less experienced users (17-18) that trend upward with experience, reaching 21 for daily users. Pattern suggests APE provides consistent benefit regardless of expertise, while Baseline performance depends heavily on prior experience.}
\end{figure*}

Figure~\ref{fig:exp_interaction} presents alignment ratings across participants with varying levels of experience using text-to-image generation models. APE shows consistently high ratings across all experience levels, while Baseline exhibits a positive trend with increasing experience.

To examine whether prior experience modulated the effect of APE on alignment ratings, we fitted an ordinal logistic regression model with condition (Baseline vs.\ APE), prior experience (modeled as an ordinal variable based on self-reported usage frequency), and their interaction. The model was fitted using maximum likelihood estimation ($N = 128$, AIC = 917.9, BIC = 1023).
The ordinal logistic regression revealed a significant main effect of condition ($\beta = 2.26$, $SE = 0.81$, $z = 2.80$, $p = 0.005$), with participants using APE reporting higher alignment ratings than those in the Baseline condition.
Prior experience was positively associated with alignment ratings, although this effect did not reach statistical significance ($\beta = 0.32$, $SE = 0.18$, $z = 1.75$, $p = .080$).
The interaction between condition and experience was not significant ($\beta = -0.29$, $SE = 0.27$, $z = -1.08$, $p = .280$), providing no evidence that the beneficial effect of APE on perceived alignment differed significantly across experience levels.

Complementarily, we conducted an exploratory analysis to further characterize how prior experience related to alignment ratings within each condition. We dichotomized participants based on self-reported usage frequency of text-to-image generation models into \textit{low-experience users} (never used through monthly; $n = 93$) and \textit{high-experience users} (weekly or more; $n = 35$).
In the Baseline condition, high-experience users reported higher alignment ratings ($M = 19.73$, $SD = 3.78$) than low-experience users ($M = 17.56$, $SD = 4.87$), corresponding to a mean difference of $2.17$ points. This difference approached conventional significance thresholds under a Welch-corrected $t$-test ($t(62) = 1.94$, $p = .059$) and was associated with a moderate effect size (Cohen’s $d = 0.48$). However, given the limited observed power (0.41), this result should be interpreted cautiously as suggesting a potential association between prior experience and higher alignment ratings when using the Baseline system.
In contrast, within the APE condition, alignment ratings were highly similar between high-experience users ($M = 21.80$, $SD = 3.52$) and low-experience users ($M = 21.86$, $SD = 3.24$), yielding a negligible mean difference of $0.06$ points ($t(62) = 0.06$, $p = .956$, Cohen’s $d = 0.02$). These exploratory results provide no clear evidence that prior experience substantially modulated APE's effectiveness given our current sample size.

\subsection{Cognitive Workload (RQ4)}

\begin{figure*}[htb]
\centering
\includegraphics[width=\linewidth]{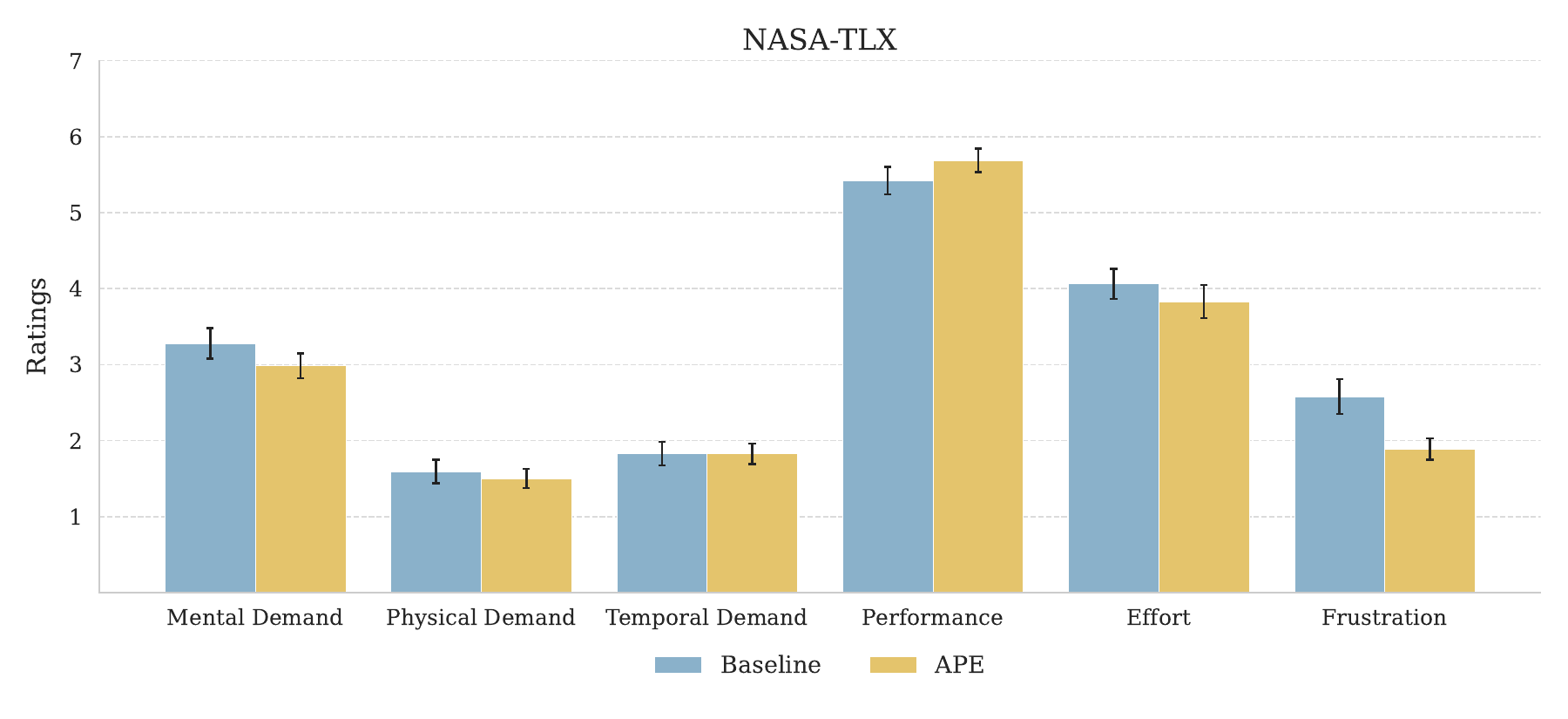}
\caption{NASA-TLX ratings (± SE) across six dimensions for Baseline and APE conditions. No significant differences were found between conditions (all $p > 0.05$).}
\label{fig:nasa_tlx}
\Description{Grouped bar chart displaying NASA-TLX ratings (1-7 scale) with standard errors for six workload dimensions, comparing Baseline (blue) vs APE (yellow). No significant differences found (all p>0.05), but patterns favor APE: Mental Demand (2.98 vs 3.28), Physical Demand (both very low at 1.5-1.6), Temporal Demand (identical at 1.83), Performance higher for APE (5.69 vs 5.42), Effort slightly lower for APE (3.83 vs 4.06), and notably lower Frustration for APE (1.89 vs 2.58, 27\% reduction). Results show APE achieves alignment improvements without increasing cognitive burden.}
\end{figure*}

\subsubsection{Quantitative Results}

Figure~\ref{fig:nasa_tlx} presents NASA-TLX ratings across six dimensions: mental demand, physical demand, temporal demand, performance, effort, and frustration. We did not find statistically significant differences between APE and Baseline across any dimension (Mann-Whitney U tests, Bonferroni-corrected, all $p > 0.05$).

These results suggest that APE achieves its alignment improvements without adding significant cognitive burdens. While one might hypothesize that answering multiple visual queries would increase mental demand or frustration compared to direct prompting, APE users reported statistically similar experiences across workload dimensions. 

\subsubsection{Qualitative Feedback}
In the Baseline condition, users generally reported an acceptable cognitive load when expressing their intent in a single prompt. However, the major source of difficulty stemmed from the need to craft keyword-rich, model-aware prompts, which often led to iterative trial-and-error and substantial mental overhead. Representative comments included: ``\textit{I had to try many variants; long prompts made results worse}'' and ``\textit{It was frustrating to want one thing and get a completely different result.}'' (See Figure~\ref{fig:ex_baseline_1})

In the APE condition, many participants emphasized that APE reduced linguistic effort by presenting lightweight visual choices instead of requiring exhaustive textual descriptions: ``\textit{The visual prompts were extremely helpful}'' and ``\textit{It helped me tease out details I had not thought of.}'' Several participants explicitly reported lower workload and greater ease of use: ``\textit{It was very easy to use the system.}'' Nonetheless, a few participants noted the desire for additional flexibility, such as the ability to skip or undo queries, or found some question sequences somewhat lengthy: ``\textit{I wish it was easier to go back though or to skip a question.}'' A small fraction also reported feeling constrained when their intent could not be fully captured by the available visual options, requiring them to specify additional preferences manually.

\subsection{Interaction Patterns and Efficiency (RQ5)}
\label{sec:interaction_efficiency}

Beyond subjective ratings, we analyzed objective behavioral measures to understand how users actually interacted with APE and the resulting efficiency. We analyzed logged interaction data from all 64 participants in the APE condition, examining feature usage patterns, query effectiveness, and iteration requirements.

\begin{figure*}[ht]
\centering
\includegraphics[width=\linewidth]{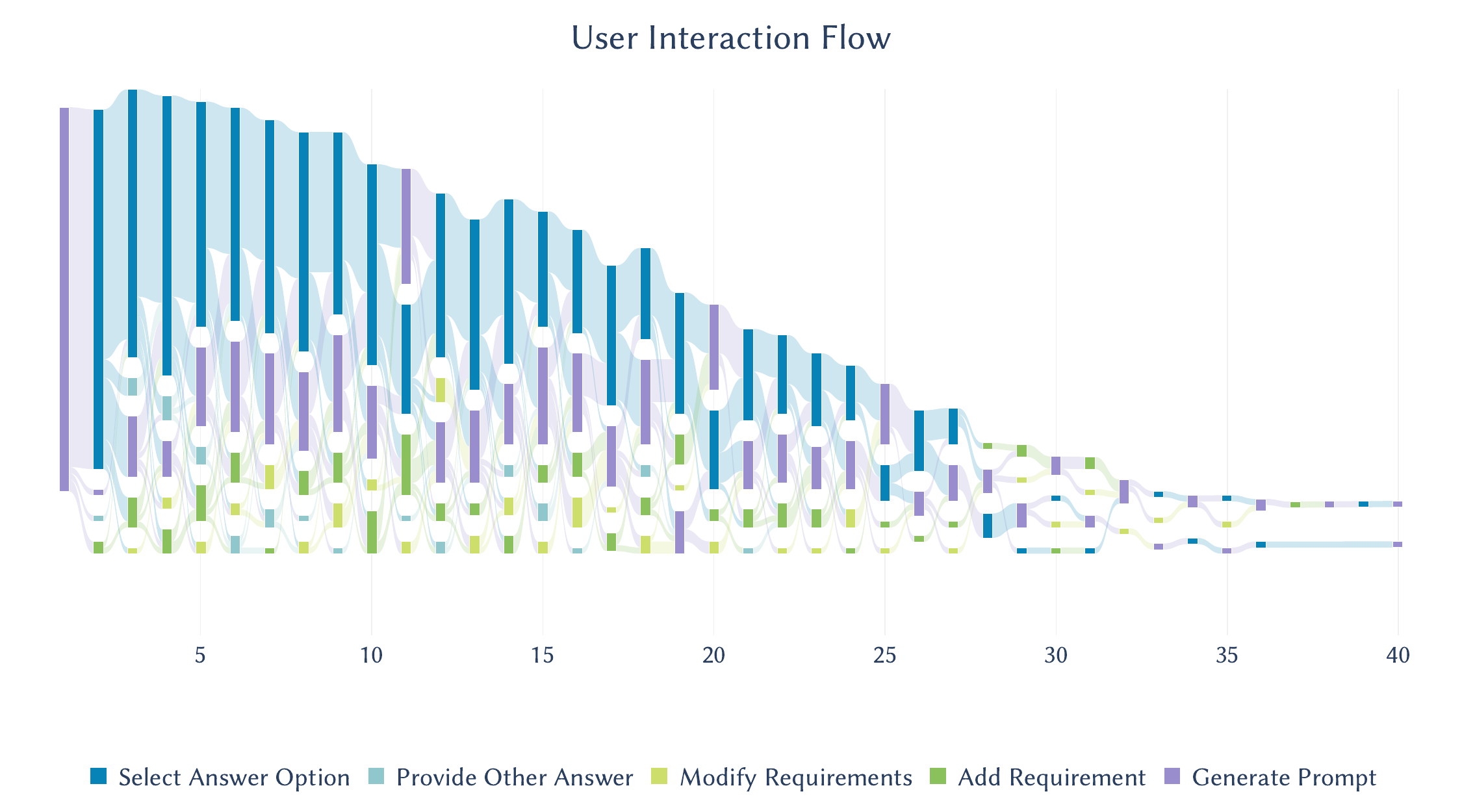}
\caption{Sankey diagram of user interaction flows across iterations (\textit{n}=64 participants, 1{,}199 actions). Flow width indicates action frequency. Interactions are front-loaded, with 68.89\% occurring within the first 15 iterations, followed by a steady decline in later iterations. Users commonly engage in several rounds of visual query responses before transitioning to image generation (purple). Transitions involving manual requirement editing are comparatively infrequent, suggesting that visual query interactions support iterative refinement prior to generation.}
\label{fig:sankey}
\Description{Sankey diagram visualizing 1{,}199 user actions across up to 40 iterations from 64 participants. Vertical bars represent iterations, and colored flows indicate transitions between interaction types, with width proportional to frequency. Five interaction types are shown: Select Answer Option (blue, 54\%), Provide Other Answer (light blue, 3\%), Modify Requirements (yellow, 5\%), Add Requirement (green, 8\%), and Generate Prompt (purple, 31\%). Early iterations are dominated by repeated answer-related actions, while transitions leading to image generation become more prevalent in later iterations. Manual requirement editing appears less frequently throughout. Overall flow width decreases over iterations, indicating reduced interaction frequency as sessions progress.}
\end{figure*}

\subsubsection{Interaction Patterns}
Across all APE sessions, we logged 1,199 total user actions, distributed as follows: 645 visual query responses by selecting from provided options ("Select Answer Option" 53.8\%), 31 visual query responses by providing ``Other'' answers ("Provide Other Answer", 2.6\%), 367 prompt generation requests ("Generate Prompt", 30.6\%), 98 requirement additions ("Add Requirement", 8.2\%), and 58 requirement modifications ("Modify Requirement", 4.8\%).

Figure~\ref{fig:sankey} visualizes interaction flows across iterations. The distribution exhibits a front-loading effect, with approximately 68.89\% of all interactions occurring within the first 15 iterations, after which interaction frequency steadily decreases. The figure shows a gradual drop-off in engagement over successive iterations.

Across iterations, users most commonly engage in repeated cycles of responding to visual outputs by selecting an answer option, followed by a transition to image generation. While answer-to-answer transitions dominate early iterations, the relative proportion of transitions leading to generation increases over time, indicating that users tend to move from exploratory visual queries to generation after several rounds of refinement. Detours into manual requirement editing are comparatively infrequent. This pattern suggests that APE’s visual query interactions help users iteratively refine intent before committing to generation, reducing reliance on repeated text-based prompt rewriting.

The rarity of certain interaction types provides additional insights. Participants selected the ``Other'' option in only 4.59\% of visual query responses, with 46 participants never using this option and only 10 participants using it more than once. This low frequency suggests that LLM-generated feature options are capable of capturing most users' preferences within the presented choices; however, it may also reflect a usability effect, where providing custom inputs requires greater effort and thus discourages use of the Other’’ option. Similarly, direct editing of the requirements panel occurred in only 13.01\% of all interactions, with the vast majority of user effort directed toward responding to system-generated queries rather than manual text refinement. 

The temporal concentration of activity also reveals differentiated user behaviors. While most (76.56\%) users converged within 25 iterations, a small subset (9.38\%) engaged beyond iteration 30, suggesting individual differences in task complexity and exploration preferences. Nevertheless, the overall pattern demonstrates that APE successfully guides users through a coherent, efficient workflow centered on visual query interaction rather than linguistic articulation.

\paragraph{Example of Interaction Trace}

\begin{figure*}[htb]
\centering
\includegraphics[width=\linewidth]{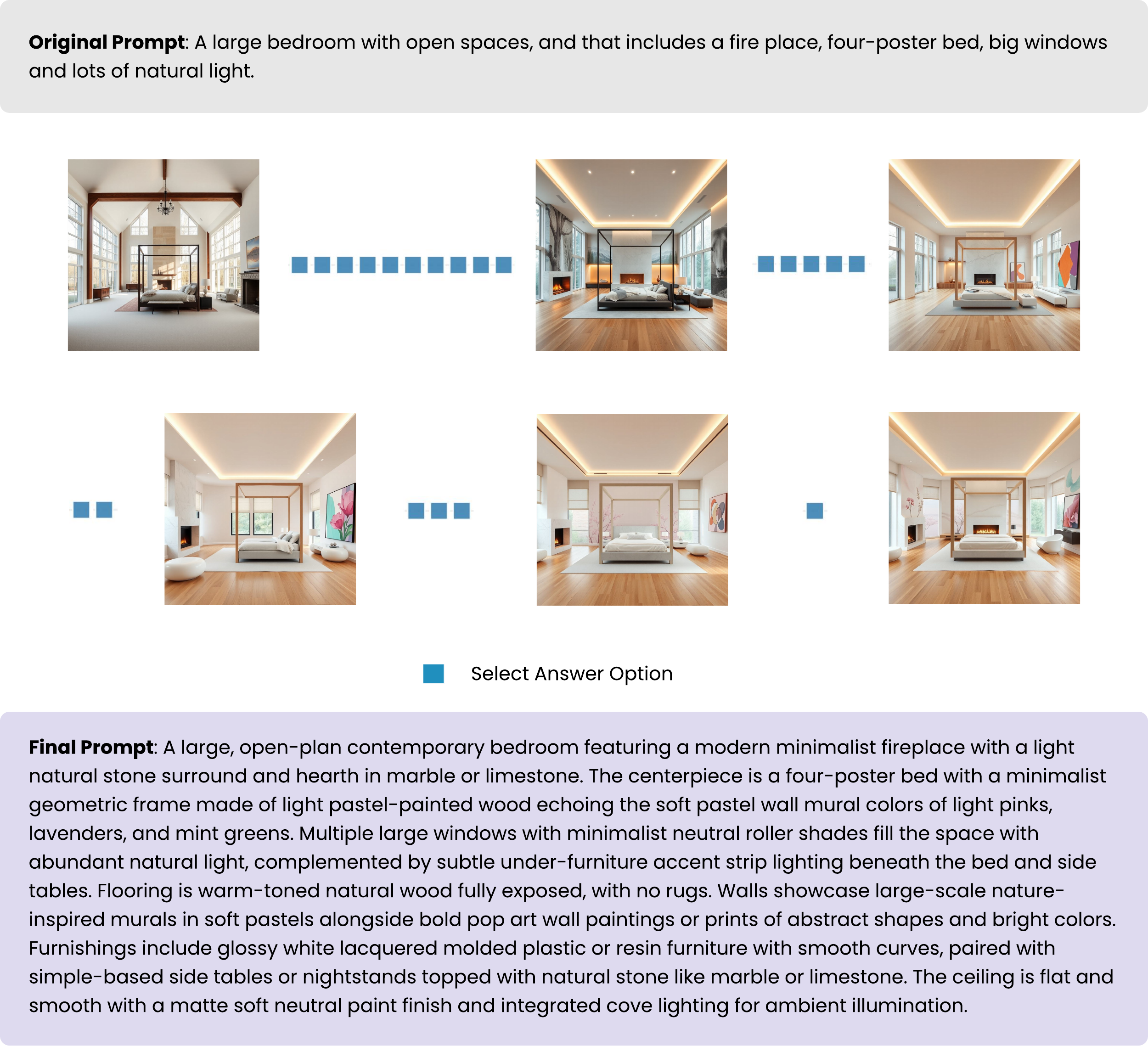}
\caption{Interaction trace from an APE participant in an interior design task. Beginning from a brief functional prompt, the participant’s intent is progressively refined across six generation rounds (I1–I6). Generated images are ordered temporally, and blue squares indicate visual query interactions. Early rounds establish high-level stylistic and compositional choices (e.g., modern minimalist style, spatial layout, and major architectural elements), followed by mid-stage specification of aesthetic direction (color palette, art style, and material coherence), and later-stage refinement of fine-grained details (window treatments, lighting configuration, and furniture materials), exemplifying APE’s hierarchical elicitation process and visually grounded intent formation.}
\label{fig:ape_trace}
\Description{A sequence of six generated bedroom images arranged in temporal order, illustrating progressive refinement during an interior design task. The first image reflects a sparse, open-plan bedroom layout based on a brief functional prompt. Subsequent images show incremental additions and refinements, including modern minimalist architectural elements, a four-poster bed, natural wood flooring, wall murals, color palette choices, lighting configurations, and furniture materials. Small blue squares between images indicate visual query interactions where the participant selected among alternatives. Each image incorporates all previously selected attributes, demonstrating cumulative and hierarchical intent refinement across the six generation rounds.}
\end{figure*}

Figure~\ref{fig:ape_trace} illustrates APE's progressive elicitation through a representative interior design task. The participant began with a concise specification: \textit{``A large bedroom with open spaces, and that includes a fireplace, four-poster bed, big windows and lots of natural light.''} While this 21-word prompt establishes the basic theme, it leaves numerous design decisions unspecified, such as aesthetic style, color palette, materials, and detailed furnishing choices.

Through six generation rounds with visual query interactions between each, APE systematically elicited the participant’s intent. Following the first generation (I1), core stylistic and structural decisions were established, including a modern minimalist style, a minimalist geometric four-poster bed frame, natural wood flooring with warm tones, nature-inspired large-scale wall murals, integrated ambient LED lighting, a fireplace with a light natural stone surround, and a flat ceiling with integrated cove lighting. Between I2 and I3, the aesthetic direction was further articulated by specifying soft pastel wall color palettes contrasted with contemporary pop art pieces featuring abstract shapes and bright colors, as well as by coordinating the bed frame material and color with the mural palette and introducing glossy white lacquered furnishings. Subsequent rounds refined material and furnishing details: window treatments and furniture material and style (I4); ceiling finish and lighting (I5); and side table material (I6).

This trace establishes broad aesthetic and compositional foundations early, then progressively adds material specifications and detail-level choices, exemplifying a hierarchical refinement pattern observed across APE sessions: The final prompt (160 words) is 8× more detailed than the initial specification, yet achieved in only six generation rounds through structured elicitation. The participant's selections reveal preference discovery—elements like the pastel-to-pop-art color contrast, stone fireplace surround, and under-furniture LED strips emerged through visual comparison rather than initial articulation, illustrating how APE facilitates intent formation beyond mere preference elicitation.

\subsubsection{Query Effectiveness}
To assess the quality of system-generated queries, we measured how often participants found them useful. On average, 53.19\% ($SE = 2.71\%$) of questions were reported as useful by the participants. While this represents roughly half of all queries, it is important to note that this metric reflects explicit usefulness ratings; queries that did not receive positive ratings may still have helped narrow the design space or clarified user preferences implicitly. This finding highlights both the effectiveness of the information-theoretic query selection strategy and the highly personalized nature of user intent.

Qualitative feedback supported this interpretation. Participants commented that APE's visual queries were generally effective at eliciting discriminative preferences, such as motifs, styles, and composition, quickly narrowing the design space: ``\textit{The visual prompts allowed me to quickly get the idea of the kind of image I wanted to generate.}'' Nonetheless, the effectiveness of queries depended on the coverage and granularity of the offered options; a few users requested more variety or noted that certain fine details could not be captured through the available options.

\subsubsection{Iteration Requirements}
We tracked how many generation attempts participants required before achieving satisfactory results. APE users needed significantly fewer iterations of image generation ($M = 4.88, SE = 0.41$) compared to Baseline users ($M = 8.06, SE = 0.85$; $U = 2675.0, p = 0.003, r = -0.31$), corresponding to a 39.5\% reduction in iteration cost. Qualitative feedback supported this efficiency gain, with one APE participant noting: ``\textit{Quite fast I found the perfect image, very close to what I imagined.}'' 

However, this advantage may be partially offset by increased latency due to additional API calls and background computation in APE. Our study employed FLUX.1, a flow-matching-based generative model for rapid inference. One Baseline participant with prior experience using slower systems (Midjourney, DALL·E) commented: ``\textit{I was impressed with how fast the images were generated,}'' suggesting FLUX's speed may diminish the perceived value of iteration reduction. Conversely, APE's additional computational overhead drew occasional criticism: ``\textit{The system could be a little faster.}'' The efficiency trade-off likely shifts considerably with slower base models. For text-to-image systems requiring 30-60 seconds per generation, reducing iterations from 8 to 5 would save 1.5-3 minutes of cumulative wait time—a more substantial practical benefit. Future implementations could employ hybrid strategies: using fast distilled models (e.g., FLUX.1-schnell) for generating visual query exemplars during elicitation, then applying the refined prompt to slower, higher-quality models for final generation. This approach would preserve APE's interaction efficiency while accommodating diverse model speed-quality trade-offs.

\section{Discussion}
\label{sec:discussion}
The central challenge in text-to-image generation is not merely model capacity but the \textit{alignment gap}: the mismatch between users' latent visual intentions and their ability to articulate these intentions through text. Our work provides evidence that reframing prompting as bidirectional communication, where systems actively elicit preferences through strategic visual queries, can significantly reduce this gap.

Across controlled technical evaluation (Section~\ref{sec:technical_eval}) and real-world user studies (Section~\ref{sec:user_study}), a consistent pattern emerges: information-theoretically optimized visual queries enable users to specify intent with greater precision than traditional prompt engineering, while requiring fewer iterations and imposing no additional cognitive burden. These benefits stem not from bypassing linguistic articulation, but from scaffolding it—transforming the task from open-ended prompt construction to guided refinement through comparative judgment. We now examine: (1) why visual query interaction proves effective for preference elicitation, (2) how transparent representations enhance user agency, (3) the tension between convergent elicitation and exploratory creativity, and (4) limitations and future directions.

\subsection{Visual Query Interaction for Intent Elicitation}

Our results validate a core design principle underlying APE: that visual query interaction provides an effective complement to textual articulation for preference elicitation. By presenting visual options that embody specific feature values, APE transforms preference specification from open-ended articulation into comparative judgment, directly addressing documented difficulties in verbally specifying rich visual concepts~\cite{mahdavigoloujeh2024It, sanchez2023Examining} by leveraging the cognitive advantage of recognition over generation~\cite{schooler1990verbal}. The significant improvements across all generative alignment dimensions suggest that this two-way communication enhances perceived alignment~\cite{kim2025Journey, shen2025Position}.

The reduction in image generation iterations represents substantial practical value in real-world use cases, consistent with prior work showing interactive query-based prompt optimization reduces search costs in image generation~\cite{hahn2025Proactive, wang2025Twin}. A generation cycle involves cognitive effort (evaluating results, deciding on modifications), time (waiting for generation), and computational cost (running the text-to-image model). By nearly halving the required iterations, APE transforms user experience from exhausting trial-and-error to structured refinement. 
This efficiency improvement appears particularly valuable for users with clear conceptual goals but limited prompt engineering strategies, though our exploratory subgroup analysis was underpowered for definitive claims about experience-based moderation.

\subsection{Transparency and User Agency in AI-Assisted Creation}

Besides query-based interaction, an important benefit emerged from APE's explicit feature requirement representation: transparency that supports both understanding and control. Participants commented positively on seeing preferences as structured requirements, with one noting: ``\textit{It was innovative and smart that I could see the filters that are being applied before creating an image and being able to edit them. In this way, I can understand that the AI has understood what I am asking for.}'' This transparency likely contributed to APE's particularly strong improvement in the instruction gap dimension, where users reported better ability to provide appropriate input. The editable requirements panel also enabled mid-process corrections—users could identify and revise specific features rather than restarting from scratch—which may explain the reduced intentionality gap as users gained confidence that their vision was accurately captured~\cite{kim2025Journey}. This contrasts with black-box prompting, where users input text without understanding model interpretation, limiting opportunities for bias detection, error correction, or trust-building through interpretability.

The mixed-initiative design~\cite{deterding2017MixedInitiative} of APE also reflects emerging principles for human-AI collaboration. In APE, the structured requirements panel serves as an explainable interface helping user to identify when the system encodes biased assumptions or makes incorrect inferences about their intent~\cite{vasconcelos2023Explanations}. The text editing capability then provides \textit{user agency} to correct these errors, consistent with research showing that editable AI outputs better support user-driven value alignment than uneditable suggestions~\cite{fan2025UserDriven}. Notably, these safeguards require users to recognize biases, which may not occur if biases align with users' own assumptions or operate subtly~\cite{bianchi2023Easily, gupta2023Bias, castleman2025Adultification}. Future work could explore proactive such as diverse sampling or bias detection like OpenBias~\cite{dinca2024OpenBias} to further strengthen these protections.

\subsection{Creative Exploration Through Structured Elicitation}

While APE optimizes for efficient convergence to user intent, an interesting pattern emerged in user feedback: some participants reported that structured queries facilitated creative exploration. Multiple participants noted discovering preferences through visual options they had not initially considered: ``\textit{The question about lighting made me realize I cared about that. I hadn't thought about it before, but seeing the options, I definitely had a preference.}'' and ``\textit{The system questions helped in getting the creativity flowing.}'' This suggests that APE's queries may serve dual purposes: efficiently gathering preference information while scaffolding preference discovery through systematic consideration of visual dimensions users might otherwise overlook.
This finding connects to research on creativity support tools showing that constraints can enhance rather than hinder creativity by providing structure for exploration~\cite{stokes2005creativity}. APE's queries function as design prompts that encourage systematic thinking about visual dimensions, potentially supporting creative development alongside efficient elicitation. 

However, we note important scope limitations: while we observed anecdotal evidence of preference discovery, APE's core design is grounded in traditional preference elicitation frameworks that assume users have pre-existing preferences to uncover~\cite{viappiani2014Preference, gonzalez2017Preferential, chu2005Preference}. Our information-theoretic query strategy explicitly targets efficient convergence to latent user intent, making APE well-suited for \textit{goal-directed tasks} where users face articulation barriers. For purely \textit{exploratory tasks} prioritizing serendipitous discovery, alternative approaches emphasizing divergent exploration over convergent elicitation may prove more appropriate. 
The tension between efficient preference elicitation and open-ended creative exploration represents an important theoretical distinction that future work should address through adaptive interaction modes.

\subsection{Limitations and Future Work}

We identify several important limitations that bound the applicability of our approach and suggest directions for future work.

Firstly, APE addresses misalignment by reducing epistemic uncertainty (the gap between user intent and system understanding) but does not address inherent limitations of image generation models (aleatoric uncertainty). The method's effectiveness is fundamentally bounded by model capacity: as generative models improve, APE's potential increases; conversely, with limited models, alignment gains may be constrained. Future work could explore synergistic approaches that integrate APE with model improvement techniques, such as user-specific fine-tuning~\cite{zeng2024IntentTuner} or self-improvement via consistency regularization~\cite{oscar2024improving}, to jointly enhance alignment and output quality.

Secondly, our information-theoretic selection optimizes for efficiency, assuming users have clear (though unarticulated) preferences and a proper model of user intention. In reality, preferences often form dynamically and correlatively. A few participants noted the desire for additional flexibility, such as the ability to skip or undo queries, or found some question sequences somewhat lengthy: ``\textit{I wish it were easier to go back, though, or to skip a question.}'' Future interfaces might provide more support for calibrating misspecified priors, and balance efficiency with exploration support, occasionally selecting queries that maximize creative insight rather than immediate information gain, though designing such balanced objectives remains an open challenge.

Thirdly, APE's effectiveness depends critically on LLM quality for feature generation, completion sampling, and prompt synthesis. The LLM persona may encode biases from pre-training data, such as demographic, cultural, or aesthetic biases~\cite{bianchi2023Easily, gupta2023Bias, castleman2025Adultification, resnik2025Large}. While our mixed-initiative design provides mitigation mechanisms (transparent panel for bias detection~\cite{vasconcelos2023Explanations}, editing for user agency~\cite{fan2025UserDriven}), these require users to recognize biases, which may not occur if biases align with users' assumptions or operate subtly. Future work could investigate proactive bias detection, diverse sampling strategies~\cite{dinca2024OpenBias}, and diversity-aware query generation.

Lastly, while the model-agnostic design of APE enables its application to a broad range of text-to-image generation models, our empirical evaluation is limited to a single state-of-the-art model. Future work could explore its generalizability across a wider set of generative architectures to assess robustness and uncover model-specific factors. In addition, investigating APE’s potential to extend across modalities (e.g. text-to-music~\cite{copet2023Simple} and text-to-video~\cite{singer2023makeavideo} generation) offers a compelling avenue for further study.
As generative models continue to advance, addressing the alignment gap becomes increasingly critical beyond improvements in model capacity alone. APE constitutes an initial step toward mitigating this gap through principled interaction design, pointing toward future creative AI systems that are more interactive, transparent, and adaptive than current text-based interfaces.

\section{Conclusion}

Aligning text-to-image generation with user intent remains a persistent challenge when users struggle to articulate rich visual concepts through text. We presented \textsc{Adaptive Prompt Elicitation} (APE), which reframes prompting as bidirectional communication through information-theoretically optimized visual queries. By eliciting preferences through strategic visual comparisons, APE transforms intent specification from unilateral articulation into guided interactive refinement. 
Evaluation across controlled benchmarks and user studies demonstrates that this interactive approach achieves stronger alignment with improved efficiency compared to manual prompting and automatic optimization. As generative models advance, principled interaction design that actively elicits user intent offers a complementary path toward alignment alongside improvements in model capacity.

\begin{acks}
This work was supported by the Research Council of Finland (flagship program: Finnish Center for Artificial Intelligence FCAI; Subjective Functions, grant 357578) and the ERC Advanced Grant (grant 101141916). We acknowledge the Aalto Science-IT project for providing computational and data storage resources. We thank Prof. Luigi Acerbi and all the reviewers for their valuable suggestions and comments.
\end{acks}

\bibliographystyle{ACM-Reference-Format}
\bibliography{reference}

%%% -*-BibTeX-*-
%%% Do NOT edit. File created by BibTeX with style
%%% ACM-Reference-Format-Journals [18-Jan-2012].

\begin{thebibliography}{92}

%%% ====================================================================
%%% NOTE TO THE USER: you can override these defaults by providing
%%% customized versions of any of these macros before the \bibliography
%%% command.  Each of them MUST provide its own final punctuation,
%%% except for \shownote{} and \showURL{}.  The latter two
%%% do not use final punctuation, in order to avoid confusing it with
%%% the Web address.
%%%
%%% To suppress output of a particular field, define its macro to expand
%%% to an empty string, or better, \unskip, like this:
%%%
%%% \newcommand{\showURL}[1]{\unskip}   % LaTeX syntax
%%%
%%% \def \showURL #1{\unskip}           % plain TeX syntax
%%%
%%% ====================================================================

\ifx \showCODEN    \undefined \def \showCODEN     #1{\unskip}     \fi
\ifx \showISBNx    \undefined \def \showISBNx     #1{\unskip}     \fi
\ifx \showISBNxiii \undefined \def \showISBNxiii  #1{\unskip}     \fi
\ifx \showISSN     \undefined \def \showISSN      #1{\unskip}     \fi
\ifx \showLCCN     \undefined \def \showLCCN      #1{\unskip}     \fi
\ifx \shownote     \undefined \def \shownote      #1{#1}          \fi
\ifx \showarticletitle \undefined \def \showarticletitle #1{#1}   \fi
\ifx \showURL      \undefined \def \showURL       {\relax}        \fi
% The following commands are used for tagged output and should be
% invisible to TeX
\providecommand\bibfield[2]{#2}
\providecommand\bibinfo[2]{#2}
\providecommand\natexlab[1]{#1}
\providecommand\showeprint[2][]{arXiv:#2}

\bibitem[Adamkiewicz et~al\mbox{.}(2025)]%
        {adamkiewicz2025PromptMap}
\bibfield{author}{\bibinfo{person}{Krzysztof Adamkiewicz}, \bibinfo{person}{Pawe{\l}~W. Wo{\'z}niak}, \bibinfo{person}{Julia Dominiak}, \bibinfo{person}{Andrzej Romanowski}, \bibinfo{person}{Jakob Karolus}, {and} \bibinfo{person}{Stanislav Frolov}.} \bibinfo{year}{2025}\natexlab{}.
\newblock \showarticletitle{{{PromptMap}}: {{An Alternative Interaction Style}} for {{AI-Based Image Generation}}}. In \bibinfo{booktitle}{\emph{Proceedings of the 30th {{International Conference}} on {{Intelligent User Interfaces}}}} \emph{(\bibinfo{series}{{{IUI}} '25})}. \bibinfo{publisher}{Association for Computing Machinery}, \bibinfo{address}{New York, NY, USA}, \bibinfo{pages}{1162--1176}.
\newblock
\showISBNx{979-8-4007-1306-4}
\href{https://doi.org/10.1145/3708359.3712150}{doi:\nolinkurl{10.1145/3708359.3712150}}


\bibitem[Bianchi et~al\mbox{.}(2023)]%
        {bianchi2023Easily}
\bibfield{author}{\bibinfo{person}{Federico Bianchi}, \bibinfo{person}{Pratyusha Kalluri}, \bibinfo{person}{Esin Durmus}, \bibinfo{person}{Faisal Ladhak}, \bibinfo{person}{Myra Cheng}, \bibinfo{person}{Debora Nozza}, \bibinfo{person}{Tatsunori Hashimoto}, \bibinfo{person}{Dan Jurafsky}, \bibinfo{person}{James Zou}, {and} \bibinfo{person}{Aylin Caliskan}.} \bibinfo{year}{2023}\natexlab{}.
\newblock \showarticletitle{Easily {{Accessible Text-to-Image Generation Amplifies Demographic Stereotypes}} at {{Large Scale}}}. In \bibinfo{booktitle}{\emph{Proceedings of the 2023 {{ACM Conference}} on {{Fairness}}, {{Accountability}}, and {{Transparency}}}} \emph{(\bibinfo{series}{{{FAccT}} '23})}. \bibinfo{publisher}{Association for Computing Machinery}, \bibinfo{address}{New York, NY, USA}, \bibinfo{pages}{1493--1504}.
\newblock
\showISBNx{979-8-4007-0192-4}
\href{https://doi.org/10.1145/3593013.3594095}{doi:\nolinkurl{10.1145/3593013.3594095}}


\bibitem[Black et~al\mbox{.}(2024)]%
        {black2024training}
\bibfield{author}{\bibinfo{person}{Kevin Black}, \bibinfo{person}{Michael Janner}, \bibinfo{person}{Yilun Du}, \bibinfo{person}{Ilya Kostrikov}, {and} \bibinfo{person}{Sergey Levine}.} \bibinfo{year}{2024}\natexlab{}.
\newblock \showarticletitle{Training Diffusion Models with Reinforcement Learning}. In \bibinfo{booktitle}{\emph{The Twelfth International Conference on Learning Representations}}.
\newblock


\bibitem[Brade et~al\mbox{.}(2023)]%
        {brade2023Promptify}
\bibfield{author}{\bibinfo{person}{Stephen Brade}, \bibinfo{person}{Bryan Wang}, \bibinfo{person}{Mauricio Sousa}, \bibinfo{person}{Sageev Oore}, {and} \bibinfo{person}{Tovi Grossman}.} \bibinfo{year}{2023}\natexlab{}.
\newblock \showarticletitle{Promptify: {{Text-to-Image Generation}} through {{Interactive Prompt Exploration}} with {{Large Language Models}}}. In \bibinfo{booktitle}{\emph{Proceedings of the 36th {{Annual ACM Symposium}} on {{User Interface Software}} and {{Technology}}}} \emph{(\bibinfo{series}{{{UIST}} '23})}. \bibinfo{publisher}{Association for Computing Machinery}, \bibinfo{address}{New York, NY, USA}, \bibinfo{pages}{1--14}.
\newblock
\showISBNx{979-8-4007-0132-0}
\href{https://doi.org/10.1145/3586183.3606725}{doi:\nolinkurl{10.1145/3586183.3606725}}


\bibitem[Brochu et~al\mbox{.}(2010)]%
        {brochu2010interactive}
\bibfield{author}{\bibinfo{person}{Eric Brochu}, \bibinfo{person}{Tyson Brochu}, {and} \bibinfo{person}{Nando de Freitas}.} \bibinfo{year}{2010}\natexlab{}.
\newblock \showarticletitle{A Bayesian interactive optimization approach to procedural animation design}. In \bibinfo{booktitle}{\emph{Proceedings of the 2010 ACM SIGGRAPH/Eurographics Symposium on Computer Animation}}. \bibinfo{pages}{103--112}.
\newblock


\bibitem[Cai et~al\mbox{.}(2025)]%
        {cai2025Bayesian}
\bibfield{author}{\bibinfo{person}{Chengkun Cai}, \bibinfo{person}{Haoliang Liu}, \bibinfo{person}{Xu Zhao}, \bibinfo{person}{Zhongyu Jiang}, \bibinfo{person}{Tianfang Zhang}, \bibinfo{person}{Zongkai Wu}, \bibinfo{person}{John Lee}, \bibinfo{person}{Jenq-Neng Hwang}, {and} \bibinfo{person}{Lei Li}.} \bibinfo{year}{2025}\natexlab{}.
\newblock \showarticletitle{Bayesian {{Optimization}} for {{Controlled Image Editing}} via {{LLMs}}}. In \bibinfo{booktitle}{\emph{Findings of the {{Association}} for {{Computational Linguistics}}: {{ACL}} 2025}}. \bibinfo{publisher}{Association for Computational Linguistics}, \bibinfo{address}{Vienna, Austria}, \bibinfo{pages}{10045--10056}.
\newblock
\showISBNx{979-8-89176-256-5}
\href{https://doi.org/10.18653/v1/2025.findings-acl.523}{doi:\nolinkurl{10.18653/v1/2025.findings-acl.523}}


\bibitem[Cao et~al\mbox{.}(2023)]%
        {cao2023BeautifulPrompt}
\bibfield{author}{\bibinfo{person}{Tingfeng Cao}, \bibinfo{person}{Chengyu Wang}, \bibinfo{person}{Bingyan Liu}, \bibinfo{person}{Ziheng Wu}, \bibinfo{person}{Jinhui Zhu}, {and} \bibinfo{person}{Jun Huang}.} \bibinfo{year}{2023}\natexlab{}.
\newblock \showarticletitle{{{BeautifulPrompt}}: {{Towards Automatic Prompt Engineering}} for {{Text-to-Image Synthesis}}}. In \bibinfo{booktitle}{\emph{Proceedings of the 2023 {{Conference}} on {{Empirical Methods}} in {{Natural Language Processing}}: {{Industry Track}}}}. \bibinfo{publisher}{Association for Computational Linguistics}, \bibinfo{address}{Singapore}, \bibinfo{pages}{1--11}.
\newblock
\href{https://doi.org/10.18653/v1/2023.emnlp-industry.1}{doi:\nolinkurl{10.18653/v1/2023.emnlp-industry.1}}


\bibitem[Castleman and Korolova(2025)]%
        {castleman2025Adultification}
\bibfield{author}{\bibinfo{person}{Jane Castleman} {and} \bibinfo{person}{Aleksandra Korolova}.} \bibinfo{year}{2025}\natexlab{}.
\newblock \showarticletitle{Adultification {{Bias}} in {{LLMs}} and {{Text-to-Image Models}}}. In \bibinfo{booktitle}{\emph{Proceedings of the 2025 {{ACM Conference}} on {{Fairness}}, {{Accountability}}, and {{Transparency}}}} \emph{(\bibinfo{series}{{{FAccT}} '25})}. \bibinfo{publisher}{Association for Computing Machinery}, \bibinfo{address}{New York, NY, USA}, \bibinfo{pages}{2751--2767}.
\newblock
\showISBNx{979-8-4007-1482-5}
\href{https://doi.org/10.1145/3715275.3732178}{doi:\nolinkurl{10.1145/3715275.3732178}}


\bibitem[Chaloner and Verdinelli(1995)]%
        {chaloner1995bayesian}
\bibfield{author}{\bibinfo{person}{Kathryn Chaloner} {and} \bibinfo{person}{Isabella Verdinelli}.} \bibinfo{year}{1995}\natexlab{}.
\newblock \showarticletitle{Bayesian Experimental Design: A Review}.
\newblock \bibinfo{journal}{\emph{Statist. Sci.}} \bibinfo{volume}{10}, \bibinfo{number}{3} (\bibinfo{year}{1995}), \bibinfo{pages}{273--304}.
\newblock


\bibitem[Chen et~al\mbox{.}(2024)]%
        {chen2024Tailored}
\bibfield{author}{\bibinfo{person}{Zijie Chen}, \bibinfo{person}{Lichao Zhang}, \bibinfo{person}{Fangsheng Weng}, \bibinfo{person}{Lili Pan}, {and} \bibinfo{person}{Zhenzhong Lan}.} \bibinfo{year}{2024}\natexlab{}.
\newblock \showarticletitle{Tailored {{Visions}}: {{Enhancing Text-to-Image Generation}} with {{Personalized Prompt Rewriting}}}. In \bibinfo{booktitle}{\emph{Proceedings of the {{IEEE}}/{{CVF Conference}} on {{Computer Vision}} and {{Pattern Recognition}}}}. \bibinfo{pages}{7727--7736}.
\newblock


\bibitem[Christiano et~al\mbox{.}(2017)]%
        {christiano2017Deep}
\bibfield{author}{\bibinfo{person}{Paul~F Christiano}, \bibinfo{person}{Jan Leike}, \bibinfo{person}{Tom Brown}, \bibinfo{person}{Miljan Martic}, \bibinfo{person}{Shane Legg}, {and} \bibinfo{person}{Dario Amodei}.} \bibinfo{year}{2017}\natexlab{}.
\newblock \showarticletitle{Deep {{Reinforcement Learning}} from {{Human Preferences}}}.
\newblock \bibinfo{journal}{\emph{Advances in {{Neural Information Processing Systems}}}}  \bibinfo{volume}{30} (\bibinfo{year}{2017}).
\newblock


\bibitem[Chu and Ghahramani(2005)]%
        {chu2005Preference}
\bibfield{author}{\bibinfo{person}{Wei Chu} {and} \bibinfo{person}{Zoubin Ghahramani}.} \bibinfo{year}{2005}\natexlab{}.
\newblock \showarticletitle{Preference Learning with {{Gaussian}} Processes}. In \bibinfo{booktitle}{\emph{Proceedings of the 22nd International Conference on {{Machine}} Learning - {{ICML}} '05}}. \bibinfo{publisher}{ACM Press}, \bibinfo{address}{Bonn, Germany}, \bibinfo{pages}{137--144}.
\newblock
\showISBNx{978-1-59593-180-1}
\href{https://doi.org/10.1145/1102351.1102369}{doi:\nolinkurl{10.1145/1102351.1102369}}


\bibitem[Chung and Adar(2023)]%
        {chung2023PromptPaint}
\bibfield{author}{\bibinfo{person}{John Joon~Young Chung} {and} \bibinfo{person}{Eytan Adar}.} \bibinfo{year}{2023}\natexlab{}.
\newblock \showarticletitle{{{PromptPaint}}: {{Steering Text-to-Image Generation Through Paint Medium-like Interactions}}}. In \bibinfo{booktitle}{\emph{Proceedings of the 36th {{Annual ACM Symposium}} on {{User Interface Software}} and {{Technology}}}} \emph{(\bibinfo{series}{{{UIST}} '23})}. \bibinfo{publisher}{Association for Computing Machinery}, \bibinfo{address}{New York, NY, USA}, \bibinfo{pages}{1--17}.
\newblock
\showISBNx{979-8-4007-0132-0}
\href{https://doi.org/10.1145/3586183.3606777}{doi:\nolinkurl{10.1145/3586183.3606777}}


\bibitem[Copet et~al\mbox{.}(2023)]%
        {copet2023Simple}
\bibfield{author}{\bibinfo{person}{Jade Copet}, \bibinfo{person}{Felix Kreuk}, \bibinfo{person}{Itai Gat}, \bibinfo{person}{Tal Remez}, \bibinfo{person}{David Kant}, \bibinfo{person}{Gabriel Synnaeve}, \bibinfo{person}{Yossi Adi}, {and} \bibinfo{person}{Alexandre Defossez}.} \bibinfo{year}{2023}\natexlab{}.
\newblock \showarticletitle{Simple and {{Controllable Music Generation}}}.
\newblock \bibinfo{journal}{\emph{Advances in Neural Information Processing Systems}}  \bibinfo{volume}{36} (\bibinfo{year}{2023}), \bibinfo{pages}{47704--47720}.
\newblock


\bibitem[Datta et~al\mbox{.}(2024)]%
        {datta2024Prompt}
\bibfield{author}{\bibinfo{person}{Siddhartha Datta}, \bibinfo{person}{Alexander Ku}, \bibinfo{person}{Deepak Ramachandran}, {and} \bibinfo{person}{Peter Anderson}.} \bibinfo{year}{2024}\natexlab{}.
\newblock \showarticletitle{Prompt {{Expansion}} for {{Adaptive Text-to-Image Generation}}}. In \bibinfo{booktitle}{\emph{Proceedings of the 62nd {{Annual Meeting}} of the {{Association}} for {{Computational Linguistics}} ({{Volume}} 1: {{Long Papers}})}}. \bibinfo{publisher}{Association for Computational Linguistics}, \bibinfo{address}{Bangkok, Thailand}, \bibinfo{pages}{3449--3476}.
\newblock
\href{https://doi.org/10.18653/v1/2024.acl-long.189}{doi:\nolinkurl{10.18653/v1/2024.acl-long.189}}


\bibitem[Deterding et~al\mbox{.}(2017)]%
        {deterding2017MixedInitiative}
\bibfield{author}{\bibinfo{person}{Sebastian Deterding}, \bibinfo{person}{Jonathan Hook}, \bibinfo{person}{Rebecca Fiebrink}, \bibinfo{person}{Marco Gillies}, \bibinfo{person}{Jeremy Gow}, \bibinfo{person}{Memo Akten}, \bibinfo{person}{Gillian Smith}, \bibinfo{person}{Antonios Liapis}, {and} \bibinfo{person}{Kate Compton}.} \bibinfo{year}{2017}\natexlab{}.
\newblock \showarticletitle{Mixed-{{Initiative Creative Interfaces}}}. In \bibinfo{booktitle}{\emph{Proceedings of the 2017 {{CHI Conference Extended Abstracts}} on {{Human Factors}} in {{Computing Systems}}}} \emph{(\bibinfo{series}{{{CHI EA}} '17})}. \bibinfo{publisher}{Association for Computing Machinery}, \bibinfo{address}{New York, NY, USA}, \bibinfo{pages}{628--635}.
\newblock
\showISBNx{978-1-4503-4656-6}
\href{https://doi.org/10.1145/3027063.3027072}{doi:\nolinkurl{10.1145/3027063.3027072}}


\bibitem[D'Inc{\`a} et~al\mbox{.}(2024)]%
        {dinca2024OpenBias}
\bibfield{author}{\bibinfo{person}{Moreno D'Inc{\`a}}, \bibinfo{person}{Elia Peruzzo}, \bibinfo{person}{Massimiliano Mancini}, \bibinfo{person}{Dejia Xu}, \bibinfo{person}{Vidit Goe}, \bibinfo{person}{Xingqian Xu}, \bibinfo{person}{Zhangyang Wang}, \bibinfo{person}{Humphrey Shi}, {and} \bibinfo{person}{Nicu Sebe}.} \bibinfo{year}{2024}\natexlab{}.
\newblock \showarticletitle{{{OpenBias}}: {{Open-Set Bias Detection}} in {{Text-to-Image Generative Models}}}. In \bibinfo{booktitle}{\emph{2024 {{IEEE}}/{{CVF Conference}} on {{Computer Vision}} and {{Pattern Recognition}} ({{CVPR}})}}. \bibinfo{pages}{12225--12235}.
\newblock
\showISSN{2575-7075}
\href{https://doi.org/10.1109/CVPR52733.2024.01162}{doi:\nolinkurl{10.1109/CVPR52733.2024.01162}}


\bibitem[Fails and Olsen(2003)]%
        {fails2003interactive}
\bibfield{author}{\bibinfo{person}{Jerry~Alan Fails} {and} \bibinfo{person}{Dan~R. Olsen}.} \bibinfo{year}{2003}\natexlab{}.
\newblock \showarticletitle{Interactive Machine Learning}. In \bibinfo{booktitle}{\emph{Proceedings of the 8th International Conference on {{Intelligent}} User Interfaces}} \emph{(\bibinfo{series}{{{IUI}} '03})}. \bibinfo{publisher}{Association for Computing Machinery}, \bibinfo{address}{New York, NY, USA}, \bibinfo{pages}{39--45}.
\newblock
\showISBNx{978-1-58113-586-2}
\href{https://doi.org/10.1145/604045.604056}{doi:\nolinkurl{10.1145/604045.604056}}


\bibitem[Fan et~al\mbox{.}(2025)]%
        {fan2025UserDriven}
\bibfield{author}{\bibinfo{person}{Xianzhe Fan}, \bibinfo{person}{Qing Xiao}, \bibinfo{person}{Xuhui Zhou}, \bibinfo{person}{Jiaxin Pei}, \bibinfo{person}{Maarten Sap}, \bibinfo{person}{Zhicong Lu}, {and} \bibinfo{person}{Hong Shen}.} \bibinfo{year}{2025}\natexlab{}.
\newblock \showarticletitle{User-{{Driven Value Alignment}}: {{Understanding Users}}' {{Perceptions}} and {{Strategies}} for {{Addressing Biased}} and {{Discriminatory Statements}} in {{AI Companions}}}. In \bibinfo{booktitle}{\emph{Proceedings of the 2025 {{CHI Conference}} on {{Human Factors}} in {{Computing Systems}}}} \emph{(\bibinfo{series}{{{CHI}} '25})}. \bibinfo{publisher}{Association for Computing Machinery}, \bibinfo{address}{New York, NY, USA}, \bibinfo{pages}{1--19}.
\newblock
\showISBNx{979-8-4007-1394-1}
\href{https://doi.org/10.1145/3706598.3713477}{doi:\nolinkurl{10.1145/3706598.3713477}}


\bibitem[Faul et~al\mbox{.}(2009)]%
        {faul2009statistical}
\bibfield{author}{\bibinfo{person}{Franz Faul}, \bibinfo{person}{Edgar Erdfelder}, \bibinfo{person}{Axel Buchner}, {and} \bibinfo{person}{Albert-Georg Lang}.} \bibinfo{year}{2009}\natexlab{}.
\newblock \showarticletitle{Statistical power analyses using G* Power 3.1: Tests for correlation and regression analyses}.
\newblock \bibinfo{journal}{\emph{Behavior research methods}} \bibinfo{volume}{41}, \bibinfo{number}{4} (\bibinfo{year}{2009}), \bibinfo{pages}{1149--1160}.
\newblock


\bibitem[Feng et~al\mbox{.}(2024)]%
        {feng2024PromptMagician}
\bibfield{author}{\bibinfo{person}{Yingchaojie Feng}, \bibinfo{person}{Xingbo Wang}, \bibinfo{person}{Kam~Kwai Wong}, \bibinfo{person}{Sijia Wang}, \bibinfo{person}{Yuhong Lu}, \bibinfo{person}{Minfeng Zhu}, \bibinfo{person}{Baicheng Wang}, {and} \bibinfo{person}{Wei Chen}.} \bibinfo{year}{2024}\natexlab{}.
\newblock \showarticletitle{{{PromptMagician}}: {{Interactive Prompt Engineering}} for {{Text-to-Image Creation}}}.
\newblock \bibinfo{journal}{\emph{IEEE Transactions on Visualization and Computer Graphics}} \bibinfo{volume}{30}, \bibinfo{number}{1} (\bibinfo{date}{Jan.} \bibinfo{year}{2024}), \bibinfo{pages}{295--305}.
\newblock
\showISSN{1941-0506}
\href{https://doi.org/10.1109/TVCG.2023.3327168}{doi:\nolinkurl{10.1109/TVCG.2023.3327168}}


\bibitem[Fu et~al\mbox{.}(2023)]%
        {fu2023DreamSim}
\bibfield{author}{\bibinfo{person}{Stephanie Fu}, \bibinfo{person}{Netanel Tamir}, \bibinfo{person}{Shobhita Sundaram}, \bibinfo{person}{Lucy Chai}, \bibinfo{person}{Richard Zhang}, \bibinfo{person}{Tali Dekel}, {and} \bibinfo{person}{Phillip Isola}.} \bibinfo{year}{2023}\natexlab{}.
\newblock \showarticletitle{{{DreamSim}}: {{Learning New Dimensions}} of {{Human Visual Similarity}} Using {{Synthetic Data}}}.
\newblock \bibinfo{journal}{\emph{Advances in Neural Information Processing Systems}}  \bibinfo{volume}{36} (\bibinfo{year}{2023}), \bibinfo{pages}{50742--50768}.
\newblock


\bibitem[Gal et~al\mbox{.}(2017)]%
        {gal2017deep}
\bibfield{author}{\bibinfo{person}{Yarin Gal}, \bibinfo{person}{Riashat Islam}, {and} \bibinfo{person}{Zoubin Ghahramani}.} \bibinfo{year}{2017}\natexlab{}.
\newblock \showarticletitle{Deep {{Bayesian Active Learning}} with {{Image Data}}}. In \bibinfo{booktitle}{\emph{Proceedings of the 34th {{International Conference}} on {{Machine Learning}}}}. \bibinfo{publisher}{PMLR}, \bibinfo{pages}{1183--1192}.
\newblock
\showISSN{2640-3498}


\bibitem[Gonz{\'a}lez et~al\mbox{.}(2017)]%
        {gonzalez2017Preferential}
\bibfield{author}{\bibinfo{person}{Javier Gonz{\'a}lez}, \bibinfo{person}{Zhenwen Dai}, \bibinfo{person}{Andreas Damianou}, {and} \bibinfo{person}{Neil~D. Lawrence}.} \bibinfo{year}{2017}\natexlab{}.
\newblock \showarticletitle{Preferential {{Bayesian Optimization}}}. In \bibinfo{booktitle}{\emph{Proceedings of the 34th {{International Conference}} on {{Machine Learning}}}}. \bibinfo{publisher}{PMLR}, \bibinfo{pages}{1282--1291}.
\newblock
\showISSN{2640-3498}


\bibitem[Guo et~al\mbox{.}(2025)]%
        {guo2024PrompTHis}
\bibfield{author}{\bibinfo{person}{Yuhan Guo}, \bibinfo{person}{Hanning Shao}, \bibinfo{person}{Can Liu}, \bibinfo{person}{Kai Xu}, {and} \bibinfo{person}{Xiaoru Yuan}.} \bibinfo{year}{2025}\natexlab{}.
\newblock \showarticletitle{{{PrompTHis}}: {{Visualizing}} the {{Process}} and {{Influence}} of {{Prompt Editing}} during {{Text-to-Image Creation}}}.
\newblock  \bibinfo{volume}{31}, \bibinfo{number}{9} (\bibinfo{year}{2025}), \bibinfo{pages}{4547--4559}.
\newblock
\href{https://doi.org/10.1109/TVCG.2024.3408255}{doi:\nolinkurl{10.1109/TVCG.2024.3408255}}


\bibitem[Gupta et~al\mbox{.}(2024)]%
        {gupta2023Bias}
\bibfield{author}{\bibinfo{person}{Shashank Gupta}, \bibinfo{person}{Vaishnavi Shrivastava}, \bibinfo{person}{Ameet Deshpande}, \bibinfo{person}{Ashwin Kalyan}, \bibinfo{person}{Peter Clark}, \bibinfo{person}{Ashish Sabharwal}, {and} \bibinfo{person}{Tushar Khot}.} \bibinfo{year}{2024}\natexlab{}.
\newblock \showarticletitle{Bias {{Runs Deep}}: {{Implicit Reasoning Biases}} in {{Persona-Assigned LLMs}}}. In \bibinfo{booktitle}{\emph{The {{Twelfth International Conference}} on {{Learning Representations}}}}.
\newblock


\bibitem[Hahn et~al\mbox{.}(2025)]%
        {hahn2025Proactive}
\bibfield{author}{\bibinfo{person}{Meera Hahn}, \bibinfo{person}{Wenjun Zeng}, \bibinfo{person}{Nithish Kannen}, \bibinfo{person}{Rich Galt}, \bibinfo{person}{Kartikeya Badola}, \bibinfo{person}{Been Kim}, {and} \bibinfo{person}{Zi Wang}.} \bibinfo{year}{2025}\natexlab{}.
\newblock \showarticletitle{Proactive {{Agents}} for {{Multi-Turn Text-to-Image Generation Under Uncertainty}}}. In \bibinfo{booktitle}{\emph{Forty-Second {{International Conference}} on {{Machine Learning}}}}.
\newblock


\bibitem[Han and Fussell(2025)]%
        {han2025Understanding}
\bibfield{author}{\bibinfo{person}{Shu-Jung Han} {and} \bibinfo{person}{Susan~R. Fussell}.} \bibinfo{year}{2025}\natexlab{}.
\newblock \showarticletitle{Understanding {{User Perceptions}} and the {{Role}} of {{AI Image Generators}} in {{Image Creation Workflows}}}. In \bibinfo{booktitle}{\emph{Proceedings of the 2025 {{CHI Conference}} on {{Human Factors}} in {{Computing Systems}}}} (New York, NY, USA, 2025-04-25) \emph{(\bibinfo{series}{{{CHI}} '25})}. \bibinfo{publisher}{Association for Computing Machinery}, \bibinfo{pages}{1--17}.
\newblock
\showISBNx{979-8-4007-1394-1}
\href{https://doi.org/10.1145/3706598.3713227}{doi:\nolinkurl{10.1145/3706598.3713227}}


\bibitem[Hao et~al\mbox{.}(2023)]%
        {hao2023Optimizing}
\bibfield{author}{\bibinfo{person}{Yaru Hao}, \bibinfo{person}{Zewen Chi}, \bibinfo{person}{Li Dong}, {and} \bibinfo{person}{Furu Wei}.} \bibinfo{year}{2023}\natexlab{}.
\newblock \showarticletitle{Optimizing {{Prompts}} for {{Text-to-Image Generation}}}.
\newblock \bibinfo{journal}{\emph{Advances in Neural Information Processing Systems}}  \bibinfo{volume}{36} (\bibinfo{year}{2023}), \bibinfo{pages}{66923--66939}.
\newblock


\bibitem[Hart and Staveland(1988)]%
        {hart1988development}
\bibfield{author}{\bibinfo{person}{Sandra~G Hart} {and} \bibinfo{person}{Lowell~E Staveland}.} \bibinfo{year}{1988}\natexlab{}.
\newblock \showarticletitle{Development of NASA-TLX (Task Load Index): Results of empirical and theoretical research}.
\newblock In \bibinfo{booktitle}{\emph{Advances in psychology}}. Vol.~\bibinfo{volume}{52}. \bibinfo{publisher}{Elsevier}, \bibinfo{pages}{139--183}.
\newblock


\bibitem[He et~al\mbox{.}(2025)]%
        {he2025automated}
\bibfield{author}{\bibinfo{person}{Yutong He}, \bibinfo{person}{Alexander Robey}, \bibinfo{person}{Naoki Murata}, \bibinfo{person}{Yiding Jiang}, \bibinfo{person}{Joshua~Nathaniel Williams}, \bibinfo{person}{George~J. Pappas}, \bibinfo{person}{Hamed Hassani}, \bibinfo{person}{Yuki Mitsufuji}, \bibinfo{person}{Ruslan Salakhutdinov}, {and} \bibinfo{person}{J~Zico Kolter}.} \bibinfo{year}{2025}\natexlab{}.
\newblock \showarticletitle{Automated Black-box Prompt Engineering for Personalized Text-to-Image Generation}.
\newblock \bibinfo{journal}{\emph{Transactions on Machine Learning Research}} (\bibinfo{year}{2025}).
\newblock
\showISSN{2835-8856}


\bibitem[Hei et~al\mbox{.}(2024)]%
        {hei2024UserFriendly}
\bibfield{author}{\bibinfo{person}{Nailei Hei}, \bibinfo{person}{Qianyu Guo}, \bibinfo{person}{Zihao Wang}, \bibinfo{person}{Yan Wang}, \bibinfo{person}{Haofen Wang}, {and} \bibinfo{person}{Wenqiang Zhang}.} \bibinfo{year}{2024}\natexlab{}.
\newblock \showarticletitle{A {{User-Friendly Framework}} for {{Generating Model-Preferred Prompts}} in {{Text-to-Image Synthesis}}}.
\newblock \bibinfo{journal}{\emph{Proceedings of the AAAI Conference on Artificial Intelligence}} \bibinfo{volume}{38}, \bibinfo{number}{3} (\bibinfo{date}{March} \bibinfo{year}{2024}), \bibinfo{pages}{2139--2147}.
\newblock
\showISSN{2374-3468}
\href{https://doi.org/10.1609/aaai.v38i3.27986}{doi:\nolinkurl{10.1609/aaai.v38i3.27986}}


\bibitem[Houlsby et~al\mbox{.}(2011)]%
        {houlsby2011bayesian}
\bibfield{author}{\bibinfo{person}{Neil Houlsby}, \bibinfo{person}{Ferenc Huszár}, \bibinfo{person}{Zoubin Ghahramani}, {and} \bibinfo{person}{Máté Lengyel}.} \bibinfo{year}{2011}\natexlab{}.
\newblock \bibinfo{title}{Bayesian Active Learning for Classification and Preference Learning}.
\newblock
\showeprint[arxiv]{1112.5745}~[stat.ML]
\urldef\tempurl%
\url{https://arxiv.org/abs/1112.5745}
\showURL{%
\tempurl}


\bibitem[Hsieh and Shannon(2005)]%
        {hsieh2005three}
\bibfield{author}{\bibinfo{person}{Hsiu-Fang Hsieh} {and} \bibinfo{person}{Sarah~E Shannon}.} \bibinfo{year}{2005}\natexlab{}.
\newblock \showarticletitle{Three approaches to qualitative content analysis}.
\newblock \bibinfo{journal}{\emph{Qualitative health research}} \bibinfo{volume}{15}, \bibinfo{number}{9} (\bibinfo{year}{2005}), \bibinfo{pages}{1277--1288}.
\newblock


\bibitem[Huang et~al\mbox{.}(2024)]%
        {huang2024Amortized}
\bibfield{author}{\bibinfo{person}{Daolang Huang}, \bibinfo{person}{Yujia Guo}, \bibinfo{person}{Luigi Acerbi}, {and} \bibinfo{person}{Samuel Kaski}.} \bibinfo{year}{2024}\natexlab{}.
\newblock \showarticletitle{Amortized {{Bayesian Experimental Design}} for {{Decision-Making}}}.
\newblock \bibinfo{journal}{\emph{Advances in Neural Information Processing Systems}}  \bibinfo{volume}{37} (\bibinfo{year}{2024}), \bibinfo{pages}{109460--109486}.
\newblock


\bibitem[Jain et~al\mbox{.}(2025)]%
        {jain2025AdaptiveSliders}
\bibfield{author}{\bibinfo{person}{Rahul Jain}, \bibinfo{person}{Amit Goel}, \bibinfo{person}{Koichiro Niinuma}, {and} \bibinfo{person}{Aakar Gupta}.} \bibinfo{year}{2025}\natexlab{}.
\newblock \showarticletitle{{{AdaptiveSliders}}: {{User-aligned Semantic Slider-based Editing}} of {{Text-to-Image Model Output}}}. In \bibinfo{booktitle}{\emph{Proceedings of the 2025 {{CHI Conference}} on {{Human Factors}} in {{Computing Systems}}}} \emph{(\bibinfo{series}{{{CHI}} '25})}. \bibinfo{publisher}{Association for Computing Machinery}, \bibinfo{address}{New York, NY, USA}, \bibinfo{pages}{1--27}.
\newblock
\showISBNx{979-8-4007-1394-1}
\href{https://doi.org/10.1145/3706598.3714292}{doi:\nolinkurl{10.1145/3706598.3714292}}


\bibitem[Ji et~al\mbox{.}(2025)]%
        {ji2025reinforcement}
\bibfield{author}{\bibinfo{person}{Kaixuan Ji}, \bibinfo{person}{Jiafan He}, {and} \bibinfo{person}{Quanquan Gu}.} \bibinfo{year}{2025}\natexlab{}.
\newblock \showarticletitle{Reinforcement Learning from Human Feedback with Active Queries}.
\newblock \bibinfo{journal}{\emph{Transactions on Machine Learning Research}} (\bibinfo{year}{2025}).
\newblock
\showISSN{2835-8856}


\bibitem[Kim et~al\mbox{.}(2025)]%
        {kim2025Journey}
\bibfield{author}{\bibinfo{person}{Soomin Kim}, \bibinfo{person}{Jinsu Eun}, \bibinfo{person}{Changhoon Oh}, {and} \bibinfo{person}{Joonhwan Lee}.} \bibinfo{year}{2025}\natexlab{}.
\newblock \showarticletitle{“{{Journey}} of {{Finding}} the {{Best Query}}”: {{Understanding}} the {{User Experience}} of {{AI Image Generation System}}}.
\newblock \bibinfo{journal}{\emph{International Journal of Human–Computer Interaction}} \bibinfo{volume}{41}, \bibinfo{number}{2} (\bibinfo{year}{2025}), \bibinfo{pages}{951--969}.
\newblock
\showISSN{1044-7318}
\href{https://doi.org/10.1080/10447318.2024.2307670}{doi:\nolinkurl{10.1080/10447318.2024.2307670}}


\bibitem[Kirsch et~al\mbox{.}(2019)]%
        {kirsch2019batchbald}
\bibfield{author}{\bibinfo{person}{Andreas Kirsch}, \bibinfo{person}{Joost van Amersfoort}, {and} \bibinfo{person}{Yarin Gal}.} \bibinfo{year}{2019}\natexlab{}.
\newblock \showarticletitle{{{BatchBALD}}: {{Efficient}} and {{Diverse Batch Acquisition}} for {{Deep Bayesian Active Learning}}}.
\newblock \bibinfo{journal}{\emph{Advances in {{Neural Information Processing Systems}}}}  \bibinfo{volume}{32} (\bibinfo{year}{2019}).
\newblock


\bibitem[Kobalczyk et~al\mbox{.}(2025)]%
        {kobalczyk2025active}
\bibfield{author}{\bibinfo{person}{Kasia Kobalczyk}, \bibinfo{person}{Nicol{\'a}s Astorga}, \bibinfo{person}{Tennison Liu}, {and} \bibinfo{person}{Mihaela van~der Schaar}.} \bibinfo{year}{2025}\natexlab{}.
\newblock \showarticletitle{Active Task Disambiguation with {LLM}s}. In \bibinfo{booktitle}{\emph{The Thirteenth International Conference on Learning Representations}}.
\newblock


\bibitem[Koyama and Goto(2022)]%
        {koyama2022BO}
\bibfield{author}{\bibinfo{person}{Yuki Koyama} {and} \bibinfo{person}{Masataka Goto}.} \bibinfo{year}{2022}\natexlab{}.
\newblock \showarticletitle{{{BO}} as {{Assistant}}: {{Using Bayesian Optimization}} for {{Asynchronously Generating Design Suggestions}}}. In \bibinfo{booktitle}{\emph{Proceedings of the 35th {{Annual ACM Symposium}} on {{User Interface Software}} and {{Technology}}}} \emph{(\bibinfo{series}{{{UIST}} '22})}. \bibinfo{publisher}{Association for Computing Machinery}, \bibinfo{address}{New York, NY, USA}, \bibinfo{pages}{1--14}.
\newblock
\showISBNx{978-1-4503-9320-1}
\href{https://doi.org/10.1145/3526113.3545664}{doi:\nolinkurl{10.1145/3526113.3545664}}


\bibitem[Koyama et~al\mbox{.}(2020)]%
        {koyama2020sequential}
\bibfield{author}{\bibinfo{person}{Yuki Koyama}, \bibinfo{person}{Issei Sato}, {and} \bibinfo{person}{Masataka Goto}.} \bibinfo{year}{2020}\natexlab{}.
\newblock \showarticletitle{Sequential Gallery for Interactive Visual Design Optimization}.
\newblock \bibinfo{journal}{\emph{ACM Trans. Graph.}} \bibinfo{volume}{39}, \bibinfo{number}{4} (\bibinfo{date}{Aug.} \bibinfo{year}{2020}), \bibinfo{pages}{88:88:1--88:88:12}.
\newblock
\showISSN{0730-0301}
\href{https://doi.org/10.1145/3386569.3392444}{doi:\nolinkurl{10.1145/3386569.3392444}}


\bibitem[Labs(2025)]%
        {labs2025FLUX1a}
\bibfield{author}{\bibinfo{person}{Black~Forest Labs}.} \bibinfo{year}{2025}\natexlab{}.
\newblock \bibinfo{title}{{{FLUX}}.1 - a Black-Forest-Labs {{Collection}}}.
\newblock
\urldef\tempurl%
\url{https://huggingface.co/collections/black-forest-labs/flux1}
\showURL{%
\tempurl}
\newblock
\shownote{Accessed: 2025-10-10}.


\bibitem[Lee et~al\mbox{.}(2025)]%
        {lee2025ThematicPlane}
\bibfield{author}{\bibinfo{person}{Daniel Lee}, \bibinfo{person}{Nikhil Sharma}, \bibinfo{person}{Donghoon Shin}, \bibinfo{person}{DaEun Choi}, \bibinfo{person}{Harsh Sharma}, \bibinfo{person}{Jeonghwan Kim}, {and} \bibinfo{person}{Heng Ji}.} \bibinfo{year}{2025}\natexlab{}.
\newblock \showarticletitle{ThematicPlane: Bridging Tacit User Intent and Latent Spaces for Image Generation}. In \bibinfo{booktitle}{\emph{Adjunct Proceedings of the 38th Annual ACM Symposium on User Interface Software and Technology}} \emph{(\bibinfo{series}{UIST Adjunct '25})}. \bibinfo{publisher}{Association for Computing Machinery}, \bibinfo{address}{New York, NY, USA}, Article \bibinfo{articleno}{120}, \bibinfo{numpages}{3}~pages.
\newblock
\showISBNx{9798400720369}
\href{https://doi.org/10.1145/3746058.3758376}{doi:\nolinkurl{10.1145/3746058.3758376}}


\bibitem[Lee et~al\mbox{.}(2023)]%
        {lee2023aligning}
\bibfield{author}{\bibinfo{person}{Kimin Lee}, \bibinfo{person}{Hao Liu}, \bibinfo{person}{Moonkyung Ryu}, \bibinfo{person}{Olivia Watkins}, \bibinfo{person}{Yuqing Du}, \bibinfo{person}{Craig Boutilier}, \bibinfo{person}{Pieter Abbeel}, \bibinfo{person}{Mohammad Ghavamzadeh}, {and} \bibinfo{person}{Shixiang~Shane Gu}.} \bibinfo{year}{2023}\natexlab{}.
\newblock \bibinfo{title}{Aligning Text-to-Image Models using Human Feedback}.
\newblock
\showeprint[arxiv]{2302.12192}~[cs.LG]
\urldef\tempurl%
\url{https://arxiv.org/abs/2302.12192}
\showURL{%
\tempurl}


\bibitem[Li et~al\mbox{.}(2025)]%
        {li2025eliciting}
\bibfield{author}{\bibinfo{person}{Belinda~Z. Li}, \bibinfo{person}{Alex Tamkin}, \bibinfo{person}{Noah Goodman}, {and} \bibinfo{person}{Jacob Andreas}.} \bibinfo{year}{2025}\natexlab{}.
\newblock \showarticletitle{Eliciting Human Preferences with Language Models}. In \bibinfo{booktitle}{\emph{The Thirteenth International Conference on Learning Representations}}.
\newblock


\bibitem[Liang et~al\mbox{.}(2025)]%
        {liang2025IDEABench}
\bibfield{author}{\bibinfo{person}{Chen Liang}, \bibinfo{person}{Lianghua Huang}, \bibinfo{person}{Jingwu Fang}, \bibinfo{person}{Huanzhang Dou}, \bibinfo{person}{Wei Wang}, \bibinfo{person}{Zhi-Fan Wu}, \bibinfo{person}{Yupeng Shi}, \bibinfo{person}{Junge Zhang}, \bibinfo{person}{Xin Zhao}, {and} \bibinfo{person}{Yu Liu}.} \bibinfo{year}{2025}\natexlab{}.
\newblock \showarticletitle{{{IDEA-Bench}}: {{How Far}} Are {{Generative Models}} from {{Professional Designing}}?}. In \bibinfo{booktitle}{\emph{Proceedings of the {{IEEE}}/{{CVF Conference}} on {{Computer Vision}} and {{Pattern Recognition}}}}. \bibinfo{pages}{18541--18551}.
\newblock


\bibitem[Lin et~al\mbox{.}(2025)]%
        {lin2025SketchFlex}
\bibfield{author}{\bibinfo{person}{Haichuan Lin}, \bibinfo{person}{Yilin Ye}, \bibinfo{person}{Jiazhi Xia}, {and} \bibinfo{person}{Wei Zeng}.} \bibinfo{year}{2025}\natexlab{}.
\newblock \showarticletitle{{{SketchFlex}}: {{Facilitating Spatial-Semantic Coherence}} in {{Text-to-Image Generation}} with {{Region-Based Sketches}}}. In \bibinfo{booktitle}{\emph{Proceedings of the 2025 {{CHI Conference}} on {{Human Factors}} in {{Computing Systems}}}} \emph{(\bibinfo{series}{{{CHI}} '25})}. \bibinfo{publisher}{Association for Computing Machinery}, \bibinfo{address}{New York, NY, USA}, \bibinfo{pages}{1--19}.
\newblock
\showISBNx{979-8-4007-1394-1}
\href{https://doi.org/10.1145/3706598.3713801}{doi:\nolinkurl{10.1145/3706598.3713801}}


\bibitem[Lin et~al\mbox{.}(2024)]%
        {lin2024Evaluating}
\bibfield{author}{\bibinfo{person}{Zhiqiu Lin}, \bibinfo{person}{Deepak Pathak}, \bibinfo{person}{Baiqi Li}, \bibinfo{person}{Jiayao Li}, \bibinfo{person}{Xide Xia}, \bibinfo{person}{Graham Neubig}, \bibinfo{person}{Pengchuan Zhang}, {and} \bibinfo{person}{Deva Ramanan}.} \bibinfo{year}{2024}\natexlab{}.
\newblock \showarticletitle{Evaluating {{Text-to-Visual Generation}} with {{Image-to-Text Generation}}}. In \bibinfo{booktitle}{\emph{European {{Conference}} on {{Computer Vision}}}}. \bibinfo{publisher}{Springer}, \bibinfo{pages}{366--384}.
\newblock


\bibitem[Liu et~al\mbox{.}(2024)]%
        {liu2024Language}
\bibfield{author}{\bibinfo{person}{Shihong Liu}, \bibinfo{person}{Samuel Yu}, \bibinfo{person}{Zhiqiu Lin}, \bibinfo{person}{Deepak Pathak}, {and} \bibinfo{person}{Deva Ramanan}.} \bibinfo{year}{2024}\natexlab{}.
\newblock \showarticletitle{Language {{Models}} as {{Black-Box Optimizers}} for {{Vision-Language Models}}}. In \bibinfo{booktitle}{\emph{Proceedings of the {{IEEE}}/{{CVF Conference}} on {{Computer Vision}} and {{Pattern Recognition}}}}. \bibinfo{pages}{12687--12697}.
\newblock


\bibitem[Liu and Chilton(2022)]%
        {liu2022Design}
\bibfield{author}{\bibinfo{person}{Vivian Liu} {and} \bibinfo{person}{Lydia~B Chilton}.} \bibinfo{year}{2022}\natexlab{}.
\newblock \showarticletitle{Design {{Guidelines}} for {{Prompt Engineering Text-to-Image Generative Models}}}. In \bibinfo{booktitle}{\emph{Proceedings of the 2022 {{CHI Conference}} on {{Human Factors}} in {{Computing Systems}}}} (New York, NY, USA, 2022-04-29) \emph{(\bibinfo{series}{{{CHI}} '22})}. \bibinfo{publisher}{Association for Computing Machinery}, \bibinfo{pages}{1--23}.
\newblock
\showISBNx{978-1-4503-9157-3}
\href{https://doi.org/10.1145/3491102.3501825}{doi:\nolinkurl{10.1145/3491102.3501825}}


\bibitem[Louie et~al\mbox{.}(2020)]%
        {louie2020novice}
\bibfield{author}{\bibinfo{person}{Ryan Louie}, \bibinfo{person}{Andy Coenen}, \bibinfo{person}{Cheng~Zhi Huang}, \bibinfo{person}{Michael Terry}, {and} \bibinfo{person}{Carrie~J. Cai}.} \bibinfo{year}{2020}\natexlab{}.
\newblock \showarticletitle{Novice-AI Music Co-Creation via AI-Steering Tools for Deep Generative Models}. In \bibinfo{booktitle}{\emph{Proceedings of the 2020 CHI Conference on Human Factors in Computing Systems}} (Honolulu, HI, USA) \emph{(\bibinfo{series}{CHI '20})}. \bibinfo{publisher}{Association for Computing Machinery}, \bibinfo{address}{New York, NY, USA}, \bibinfo{pages}{1–13}.
\newblock
\showISBNx{9781450367080}
\href{https://doi.org/10.1145/3313831.3376739}{doi:\nolinkurl{10.1145/3313831.3376739}}


\bibitem[Ma et~al\mbox{.}(2019)]%
        {ma2019smarteye}
\bibfield{author}{\bibinfo{person}{Shuai Ma}, \bibinfo{person}{Zijun Wei}, \bibinfo{person}{Feng Tian}, \bibinfo{person}{Xiangmin Fan}, \bibinfo{person}{Jianming Zhang}, \bibinfo{person}{Xiaohui Shen}, \bibinfo{person}{Zhe Lin}, \bibinfo{person}{Jin Huang}, \bibinfo{person}{Radom{\'\i}r M{\v{e}}ch}, \bibinfo{person}{Dimitris Samaras}, {et~al\mbox{.}}} \bibinfo{year}{2019}\natexlab{}.
\newblock \showarticletitle{SmartEye: assisting instant photo taking via integrating user preference with deep view proposal network}. In \bibinfo{booktitle}{\emph{Proceedings of the 2019 CHI conference on human factors in computing systems}}. \bibinfo{pages}{1--12}.
\newblock


\bibitem[Mahdavi~Goloujeh et~al\mbox{.}(2024)]%
        {mahdavigoloujeh2024It}
\bibfield{author}{\bibinfo{person}{Atefeh Mahdavi~Goloujeh}, \bibinfo{person}{Anne Sullivan}, {and} \bibinfo{person}{Brian Magerko}.} \bibinfo{year}{2024}\natexlab{}.
\newblock \showarticletitle{Is {{It AI}} or {{Is It Me}}? {{Understanding Users}}’ {{Prompt Journey}} with {{Text-to-Image Generative AI Tools}}}. In \bibinfo{booktitle}{\emph{Proceedings of the 2024 {{CHI Conference}} on {{Human Factors}} in {{Computing Systems}}}} (New York, NY, USA, 2024-05-11) \emph{(\bibinfo{series}{{{CHI}} '24})}. \bibinfo{publisher}{Association for Computing Machinery}, \bibinfo{pages}{1--13}.
\newblock
\showISBNx{979-8-4007-0330-0}
\href{https://doi.org/10.1145/3613904.3642861}{doi:\nolinkurl{10.1145/3613904.3642861}}


\bibitem[Mahmud et~al\mbox{.}(2025)]%
        {mahmud2025MAPLEa}
\bibfield{author}{\bibinfo{person}{Saaduddin Mahmud}, \bibinfo{person}{Mason Nakamura}, {and} \bibinfo{person}{Shlomo Zilberstein}.} \bibinfo{year}{2025}\natexlab{}.
\newblock \showarticletitle{{{MAPLE}}: {{A Framework}} for {{Active Preference Learning Guided}} by {{Large Language Models}}}.
\newblock \bibinfo{journal}{\emph{Proceedings of the AAAI Conference on Artificial Intelligence}} \bibinfo{volume}{39}, \bibinfo{number}{26} (\bibinfo{date}{April} \bibinfo{year}{2025}), \bibinfo{pages}{27518--27528}.
\newblock
\showISSN{2374-3468}
\href{https://doi.org/10.1609/aaai.v39i26.34964}{doi:\nolinkurl{10.1609/aaai.v39i26.34964}}


\bibitem[Ma{\~n}as et~al\mbox{.}(2024)]%
        {oscar2024improving}
\bibfield{author}{\bibinfo{person}{Oscar Ma{\~n}as}, \bibinfo{person}{Pietro Astolfi}, \bibinfo{person}{Melissa Hall}, \bibinfo{person}{Candace Ross}, \bibinfo{person}{Jack Urbanek}, \bibinfo{person}{Adina Williams}, \bibinfo{person}{Aishwarya Agrawal}, \bibinfo{person}{Adriana Romero-Soriano}, {and} \bibinfo{person}{Michal Drozdzal}.} \bibinfo{year}{2024}\natexlab{}.
\newblock \showarticletitle{Improving Text-to-Image Consistency via Automatic Prompt Optimization}.
\newblock \bibinfo{journal}{\emph{Transactions on Machine Learning Research}} (\bibinfo{year}{2024}).
\newblock
\showISSN{2835-8856}


\bibitem[Mann and Whitney(1947)]%
        {mann1947Test}
\bibfield{author}{\bibinfo{person}{Henry~B Mann} {and} \bibinfo{person}{Donald~R Whitney}.} \bibinfo{year}{1947}\natexlab{}.
\newblock \showarticletitle{On a {{Test}} of {{Whether}} One of {{Two Random Variables}} Is {{Stochastically Larger}} than the {{Other}}}.
\newblock \bibinfo{journal}{\emph{The Annals of Mathematical Statistics}} \bibinfo{volume}{18}, \bibinfo{number}{1} (\bibinfo{year}{1947}), \bibinfo{pages}{50--60}.
\newblock
\showISSN{00034851}
\showeprint[jstor]{2236101}


\bibitem[OpenAI(2025)]%
        {2025Vector}
\bibfield{author}{\bibinfo{person}{OpenAI}.} \bibinfo{year}{2025}\natexlab{}.
\newblock \bibinfo{title}{Vector embeddings}.
\newblock
\urldef\tempurl%
\url{https://platform.openai.com/docs/guides/embeddings}
\showURL{%
\tempurl}
\newblock
\shownote{Accessed: 2025-10-10}.


\bibitem[Oppenlaender(2024)]%
        {oppenlaender2024Taxonomy}
\bibfield{author}{\bibinfo{person}{Jonas Oppenlaender}.} \bibinfo{year}{2024}\natexlab{}.
\newblock \showarticletitle{A Taxonomy of Prompt Modifiers for Text-to-Image Generation}.
\newblock \bibinfo{journal}{\emph{Behaviour \& Information Technology}} \bibinfo{volume}{43}, \bibinfo{number}{15} (\bibinfo{date}{Nov.} \bibinfo{year}{2024}), \bibinfo{pages}{3763--3776}.
\newblock
\showISSN{0144-929X}
\href{https://doi.org/10.1080/0144929X.2023.2286532}{doi:\nolinkurl{10.1080/0144929X.2023.2286532}}


\bibitem[Oquab et~al\mbox{.}(2024)]%
        {oquab2024DINOv2}
\bibfield{author}{\bibinfo{person}{Maxime Oquab}, \bibinfo{person}{Timoth{\'e}e Darcet}, \bibinfo{person}{Th{\'e}o Moutakanni}, \bibinfo{person}{Huy~V. Vo}, \bibinfo{person}{Marc Szafraniec}, \bibinfo{person}{Vasil Khalidov}, \bibinfo{person}{Pierre Fernandez}, \bibinfo{person}{Daniel Haziza}, \bibinfo{person}{Francisco Massa}, \bibinfo{person}{Alaaeldin {El-Nouby}}, \bibinfo{person}{Mido Assran}, \bibinfo{person}{Nicolas Ballas}, \bibinfo{person}{Wojciech Galuba}, \bibinfo{person}{Russell Howes}, \bibinfo{person}{Po-Yao Huang}, \bibinfo{person}{Shang-Wen Li}, \bibinfo{person}{Ishan Misra}, \bibinfo{person}{Michael Rabbat}, \bibinfo{person}{Vasu Sharma}, \bibinfo{person}{Gabriel Synnaeve}, \bibinfo{person}{Hu Xu}, \bibinfo{person}{Herve Jegou}, \bibinfo{person}{Julien Mairal}, \bibinfo{person}{Patrick Labatut}, \bibinfo{person}{Armand Joulin}, {and} \bibinfo{person}{Piotr Bojanowski}.} \bibinfo{year}{2024}\natexlab{}.
\newblock \showarticletitle{{{DINOv2}}: {{Learning Robust Visual Features}} without {{Supervision}}}.
\newblock \bibinfo{journal}{\emph{Transactions on Machine Learning Research}} (\bibinfo{year}{2024}).
\newblock
\showISSN{2835-8856}


\bibitem[Ouyang et~al\mbox{.}(2022)]%
        {ouyang2022training}
\bibfield{author}{\bibinfo{person}{Long Ouyang}, \bibinfo{person}{Jeffrey Wu}, \bibinfo{person}{Xu Jiang}, \bibinfo{person}{Diogo Almeida}, \bibinfo{person}{Carroll Wainwright}, \bibinfo{person}{Pamela Mishkin}, \bibinfo{person}{Chong Zhang}, \bibinfo{person}{Sandhini Agarwal}, \bibinfo{person}{Katarina Slama}, \bibinfo{person}{Alex Ray}, \bibinfo{person}{John Schulman}, \bibinfo{person}{Jacob Hilton}, \bibinfo{person}{Fraser Kelton}, \bibinfo{person}{Luke Miller}, \bibinfo{person}{Maddie Simens}, \bibinfo{person}{Amanda Askell}, \bibinfo{person}{Peter Welinder}, \bibinfo{person}{Paul~F. Christiano}, \bibinfo{person}{Jan Leike}, {and} \bibinfo{person}{Ryan Lowe}.} \bibinfo{year}{2022}\natexlab{}.
\newblock \showarticletitle{Training Language Models to Follow Instructions with Human Feedback}.
\newblock \bibinfo{journal}{\emph{Advances in Neural Information Processing Systems}}  \bibinfo{volume}{35} (\bibinfo{year}{2022}), \bibinfo{pages}{27730--27744}.
\newblock


\bibitem[Palani and Ramos(2024)]%
        {palani2024Evolving}
\bibfield{author}{\bibinfo{person}{Srishti Palani} {and} \bibinfo{person}{Gonzalo Ramos}.} \bibinfo{year}{2024}\natexlab{}.
\newblock \showarticletitle{Evolving {{Roles}} and {{Workflows}} of {{Creative Practitioners}} in the {{Age}} of {{Generative AI}}}. In \bibinfo{booktitle}{\emph{Proceedings of the 16th {{Conference}} on {{Creativity}} \& {{Cognition}}}} \emph{(\bibinfo{series}{C\&C '24})}. \bibinfo{publisher}{Association for Computing Machinery}, \bibinfo{address}{New York, NY, USA}, \bibinfo{pages}{170--184}.
\newblock
\showISBNx{979-8-4007-0485-7}
\href{https://doi.org/10.1145/3635636.3656190}{doi:\nolinkurl{10.1145/3635636.3656190}}


\bibitem[Rafailov et~al\mbox{.}(2023)]%
        {rafailov2023direct}
\bibfield{author}{\bibinfo{person}{Rafael Rafailov}, \bibinfo{person}{Archit Sharma}, \bibinfo{person}{Eric Mitchell}, \bibinfo{person}{Christopher~D. Manning}, \bibinfo{person}{Stefano Ermon}, {and} \bibinfo{person}{Chelsea Finn}.} \bibinfo{year}{2023}\natexlab{}.
\newblock \showarticletitle{Direct {{Preference Optimization}}: {{Your Language Model}} Is {{Secretly}} a {{Reward Model}}}. In \bibinfo{booktitle}{\emph{Advances in {{Neural Information Processing Systems}}}}, Vol.~\bibinfo{volume}{36}. \bibinfo{pages}{53728--53741}.
\newblock


\bibitem[Rainforth et~al\mbox{.}(2024)]%
        {rainforth2023modern}
\bibfield{author}{\bibinfo{person}{Tom Rainforth}, \bibinfo{person}{Adam Foster}, \bibinfo{person}{Desi~R Ivanova}, {and} \bibinfo{person}{Freddie Bickford~Smith}.} \bibinfo{year}{2024}\natexlab{}.
\newblock \showarticletitle{Modern Bayesian Experimental Design}.
\newblock \bibinfo{journal}{\emph{Statist. Sci.}} \bibinfo{volume}{39}, \bibinfo{number}{1} (\bibinfo{year}{2024}), \bibinfo{pages}{100--114}.
\newblock


\bibitem[Resnik(2025)]%
        {resnik2025Large}
\bibfield{author}{\bibinfo{person}{Philip Resnik}.} \bibinfo{year}{2025}\natexlab{}.
\newblock \showarticletitle{Large {{Language Models Are Biased Because They Are Large Language Models}}}.
\newblock \bibinfo{journal}{\emph{Computational Linguistics}} \bibinfo{volume}{51}, \bibinfo{number}{3} (\bibinfo{date}{Sept.} \bibinfo{year}{2025}), \bibinfo{pages}{885--906}.
\newblock
\showISSN{0891-2017}
\href{https://doi.org/10.1162/coli_a_00558}{doi:\nolinkurl{10.1162/coli_a_00558}}


\bibitem[Rezwana and Maher(2023)]%
        {rezwana2022designing}
\bibfield{author}{\bibinfo{person}{Jeba Rezwana} {and} \bibinfo{person}{Mary~Lou Maher}.} \bibinfo{year}{2023}\natexlab{}.
\newblock \showarticletitle{Designing Creative AI Partners with COFI: A Framework for Modeling Interaction in Human-AI Co-Creative Systems}.
\newblock \bibinfo{journal}{\emph{ACM Trans. Comput.-Hum. Interact.}} \bibinfo{volume}{30}, \bibinfo{number}{5}, Article \bibinfo{articleno}{67} (\bibinfo{date}{Sept.} \bibinfo{year}{2023}), \bibinfo{numpages}{28}~pages.
\newblock
\showISSN{1073-0516}
\href{https://doi.org/10.1145/3519026}{doi:\nolinkurl{10.1145/3519026}}


\bibitem[Rombach et~al\mbox{.}(2022)]%
        {rombach2022high}
\bibfield{author}{\bibinfo{person}{Robin Rombach}, \bibinfo{person}{Andreas Blattmann}, \bibinfo{person}{Dominik Lorenz}, \bibinfo{person}{Patrick Esser}, {and} \bibinfo{person}{Bj{\"o}rn Ommer}.} \bibinfo{year}{2022}\natexlab{}.
\newblock \showarticletitle{High-Resolution Image Synthesis with Latent Diffusion Models}. In \bibinfo{booktitle}{\emph{Proceedings of the IEEE/CVF Conference on Computer Vision and Pattern Recognition (CVPR)}}. \bibinfo{pages}{10684--10695}.
\newblock


\bibitem[Saharia et~al\mbox{.}(2022)]%
        {saharia2022photorealistic}
\bibfield{author}{\bibinfo{person}{Chitwan Saharia}, \bibinfo{person}{William Chan}, \bibinfo{person}{Saurabh Saxena}, \bibinfo{person}{Lala Li}, \bibinfo{person}{Jay Whang}, \bibinfo{person}{Emily~L Denton}, \bibinfo{person}{Kamyar Ghasemipour}, \bibinfo{person}{Raphael Gontijo~Lopes}, \bibinfo{person}{Burcu Karagol~Ayan}, \bibinfo{person}{Tim Salimans}, \bibinfo{person}{Jonathan Ho}, \bibinfo{person}{David~J Fleet}, {and} \bibinfo{person}{Mohammad Norouzi}.} \bibinfo{year}{2022}\natexlab{}.
\newblock \showarticletitle{Photorealistic Text-to-Image Diffusion Models with Deep Language Understanding}.
\newblock \bibinfo{journal}{\emph{Advances in Neural Information Processing Systems}}  \bibinfo{volume}{35} (\bibinfo{year}{2022}), \bibinfo{pages}{36479--36494}.
\newblock


\bibitem[Sanchez(2023)]%
        {sanchez2023Examining}
\bibfield{author}{\bibinfo{person}{Téo Sanchez}.} \bibinfo{year}{2023}\natexlab{}.
\newblock \showarticletitle{Examining the {{Text-to-Image Community}} of {{Practice}}: {{Why}} and {{How}} Do {{People Prompt Generative AIs}}?}. In \bibinfo{booktitle}{\emph{Proceedings of the 15th {{Conference}} on {{Creativity}} and {{Cognition}}}} (New York, NY, USA, 2023-06-19) \emph{(\bibinfo{series}{C\&C '23})}. \bibinfo{publisher}{Association for Computing Machinery}, \bibinfo{pages}{43--61}.
\newblock
\showISBNx{979-8-4007-0180-1}
\href{https://doi.org/10.1145/3591196.3593051}{doi:\nolinkurl{10.1145/3591196.3593051}}


\bibitem[Schooler and Engstler-Schooler(1990)]%
        {schooler1990verbal}
\bibfield{author}{\bibinfo{person}{Jonathan~W Schooler} {and} \bibinfo{person}{Tonya~Y Engstler-Schooler}.} \bibinfo{year}{1990}\natexlab{}.
\newblock \showarticletitle{Verbal overshadowing of visual memories: Some things are better left unsaid}.
\newblock \bibinfo{journal}{\emph{Cognitive Psychology}} \bibinfo{volume}{22}, \bibinfo{number}{1} (\bibinfo{year}{1990}), \bibinfo{pages}{36--71}.
\newblock
\showISSN{0010-0285}
\href{https://doi.org/10.1016/0010-0285(90)90003-M}{doi:\nolinkurl{10.1016/0010-0285(90)90003-M}}


\bibitem[Schulman et~al\mbox{.}(2017)]%
        {schulman2017Proximal}
\bibfield{author}{\bibinfo{person}{John Schulman}, \bibinfo{person}{Filip Wolski}, \bibinfo{person}{Prafulla Dhariwal}, \bibinfo{person}{Alec Radford}, {and} \bibinfo{person}{Oleg Klimov}.} \bibinfo{year}{2017}\natexlab{}.
\newblock \bibinfo{title}{Proximal Policy Optimization Algorithms}.
\newblock
\showeprint[arxiv]{1707.06347}~[cs.LG]
\urldef\tempurl%
\url{https://arxiv.org/abs/1707.06347}
\showURL{%
\tempurl}


\bibitem[Shen et~al\mbox{.}(2025)]%
        {shen2025Position}
\bibfield{author}{\bibinfo{person}{Hua Shen}, \bibinfo{person}{Tiffany Knearem}, \bibinfo{person}{Reshmi Ghosh}, \bibinfo{person}{Kenan Alkiek}, \bibinfo{person}{Kundan Krishna}, \bibinfo{person}{Yachuan Liu}, \bibinfo{person}{Savvas Petridis}, \bibinfo{person}{Yi-Hao Peng}, \bibinfo{person}{Li Qiwei}, \bibinfo{person}{Chenglei Si}, \bibinfo{person}{Yutong Xie}, \bibinfo{person}{Jeffrey~P. Bigham}, \bibinfo{person}{Frank Bentley}, \bibinfo{person}{Joyce Chai}, \bibinfo{person}{Zachary~Chase Lipton}, \bibinfo{person}{Qiaozhu Mei}, \bibinfo{person}{Michael Terry}, \bibinfo{person}{Diyi Yang}, \bibinfo{person}{Meredith~Ringel Morris}, \bibinfo{person}{Paul Resnick}, {and} \bibinfo{person}{David Jurgens}.} \bibinfo{year}{2025}\natexlab{}.
\newblock \showarticletitle{Position: {{Towards Bidirectional Human-AI Alignment}}}. In \bibinfo{booktitle}{\emph{The {{Thirty-Ninth Annual Conference}} on {{Neural Information Processing Systems Position Paper Track}}}}.
\newblock


\bibitem[Shi et~al\mbox{.}(2025)]%
        {shi2025Brickify}
\bibfield{author}{\bibinfo{person}{Xinyu Shi}, \bibinfo{person}{Yinghou Wang}, \bibinfo{person}{Ryan Rossi}, {and} \bibinfo{person}{Jian Zhao}.} \bibinfo{year}{2025}\natexlab{}.
\newblock \showarticletitle{Brickify: {{Enabling Expressive Design Intent Specification}} through {{Direct Manipulation}} on {{Design Tokens}}}. In \bibinfo{booktitle}{\emph{Proceedings of the 2025 {{CHI Conference}} on {{Human Factors}} in {{Computing Systems}}}} \emph{(\bibinfo{series}{{{CHI}} '25})}. \bibinfo{publisher}{Association for Computing Machinery}, \bibinfo{address}{New York, NY, USA}, \bibinfo{pages}{1--20}.
\newblock
\showISBNx{979-8-4007-1394-1}
\href{https://doi.org/10.1145/3706598.3714087}{doi:\nolinkurl{10.1145/3706598.3714087}}


\bibitem[Singer et~al\mbox{.}(2023)]%
        {singer2023makeavideo}
\bibfield{author}{\bibinfo{person}{Uriel Singer}, \bibinfo{person}{Adam Polyak}, \bibinfo{person}{Thomas Hayes}, \bibinfo{person}{Xi Yin}, \bibinfo{person}{Jie An}, \bibinfo{person}{Songyang Zhang}, \bibinfo{person}{Qiyuan Hu}, \bibinfo{person}{Harry Yang}, \bibinfo{person}{Oron Ashual}, \bibinfo{person}{Oran Gafni}, \bibinfo{person}{Devi Parikh}, \bibinfo{person}{Sonal Gupta}, {and} \bibinfo{person}{Yaniv Taigman}.} \bibinfo{year}{2023}\natexlab{}.
\newblock \showarticletitle{Make-A-Video: Text-to-Video Generation without Text-Video Data}. In \bibinfo{booktitle}{\emph{The Eleventh International Conference on Learning Representations}}.
\newblock


\bibitem[Son et~al\mbox{.}(2022)]%
        {son2022BIGexplore}
\bibfield{author}{\bibinfo{person}{Kihoon Son}, \bibinfo{person}{Kyungmin Kim}, {and} \bibinfo{person}{Kyung~Hoon Hyun}.} \bibinfo{year}{2022}\natexlab{}.
\newblock \showarticletitle{{{BIGexplore}}: {{Bayesian Information Gain Framework}} for {{Information Exploration}}}. In \bibinfo{booktitle}{\emph{Proceedings of the 2022 {{CHI Conference}} on {{Human Factors}} in {{Computing Systems}}}} \emph{(\bibinfo{series}{{{CHI}} '22})}. \bibinfo{publisher}{Association for Computing Machinery}, \bibinfo{address}{New York, NY, USA}, \bibinfo{pages}{1--16}.
\newblock
\showISBNx{978-1-4503-9157-3}
\href{https://doi.org/10.1145/3491102.3517729}{doi:\nolinkurl{10.1145/3491102.3517729}}


\bibitem[Stokes(2005)]%
        {stokes2005creativity}
\bibfield{author}{\bibinfo{person}{Patricia~D Stokes}.} \bibinfo{year}{2005}\natexlab{}.
\newblock \bibinfo{booktitle}{\emph{Creativity from Constraints: The Psychology of Breakthrough}}.
\newblock \bibinfo{publisher}{Springer Publishing Company}.
\newblock


\bibitem[Subramonyam et~al\mbox{.}(2024)]%
        {subramonyam2024bridging}
\bibfield{author}{\bibinfo{person}{Hari Subramonyam}, \bibinfo{person}{Roy Pea}, \bibinfo{person}{Christopher Pondoc}, \bibinfo{person}{Maneesh Agrawala}, {and} \bibinfo{person}{Colleen Seifert}.} \bibinfo{year}{2024}\natexlab{}.
\newblock \showarticletitle{Bridging the {{Gulf}} of {{Envisioning}}: {{Cognitive Challenges}} in {{Prompt Based Interactions}} with {{LLMs}}}. In \bibinfo{booktitle}{\emph{Proceedings of the 2024 {{CHI Conference}} on {{Human Factors}} in {{Computing Systems}}}} \emph{(\bibinfo{series}{{{CHI}} '24})}. \bibinfo{publisher}{Association for Computing Machinery}, \bibinfo{address}{New York, NY, USA}, \bibinfo{pages}{1--19}.
\newblock
\showISBNx{979-8-4007-0330-0}
\href{https://doi.org/10.1145/3613904.3642754}{doi:\nolinkurl{10.1145/3613904.3642754}}


\bibitem[Tatsukawa et~al\mbox{.}(2025)]%
        {tatsukawa2025FontCraft}
\bibfield{author}{\bibinfo{person}{Yuki Tatsukawa}, \bibinfo{person}{I-Chao Shen}, \bibinfo{person}{Mustafa~Doga Dogan}, \bibinfo{person}{Anran Qi}, \bibinfo{person}{Yuki Koyama}, \bibinfo{person}{Ariel Shamir}, {and} \bibinfo{person}{Takeo Igarashi}.} \bibinfo{year}{2025}\natexlab{}.
\newblock \showarticletitle{{{FontCraft}}: {{Multimodal Font Design Using Interactive Bayesian Optimization}}}. In \bibinfo{booktitle}{\emph{Proceedings of the 2025 {{CHI Conference}} on {{Human Factors}} in {{Computing Systems}}}} \emph{(\bibinfo{series}{{{CHI}} '25})}. \bibinfo{publisher}{Association for Computing Machinery}, \bibinfo{address}{New York, NY, USA}, \bibinfo{pages}{1--14}.
\newblock
\showISBNx{979-8-4007-1394-1}
\href{https://doi.org/10.1145/3706598.3713863}{doi:\nolinkurl{10.1145/3706598.3713863}}


\bibitem[Tian et~al\mbox{.}(2024)]%
        {tian2024Visual}
\bibfield{author}{\bibinfo{person}{Keyu Tian}, \bibinfo{person}{Yi Jiang}, \bibinfo{person}{Zehuan Yuan}, \bibinfo{person}{Bingyue Peng}, {and} \bibinfo{person}{Liwei Wang}.} \bibinfo{year}{2024}\natexlab{}.
\newblock \showarticletitle{Visual {{Autoregressive Modeling}}: {{Scalable Image Generation}} via {{Next-Scale Prediction}}}.
\newblock \bibinfo{journal}{\emph{Advances in Neural Information Processing Systems}}  \bibinfo{volume}{37} (\bibinfo{year}{2024}), \bibinfo{pages}{84839--84865}.
\newblock


\bibitem[Vasconcelos et~al\mbox{.}(2023)]%
        {vasconcelos2023Explanations}
\bibfield{author}{\bibinfo{person}{Helena Vasconcelos}, \bibinfo{person}{Matthew J{\"o}rke}, \bibinfo{person}{Madeleine {Grunde-McLaughlin}}, \bibinfo{person}{Tobias Gerstenberg}, \bibinfo{person}{Michael~S. Bernstein}, {and} \bibinfo{person}{Ranjay Krishna}.} \bibinfo{year}{2023}\natexlab{}.
\newblock \showarticletitle{Explanations {{Can Reduce Overreliance}} on {{AI Systems During Decision-Making}}}.
\newblock \bibinfo{journal}{\emph{Proc. ACM Hum.-Comput. Interact.}} \bibinfo{volume}{7}, \bibinfo{number}{CSCW1} (\bibinfo{date}{April} \bibinfo{year}{2023}), \bibinfo{pages}{129:1--129:38}.
\newblock
\href{https://doi.org/10.1145/3579605}{doi:\nolinkurl{10.1145/3579605}}


\bibitem[Viappiani(2014)]%
        {viappiani2014Preference}
\bibfield{author}{\bibinfo{person}{Paolo Viappiani}.} \bibinfo{year}{2014}\natexlab{}.
\newblock \showarticletitle{Preference {{Modeling}} and {{Preference Elicitation}}: {{An Overview}}}. In \bibinfo{booktitle}{\emph{Proceedings of the {{First International Workshop}} on {{Decision Making}} and {{Recommender Systems}} ({{DMRS2014}})}}, Vol.~\bibinfo{volume}{1278}. \bibinfo{publisher}{CEUR Workshop Proceedings}, \bibinfo{address}{Bolzano, Italy}, \bibinfo{pages}{19--24}.
\newblock


\bibitem[Wang et~al\mbox{.}(2025)]%
        {wang2025Twin}
\bibfield{author}{\bibinfo{person}{Jianhui Wang}, \bibinfo{person}{Yangfan He}, \bibinfo{person}{Yan Zhong}, \bibinfo{person}{Xinyuan Song}, \bibinfo{person}{Jiayi Su}, \bibinfo{person}{Yuheng Feng}, \bibinfo{person}{Ruoyu Wang}, \bibinfo{person}{Hongyang He}, \bibinfo{person}{Wenyu Zhu}, \bibinfo{person}{Xinhang Yuan}, \bibinfo{person}{Miao Zhang}, \bibinfo{person}{Keqin Li}, \bibinfo{person}{Jiaqi Chen}, \bibinfo{person}{Tianyu Shi}, {and} \bibinfo{person}{Xueqian Wang}.} \bibinfo{year}{2025}\natexlab{}.
\newblock \showarticletitle{Twin {{Co-Adaptive Dialogue}} for {{Progressive Image Generation}}}. In \bibinfo{booktitle}{\emph{Proceedings of the 33rd {{ACM International Conference}} on {{Multimedia}}}} \emph{(\bibinfo{series}{{{MM}} '25})}. \bibinfo{publisher}{Association for Computing Machinery}, \bibinfo{address}{New York, NY, USA}, \bibinfo{pages}{3645--3653}.
\newblock
\showISBNx{979-8-4007-2035-2}
\href{https://doi.org/10.1145/3746027.3755141}{doi:\nolinkurl{10.1145/3746027.3755141}}


\bibitem[Wang et~al\mbox{.}(2024)]%
        {wang2024multilingual}
\bibfield{author}{\bibinfo{person}{Liang Wang}, \bibinfo{person}{Nan Yang}, \bibinfo{person}{Xiaolong Huang}, \bibinfo{person}{Linjun Yang}, \bibinfo{person}{Rangan Majumder}, {and} \bibinfo{person}{Furu Wei}.} \bibinfo{year}{2024}\natexlab{}.
\newblock \bibinfo{title}{Multilingual E5 Text Embeddings: A Technical Report}.
\newblock
\showeprint[arxiv]{2402.05672}~[cs.CL]
\urldef\tempurl%
\url{https://arxiv.org/abs/2402.05672}
\showURL{%
\tempurl}


\bibitem[Wang et~al\mbox{.}(2024)]%
        {wang2024PromptCharm}
\bibfield{author}{\bibinfo{person}{Zhijie Wang}, \bibinfo{person}{Yuheng Huang}, \bibinfo{person}{Da Song}, \bibinfo{person}{Lei Ma}, {and} \bibinfo{person}{Tianyi Zhang}.} \bibinfo{year}{2024}\natexlab{}.
\newblock \showarticletitle{{{PromptCharm}}: {{Text-to-Image Generation}} through {{Multi-modal Prompting}} and {{Refinement}}}. In \bibinfo{booktitle}{\emph{Proceedings of the 2024 {{CHI Conference}} on {{Human Factors}} in {{Computing Systems}}}} (New York, NY, USA, 2024-05-11) \emph{(\bibinfo{series}{{{CHI}} '24})}. \bibinfo{publisher}{Association for Computing Machinery}, \bibinfo{pages}{1--21}.
\newblock
\showISBNx{979-8-4007-0330-0}
\href{https://doi.org/10.1145/3613904.3642803}{doi:\nolinkurl{10.1145/3613904.3642803}}


\bibitem[Wang et~al\mbox{.}(2023)]%
        {wang2023DiffusionDB}
\bibfield{author}{\bibinfo{person}{Zijie~J. Wang}, \bibinfo{person}{Evan Montoya}, \bibinfo{person}{David Munechika}, \bibinfo{person}{Haoyang Yang}, \bibinfo{person}{Benjamin Hoover}, {and} \bibinfo{person}{Duen~Horng Chau}.} \bibinfo{year}{2023}\natexlab{}.
\newblock \showarticletitle{{{DiffusionDB}}: {{A Large-scale Prompt Gallery Dataset}} for {{Text-to-Image Generative Models}}}. In \bibinfo{booktitle}{\emph{Proceedings of the 61st {{Annual Meeting}} of the {{Association}} for {{Computational Linguistics}} ({{Volume}} 1: {{Long Papers}})}}. \bibinfo{publisher}{Association for Computational Linguistics}, \bibinfo{address}{Toronto, Canada}, \bibinfo{pages}{893--911}.
\newblock
\href{https://doi.org/10.18653/v1/2023.acl-long.51}{doi:\nolinkurl{10.18653/v1/2023.acl-long.51}}


\bibitem[Wei et~al\mbox{.}(2022)]%
        {wei2022Chainofthought}
\bibfield{author}{\bibinfo{person}{Jason Wei}, \bibinfo{person}{Xuezhi Wang}, \bibinfo{person}{Dale Schuurmans}, \bibinfo{person}{Maarten Bosma}, \bibinfo{person}{brian ichter}, \bibinfo{person}{Fei Xia}, \bibinfo{person}{Ed Chi}, \bibinfo{person}{Quoc~V Le}, {and} \bibinfo{person}{Denny Zhou}.} \bibinfo{year}{2022}\natexlab{}.
\newblock \showarticletitle{Chain-of-Thought Prompting Elicits Reasoning in Large Language Models}.
\newblock \bibinfo{journal}{\emph{Advances in Neural Information Processing Systems}}  \bibinfo{volume}{35} (\bibinfo{year}{2022}), \bibinfo{pages}{24824--24837}.
\newblock


\bibitem[Xie et~al\mbox{.}(2023)]%
        {xie2023Prompt}
\bibfield{author}{\bibinfo{person}{Yutong Xie}, \bibinfo{person}{Zhaoying Pan}, \bibinfo{person}{Jinge Ma}, \bibinfo{person}{Luo Jie}, {and} \bibinfo{person}{Qiaozhu Mei}.} \bibinfo{year}{2023}\natexlab{}.
\newblock \showarticletitle{A {{Prompt Log Analysis}} of {{Text-to-Image Generation Systems}}}. In \bibinfo{booktitle}{\emph{Proceedings of the {{ACM Web Conference}} 2023}} \emph{(\bibinfo{series}{{{WWW}} '23})}. \bibinfo{publisher}{Association for Computing Machinery}, \bibinfo{address}{New York, NY, USA}, \bibinfo{pages}{3892--3902}.
\newblock
\showISBNx{978-1-4503-9416-1}
\href{https://doi.org/10.1145/3543507.3587430}{doi:\nolinkurl{10.1145/3543507.3587430}}


\bibitem[Xu et~al\mbox{.}(2023)]%
        {xu2023ImageReward}
\bibfield{author}{\bibinfo{person}{Jiazheng Xu}, \bibinfo{person}{Xiao Liu}, \bibinfo{person}{Yuchen Wu}, \bibinfo{person}{Yuxuan Tong}, \bibinfo{person}{Qinkai Li}, \bibinfo{person}{Ming Ding}, \bibinfo{person}{Jie Tang}, {and} \bibinfo{person}{Yuxiao Dong}.} \bibinfo{year}{2023}\natexlab{}.
\newblock \showarticletitle{{{ImageReward}}: {{Learning}} and {{Evaluating Human Preferences}} for {{Text-to-Image Generation}}}.
\newblock \bibinfo{journal}{\emph{Advances in Neural Information Processing Systems}}  \bibinfo{volume}{36} (\bibinfo{year}{2023}), \bibinfo{pages}{15903--15935}.
\newblock


\bibitem[Yamamoto et~al\mbox{.}(2022)]%
        {yamamoto2022Photographic}
\bibfield{author}{\bibinfo{person}{Kenta Yamamoto}, \bibinfo{person}{Yuki Koyama}, {and} \bibinfo{person}{Yoichi Ochiai}.} \bibinfo{year}{2022}\natexlab{}.
\newblock \showarticletitle{Photographic {{Lighting Design}} with {{Photographer-in-the-Loop Bayesian Optimization}}}. In \bibinfo{booktitle}{\emph{Proceedings of the 35th {{Annual ACM Symposium}} on {{User Interface Software}} and {{Technology}}}} \emph{(\bibinfo{series}{{{UIST}} '22})}. \bibinfo{publisher}{Association for Computing Machinery}, \bibinfo{address}{New York, NY, USA}, \bibinfo{pages}{1--11}.
\newblock
\showISBNx{978-1-4503-9320-1}
\href{https://doi.org/10.1145/3526113.3545690}{doi:\nolinkurl{10.1145/3526113.3545690}}


\bibitem[Zamfirescu-Pereira et~al\mbox{.}(2023)]%
        {zamfirescu2023johnny}
\bibfield{author}{\bibinfo{person}{J.D. Zamfirescu-Pereira}, \bibinfo{person}{Richmond~Y. Wong}, \bibinfo{person}{Bjoern Hartmann}, {and} \bibinfo{person}{Qian Yang}.} \bibinfo{year}{2023}\natexlab{}.
\newblock \showarticletitle{Why {Johnny} Can't Prompt: How Non-{AI} Experts Try (and Fail) to Design {LLM} Prompts}. In \bibinfo{booktitle}{\emph{Proceedings of the 2023 CHI Conference on Human Factors in Computing Systems}} (Hamburg, Germany) \emph{(\bibinfo{series}{CHI '23})}. \bibinfo{publisher}{Association for Computing Machinery}, \bibinfo{address}{New York, NY, USA}, Article \bibinfo{articleno}{437}, \bibinfo{numpages}{21}~pages.
\newblock
\showISBNx{9781450394215}
\href{https://doi.org/10.1145/3544548.3581388}{doi:\nolinkurl{10.1145/3544548.3581388}}


\bibitem[Zeng et~al\mbox{.}(2024)]%
        {zeng2024IntentTuner}
\bibfield{author}{\bibinfo{person}{Xingchen Zeng}, \bibinfo{person}{Ziyao Gao}, \bibinfo{person}{Yilin Ye}, {and} \bibinfo{person}{Wei Zeng}.} \bibinfo{year}{2024}\natexlab{}.
\newblock \showarticletitle{{{IntentTuner}}: {{An Interactive Framework}} for {{Integrating Human Intentions}} in {{Fine-tuning Text-to-Image Generative Models}}}. In \bibinfo{booktitle}{\emph{Proceedings of the 2024 {{CHI Conference}} on {{Human Factors}} in {{Computing Systems}}}} \emph{(\bibinfo{series}{{{CHI}} '24})}. \bibinfo{publisher}{Association for Computing Machinery}, \bibinfo{address}{New York, NY, USA}, \bibinfo{pages}{1--18}.
\newblock
\showISBNx{979-8-4007-0330-0}
\href{https://doi.org/10.1145/3613904.3642165}{doi:\nolinkurl{10.1145/3613904.3642165}}


\bibitem[Zhou et~al\mbox{.}(2024)]%
        {zhou24StyleFactory}
\bibfield{author}{\bibinfo{person}{Mingxu Zhou}, \bibinfo{person}{Dengming Zhang}, \bibinfo{person}{Weitao You}, \bibinfo{person}{Ziqi Yu}, \bibinfo{person}{Yifei Wu}, \bibinfo{person}{Chenghao Pan}, \bibinfo{person}{Huiting Liu}, \bibinfo{person}{Tianyu Lao}, {and} \bibinfo{person}{Pei Chen}.} \bibinfo{year}{2024}\natexlab{}.
\newblock \showarticletitle{StyleFactory: Towards Better Style Alignment in Image Creation through Style-Strength-Based Control and Evaluation}. In \bibinfo{booktitle}{\emph{Proceedings of the 37th Annual ACM Symposium on User Interface Software and Technology}} (Pittsburgh, PA, USA) \emph{(\bibinfo{series}{UIST '24})}. \bibinfo{publisher}{Association for Computing Machinery}, \bibinfo{address}{New York, NY, USA}, Article \bibinfo{articleno}{107}, \bibinfo{numpages}{15}~pages.
\newblock
\showISBNx{9798400706288}
\href{https://doi.org/10.1145/3654777.3676370}{doi:\nolinkurl{10.1145/3654777.3676370}}


\end{thebibliography}

\appendix

\section{User Study Details}
\subsection{Design Tasks}

The eight text-to-image generation scenarios used in the user study are summarized in Table~\ref{tab:tasks}.

\begin{table*}[t]
\centering
\renewcommand{\arraystretch}{1.45}
\caption{Text-to-image generation scenarios used in the user study.}
\label{tab:tasks}
\begin{tabular}{p{5cm} p{\linewidth-5cm}}
\toprule
\textbf{Scenario} & \textbf{Description} \\
\midrule
Interior Design & Imagine and create an image of your ideal room. \\
Painting & Visualize your dream travel destination as a painting. \\
Photography & Imagine and generate a photo of a dish or drink you would love to have right now. \\
App Icon & Create an icon for an app you use or would like to invent. \\
Poster Design & Design a poster for an event or celebration. \\
Logo Design & Design a logo for a project, team, or group. \\
Fashion Design & Imagine and design an outfit you would like to wear. \\
Architectural Style & Visualize a building you would like to visit. \\
\bottomrule
\end{tabular}
\end{table*}

\subsection{Pre-Study Questionnaire}

Before starting the design tasks, participants completed a pre-study questionnaire to gather information about their background and experience with AI tools and design activities. The participants' response are summarized in Table~\ref{tab:pre}.

\renewcommand{\arraystretch}{1.45}
\begin{table*}[htbp]
\centering
\caption{Frequency of AI tool usage and design activities.}
\label{tab:pre}
\begin{tabularx}{\linewidth}{p{12em} p{20em} p{12em} c
}
\toprule
\textbf{Theme} & \textbf{Question} & \textbf{Usage Frequency} & \textbf{Count} \\
\midrule

\multirow{5}{*}{LLM Usage} 
& \multirow{5}{=}{How often do you use large language models (LLMs) such as ChatGPT, Claude, or DeepSeek?}
& Never & 2 \\
& & Less than once a month & 3 \\
& & 1--3 times per month & 11 \\
& & 1--6 times per week & 42 \\
& & Daily or almost daily & 70 \\

\midrule
\multirow{5}{*}{Text-to-Image Generation} 
& \multirow{5}{=}{What is your experience with text-to-image generation (e.g., Midjourney, Stable Diffusion, DALL·E, FLUX)?}
& Never & 23 \\
& & I've tried once or twice & 34 \\
& & 1--3 times per month & 36 \\
& & 1--6 times per week & 26 \\
& & Daily or almost daily & 9 \\

\midrule
\multirow{6}{*}{Design Activities} 
& \multirow{6}{=}{How often do you engage in design-related activities (graphic design, visual arts, UX/UI)?}
& Never & 21 \\
& & Once a year or less & 24 \\
& & A few times a year & 31 \\
& & Weekly & 15 \\
& & Monthly & 25 \\
& & Daily or almost daily & 12 \\

\bottomrule
\end{tabularx}
\end{table*}

\subsection{Post-Study Questionnaire}
In this section, we detail the design of post-study questionnaires and participants' response.

\subsubsection{Alignment Questionnaire}
\label{sec:alignment_ques}

Participants rated their agreement with the following statements on a 7-point Likert scale:

\begin{figure}[h]
\centering
\includegraphics[width=\linewidth]{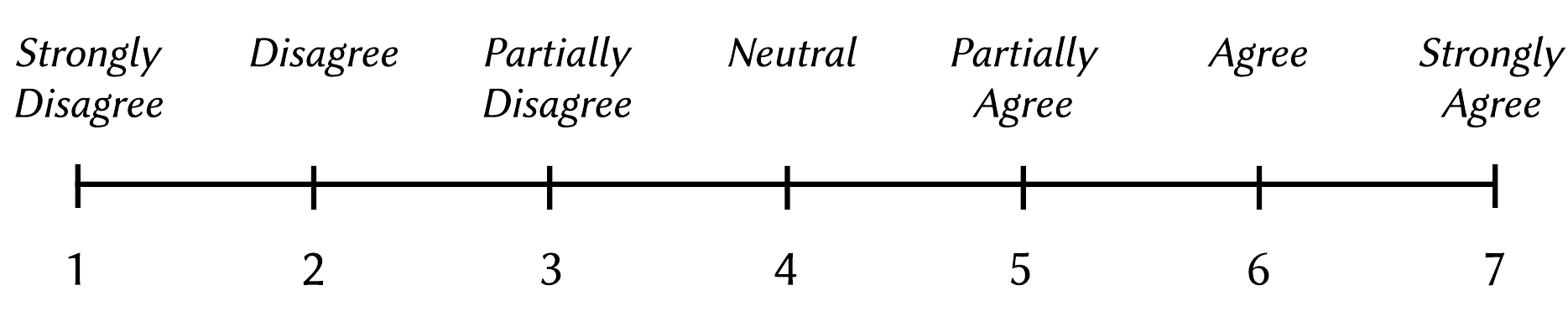}
\label{fig:likert}
\Description{Horizontal 7-point Likert scale with numeric labels from 1 to 7 marked by evenly spaced tick marks. Verbal anchors are shown above the scale: Strongly Disagree (1), Disagree (2), Partially Disagree (3), Neutral (4), Partially Agree (5), Agree (6), and Strongly Agree (7).}
\end{figure}

\textbf{Capability Gap}
\begin{itemize}
    \item[+] ``The system supported me in working out how to generate the best possible image based on what I wanted.''
    \item[--] ``The system led me away from what I wanted, and the image did not match what I had aimed for.''
\end{itemize}

\textbf{Instruction Gap}
\begin{itemize}
    \item[+] ``The system helped me provide the right input to generate the image as I intended.''  
    \item[--] ``I was disappointed by how the system interpreted my input when generating the image.''  
\end{itemize}

\textbf{Intentionality Gap}
\begin{itemize}
    \item[+] ``The system helped me clarify what I wanted from the image.''  
    \item[--] ``I was unsure how (or if) the system could generate an image more in line with my intention.''  
\end{itemize}

\textbf{Envision Overall}
\begin{itemize}
    \item[+] ``The system made me think of a wider range of possible images that fit my goal.''  
    \item[--] ``The system made it harder for me to imagine alternative images that might better fit my goal.''  
\end{itemize}

\subsubsection{Post-Study Questionnaire Results}

\begin{figure*}[htb]
\centering
\includegraphics[width=\linewidth]{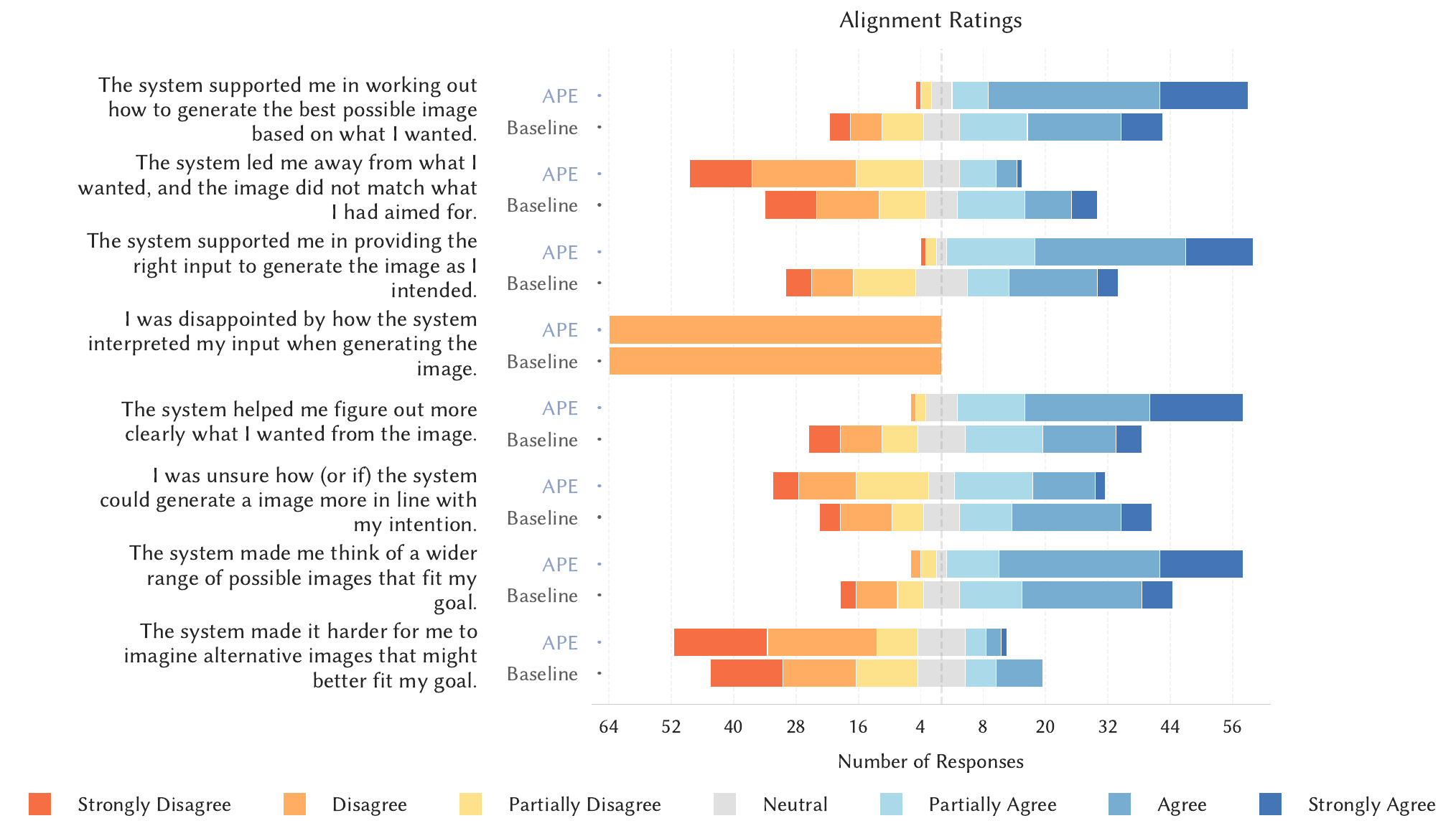}
\caption{Alignment ratings by participants}
\label{fig:ge}
\Description{Diverging horizontal bar chart showing response distributions for eight alignment statements (7-point Likert scale) comparing Baseline and APE (n=64 each). Negative ratings extend left from center; positive ratings extend right. Color progresses from red (Strongly Disagree) through yellow (neutral) to blue (Strongly Agree). For positive statements, APE shows notably more responses in "Agree/Strongly Agree" categories, particularly for "system supported me in providing right input" and "helped me clarify what I wanted." For negative statements, pattern reverses: APE shows more "Strongly Disagree/Disagree" responses, indicating users did not experience problems, while Baseline shows more neutral and agreement. Visualization confirms APE's consistent advantage across all alignment dimensions.}
\end{figure*}

\begin{figure*}[htb]
\centering
\includegraphics[width=0.9\linewidth]{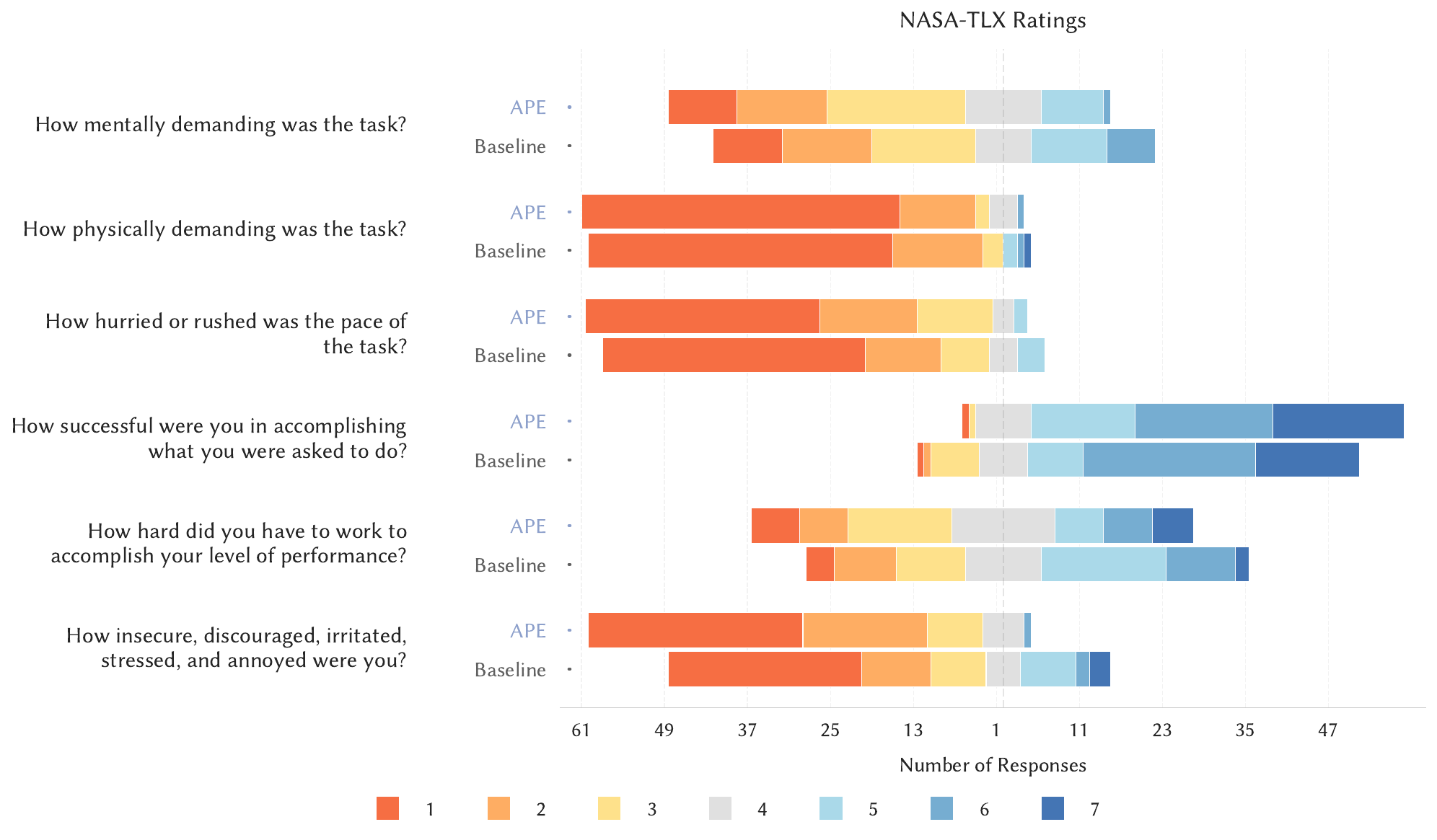}
\caption{NASA TLX ratings by participants}
\label{fig:nasa}
\Description{Diverging horizontal bar chart showing NASA-TLX response distributions (7-point scales) for six workload dimensions, comparing Baseline and APE (n=64 each). Color ranges from red (low rating) through yellow (neutral) to blue (high). Mental Demand: both show moderate demand (2-4 range), similar distributions. Physical Demand: both very low (concentrated at rating 1). Temporal Demand: nearly identical, concentrated at low levels. Performance: both show high perceived success (ratings 5-7), APE slightly higher. Effort: similar moderate distributions (3-5 range). Frustration: APE shows notably more low-frustration responses (ratings 1-2) versus Baseline's mid-range responses, consistent with 27\% lower frustration finding. Overall distributions confirm APE achieves alignment benefits without increasing cognitive workload.}
\end{figure*}

Figure~\ref{fig:ge} shows the distribution of user ratings in the alignment questionnaire.
Figure~\ref{fig:nasa} shows the distribution of user ratings in the NASA-TLX questionnaire.

\section{Generated Images}

\subsection{Technical Evaluation}

\begin{figure*}[htbp]
\centering
\includegraphics[width=\linewidth]{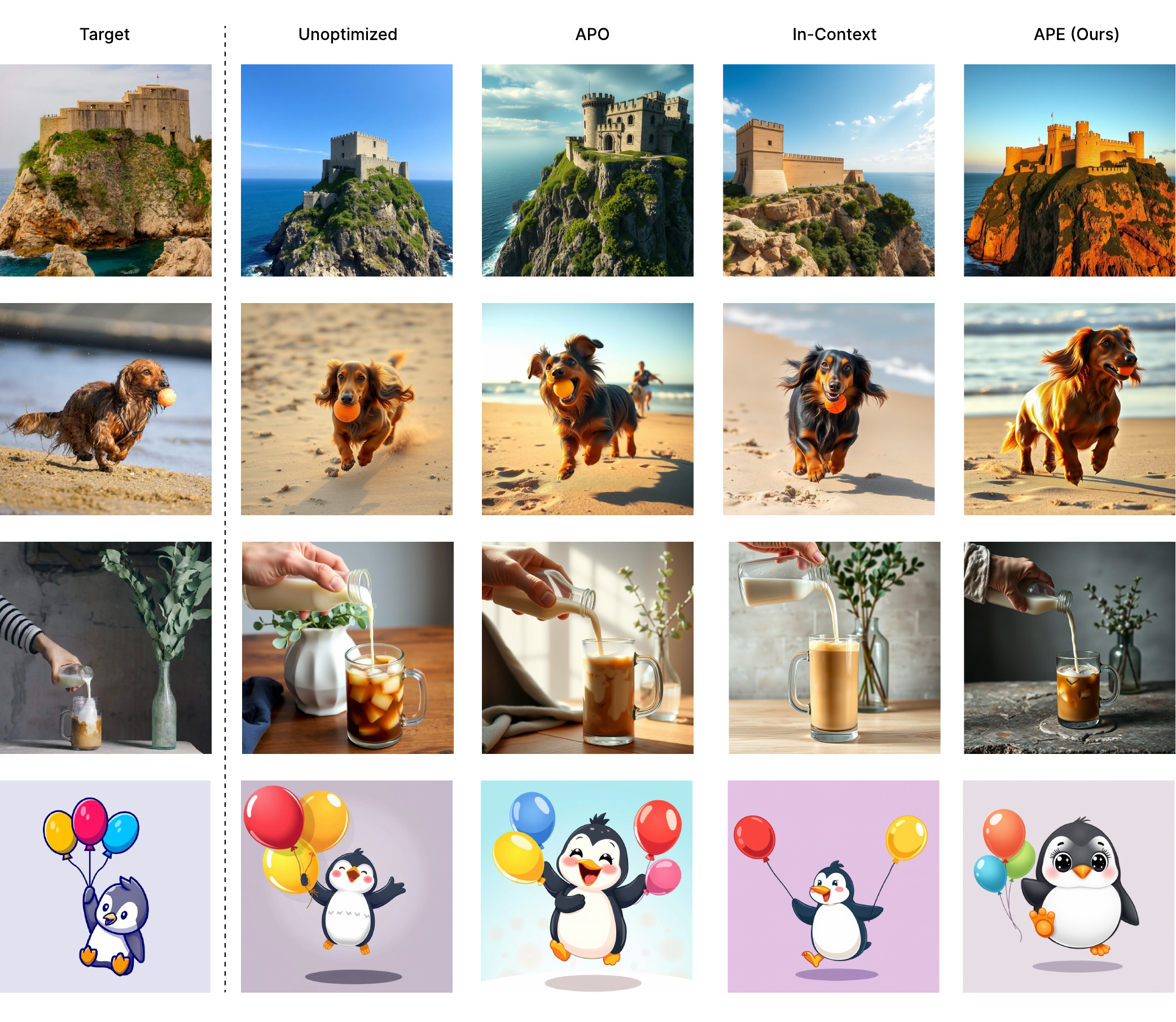}
\caption{Examples of resulting images across methods in the technical evaluation. For each row, we show the target reference image (left) followed by outputs from four approaches: Unoptimized, APO, In-Context Query, and APE (our method). APE consistently produces images with stronger visual alignment to target references, capturing key features such as \textbf{composition and color palette} (row 1), \textbf{perspective and motion dynamics} (row 2), \textbf{tonal atmosphere and spatial arrangement} (row 3), and \textbf{pose} (row 4).}
\label{fig:designbench_expamples}
\Description{4×5 grid comparing image quality across methods. Each row shows one test case with columns: Target reference, Unoptimized, APO, In-Context, and APE outputs. Row 1 (coastal fortress): APE successfully reproduces dramatic cliff positioning, warm orange lighting, and blue sea matching target atmosphere, while other methods lose key tonal or compositional elements. Row 2 (running dog): APE accurately captures low-angle perspective, running motion with dynamic fur, and beach environment; others show incorrect postures. Row 3 (pouring coffee): APE matches cool color palette, minimalist composition, and plant element; others have incorrect lighting or styling. Row 4 (penguin): APE reproduces flat 2D illustration style with correct balloon arrangement; others create 3D renderings or wrong details. Demonstrates APE's consistent ability to capture composition, perspective, color, style, and subject details that other methods miss.}
\end{figure*}

Some examples of resulting images in the technical evaluation are shown in Figure~\ref{fig:designbench_expamples}.

\subsection{User Evaluation}
\label{sec:user_eval_examples}
\begin{figure*}[htbp]
\centering
\includegraphics[width=\linewidth]{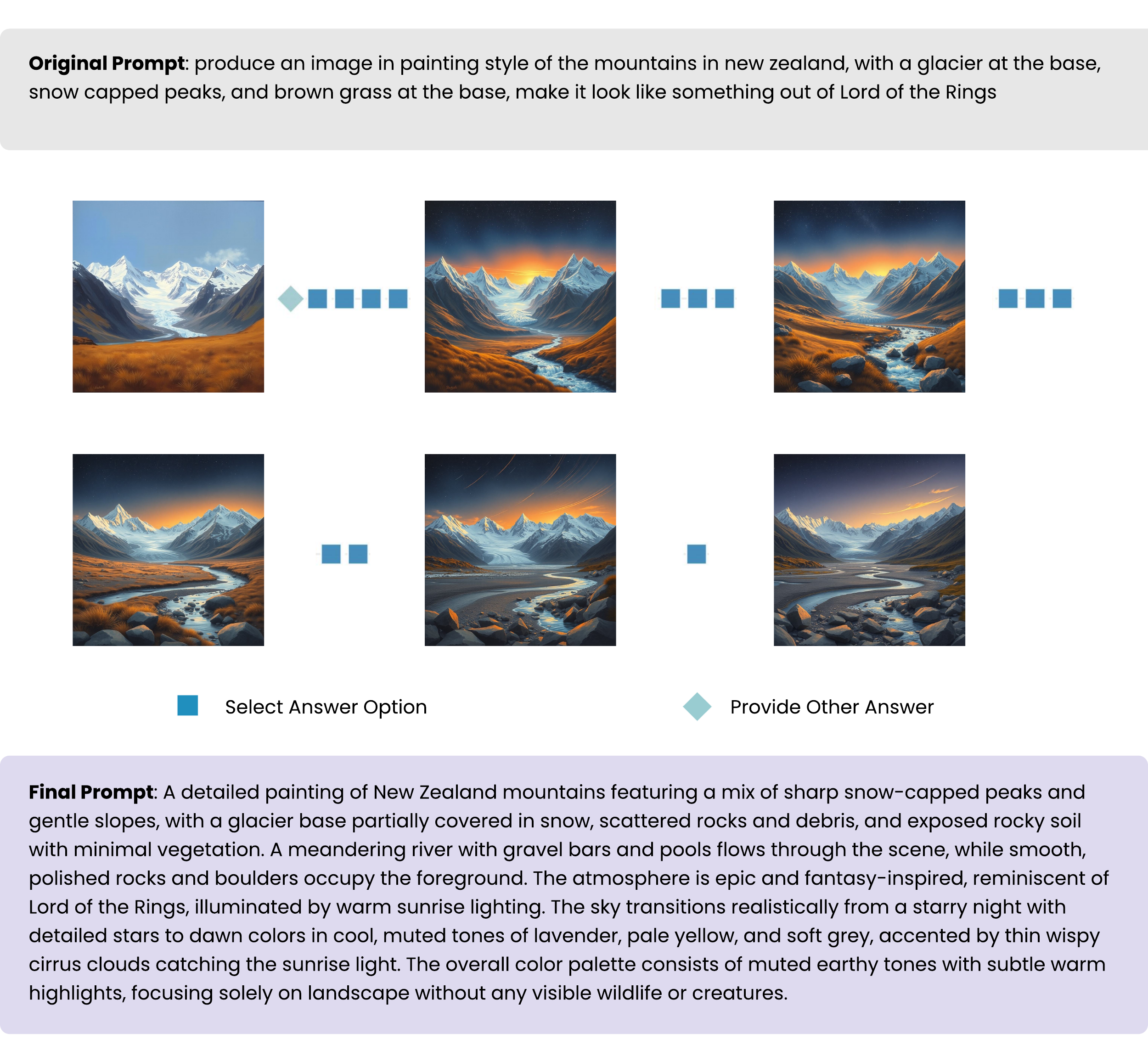}
\caption{Examples of generated images from an APE participant. The original prompt (top) provides a general theme; the user progressively clarifies their intent through visual queries (square: selected options; diamond: custom answer) between generations (I1–I6). Requirements accumulate from foundational elements (I2: sky, mountains, water) to compositional details (I3: vegetation, foreground) to aesthetic refinements and fine-grained features (I4–I6: color palette, atmospheric conditions, water features, vegetation details, glacier appearance). The visual progression demonstrates convergence from an underspecified concept to a detailed, aligned output, illustrating how structured elicitation scaffolds preference articulation.}
\label{fig:ex_ape_1}
\Description{Complete APE session for Lord of the Rings-inspired New Zealand mountain landscape. Original 38-word prompt establishes basic theme (mountains, glacier, snow peaks, LOTR style). Five iterations (I1-I6) show progressive refinement with visual query responses marked by blue squares (selections) and diamonds (custom answers). Features accumulate: I2 sets foundational elements (sunrise, mixed peaks, river); I3 sets compositional details (minimal foreground, polished rocks); I4 refines vegetation and materials (sparse tussock grass, muted earthy palette); I5-I6 specifies atmosphere and fine-grained features (cool sky tones with wispy clouds, glacier appearance, meandering river with gravel bars and pools, exposed rocky soil with minimal plant life). Visual progression shows images evolving from bright midday to moody dawn fantasy atmosphere. Final 132-word prompt integrates all specifications.}
\end{figure*}

\begin{figure*}[htbp]
\centering
\includegraphics[width=\linewidth]{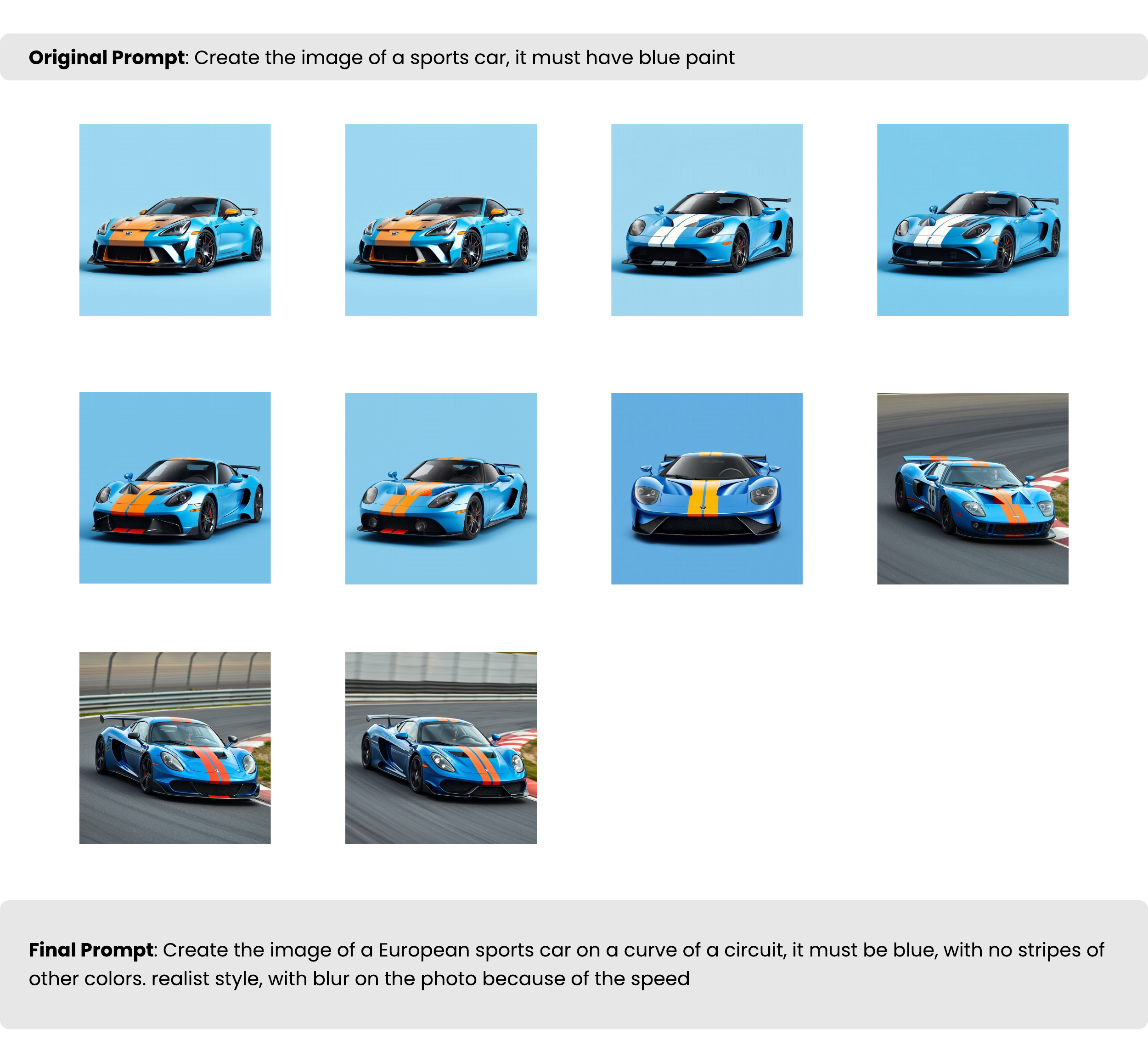}
\caption{Examples of generated images from a Baseline participant. Across 10 generation attempts, the participant struggled to eliminate unwanted orange stripes despite progressively explicit constraints (``no stripes of other colors''). The final prompt grew from 12 to 35 words through iterative additions of negative specifications and contextual details (``European,'' ``curve of a circuit,'' ``blur because of the speed''), yet failed to approach the intended blue-only design. This illustrates the inefficiency of manual refinement: users must discover through trial which constraints matter, with no systematic guidance on resolving persistent misalignments.}
\label{fig:ex_baseline_1}
\Description{Ten-iteration Baseline condition sequence showing persistent struggle to create blue-only sports car. Original 12-word prompt: "Create the image of a sports car, it must have blue paint." All ten generated images show blue cars with unwanted orange racing stripes despite participant's intent for pure blue. Refinement attempts: images 1-4 try plain backgrounds; images 5-7 add explicit negative constraint "no stripes of other colors" but stripes persist; images 8-10 add circuit background, motion blur, and realism descriptors, but orange stripes remain. Final 35-word prompt includes "European sports car on circuit curve, must be blue, with no stripes of other colors, realist style, with blur" yet never achieves blue-only design. Participant feedback: "I keep adding [to the] prompt, but the generated image is not getting better." Illustrates manual prompting limitations: lack of systematic guidance, inability to diagnose why constraints ignored, inefficient trial-and-error, no guarantee detail improves alignment.}
\end{figure*}

Examples of generated images from participants in APE and Baseline conditions are shown in Figure~\ref{fig:ex_ape_1} and Figure~\ref{fig:ex_baseline_1}, respectively.

\end{document}